\documentclass[journal,twoside]{IEEEtran}
\usepackage{cite}
\usepackage{amsmath,amssymb,amsfonts}

\usepackage{amsthm} 
\usepackage{graphicx}
\usepackage{algorithm}
\usepackage{algpseudocode}
\usepackage{url}

\def\BibTeX{{\rm B\kern-.05em{\sc i\kern-.025em b}\kern-.08em
    T\kern-.1667em\lower.7ex\hbox{E}\kern-.125emX}}
\usepackage{lipsum}
\usepackage{subcaption}

\newtheorem{remark}{Remark}

\usepackage{bm}
\usepackage{physics}

\newcommand{\qdot}{\Dot{q}}
\newcommand{\state}{\bm{x}}

\theoremstyle{definition}

\begin{document}
\title{Data-Driven Adaptive PID Control Based on Physics-Informed Neural Networks}

\author{Junsei~Ito, and Yasuaki~Wasa, \IEEEmembership{Member, IEEE}
\thanks{This work was supported in part by JST ACT-X, Japan, Grant No. JPMJAX25C3 and Waseda University Grant for Special Research Project No. 2024C-484.}
\thanks{J. Ito is with the Department of Electrical Engineering and Bioscience, Waseda University, Tokyo, 169-8555 Japan (e-mail: distinction0625@moegi.waseda.jp).}
\thanks{Y. Wasa is with the Department of Electrical Engineering and Bioscience, Waseda University, Tokyo, 169-8555 Japan (e-mail: wasa@waseda.jp).}}

\maketitle

\IEEEpubid{\raisebox{-7mm}{This work has been submitted to the IEEE Transactions on Control Systems Technology for possible publication.}}
\IEEEpubidadjcol

\begin{abstract}
This article proposes a data-driven PID controller design based on the principle of adaptive gain optimization, leveraging Physics-Informed Neural Networks (PINNs) generated for predictive modeling purposes.
The proposed control design method utilizes gradients of the PID gain optimization, achieved through the automatic differentiation of PINNs, to apply model predictive control using a cost function based on tracking error and control inputs. 
By optimizing PINNs-based PID gains, the method achieves adaptive gain tuning that ensures stability while accounting for system nonlinearities. 
The proposed method features a systematic framework for integrating PINNs-based models of dynamical control systems into closed-loop control systems, enabling direct application to PID control design.
A series of numerical experiments is conducted to demonstrate the effectiveness of the proposed method from the control perspectives based on both time and frequency domains.
\end{abstract}


\section{Introduction}
\label{sec:introduction}

\IEEEPARstart{D}{ata-driven} control is a powerful approach for guaranteeing stabilization and performance of complex systems in various engineering applications \cite{SI_CSM23}. 
One of the promising data-driven control methods is model-based tuning to approach a desired reference dynamical system such as virtual reference feedback tuning (VRFT) \cite{CLS02} and fictitious reference iterative tuning (FRIT) \cite{SKF04}, and data-driven tuning such as iterative feedback \cite{HGGL98}, real-time optimization and reinforcement learning (see \cite{SI_CSM23} and the reference therein). 
Designing and tuning a proportional-integral-derivative (PID) controller, known as a traditional data-driven control design method, is widely utilized in real-world applications \cite{HCL05,AH06,AH95}.
However, there are few studies on PID control design for multiple-input multiple-output (MIMO) systems and complex nonlinear systems.

Model predictive control (MPC) has become widely accepted in industrial applications as a real-time optimization-based control design method for MIMO systems, and the MPC theory has been studied \cite{SABA21,HWMZ20}.
The primary problem in achieving theoretically guaranteed control performance of the MPC is to obtain a high-accuracy, approximate dynamical model that includes nonlinearity and quantifies uncertainty from a time-series dataset. 
In other words, it is necessary to establish a novel data-driven model for prediction and control using the limited data available \cite{SABA21}.
Coulson et al. \cite{CLD19} present a model-free, data-driven MPC approach for unknown linear time-invariant (LTI) systems. 
Still, it is difficult to apply the method presented in \cite{CLD19} directly to general nonlinear systems.
To overcome the issue, physics-informed machine learning (PIML) is being utilized in the physics fields, such as computational fluid dynamics \cite{VB22} and materials science \cite{RMN23}.
However, as the dynamical model in physics is an autonomous system without external control input, it is challenging to apply these techniques directly to control systems.
In the systems and control field, the recent survey of PIML can be found in the tutorial paper \cite{ACC2023physics}.
Almost all PIML control methods, including \cite{ACC2023physics} and the references therein, employ kernel method-based modeling with Gaussian processes \cite{EH24,Hatanaka23} or a neural network approximated controller that satisfies specific control performance \cite{ASF24}.
The initial results of nonlinear MPC using neural networks are summarized in \cite{PSJG00}.

While many papers \cite{ACC2023physics,EH24,Hatanaka23,ASF24,PSJG00} utilize machine learning (ML) methods to obtain control strategies directly, this paper successfully incorporates ML methods into the dynamical model, incorporating information with physical meaning, to get the standard MPC-based control design.
Moriyasu et al. \cite{MKK25} propose a learning method for exactly linearizable dynamical models by using the structural similarity of a structured Hammerstein–Wiener model. 
The model-less or model-free approach to obtaining a highly accurate dynamical model generally requires a massive training dataset. 
Meanwhile, preparing the massive training dataset in real systems disturbs practical implementation.
To overcome the implementation issue of preparing a massive training dataset in the robotics and mechatronics fields, the concept of Sim2Real, based on digital twin technologies, has been realized \cite{Hofer21}.
However, a serious technical problem remains: the theory-practice (model-data) gap, also known as the reality gap.
Overall, to obtain the controller guaranteeing the theoretical control performance, there is a trade-off and dilemma between model matching, such as first principles modeling and reference model \cite{CLS02,SKF04}, and data matching based on machine learning techniques and black-box modeling to deal with the uncertainty and complexity of real systems.

To efficiently reduce the theory-practice (model-data) gap with limited data available, one promising approach is to utilize Physics-Informed Neural Networks (PINNs) \cite{MR19}. 
PINNs offer a framework for integrating the information of the nominal physical laws described by (partial/ordinary) differential equations, including state-space models and Lagrangian-based energy models, into neural networks, bridging the gap between traditional physics-based mathematical models and purely data-driven approaches.
Therefore, to grasp the hidden nonlinearity and uncertainty of real systems, PINNs integrate the data-enabled gap between known physical laws and neural network models into the loss functions by using the automatic differentiation of neural networks.
Following the pioneering paper \cite{MR19}, modeling using PINNs has garnered significant attention in the fields of computational science and related interdisciplinary fields, including physics, materials science, biomedical applications, and control engineering (see \cite{TOVZDWK24} and the references therein). 
As Brunton et al. \cite{BZKF25} pointed out, one of the future issues is to obtain an effective control strategy using automatic differentiation of ML models for nonlinear MPC. 
Liu et al. \cite{LBK24} present a PINNs model for Lagrangian dynamics-based robotic systems and control.
Liu et al. \cite{LZHLXZ24} propose MPC of an autonomous underwater vehicle with control barrier functions (CBF) by using PINN-based dynamics.
Drgo\v na et al. \cite{Drgona22b} present PINN-based MPC for buildings. 
However, many papers \cite{TOVZDWK24,LBK24,LZHLXZ24,Drgona22b} handle autonomous systems without external control input or a dynamical control system with a fixed control policy. 
Therefore, there remains the practical issue of utilizing the PINNs for optimization-based control and standard control design.

To address the aforementioned challenges, this paper proposes a data-driven approach to implement an adaptive PID control using PINNs. 
The MPC-based utilization of the PINNs is presented in \cite{NKFB22}. 
Nicodemus et al. \cite{NKFB22} present an MPC with PINNs-based dynamical model for a two-degrees-of-freedom (2-DOF) robot manipulator system. 
Meanwhile, although PID control is also a significant demand for industrial applications \cite{HCL05,AH06,AH95}, PINNs-based PID control design is not established to the best of our knowledge. 
Therefore, this paper presents an adaptive PID gain optimization in the context of MPC, leveraging compatibility with the gradient descent algorithm in optimization and the automatic differentiation of neural networks. 
Theoretical stability analysis of fixed PID control gains can be found in \cite{AM84,Alvarez00}. 
From the viewpoint of PID control design, the optimization-based approach of PID control can be found in \cite{AH06}.
The proposed PINNs-based adaptive PID control design can adaptively address nonlinearities in dynamical systems without requiring explicit feedforward compensation.
The typical gain tuning control methods are known as adaptive control \cite{Anu23}, based on iterative learning rules for unpredictable variations, and gain-scheduled control \cite{AH95}, based on scheduling variables for predictable variations. 
The proposed control design is similar to adaptive control in its ability to handle unpredictable variations and is characterized by ease of design.
With reference to these traditional control designs, this paper analyzes the control design performance of the proposed method in standard control fields, examining it from both the time and frequency domains.
Evaluating the control design performance, such as the stability margin, is not considered in machine learning fields \cite{MR19,TOVZDWK24,LBK24,LZHLXZ24,Drgona22b} and the previous MPC-based utilization \cite{NKFB22}.

The technical contributions of this paper are summarized as follows:
\begin{itemize}
\item A systematic framework for integrating PINNs-based models of dynamical control systems into closed-loop control systems is established, enabling direct application to PID control design;
\item An adaptive PID gain tuning method for solving a general nonlinear optimal control problem through automatic differentiation of PINNs is proposed;
\item To solve an optimal control problem with a long-term prediction horizon while guaranteeing a high-accuracy PINNs-based model, an MPC with the control input restricted to a sampled signal with zero-order hold (ZOH) is implemented, as outlined in \cite{NKFB22,AC24}; 
\item From the control perspectives based on both time and frequency domains, the quantitative effectiveness of stability, convergence, and robustness of the proposed control is evaluated through simulations with two typical case studies: 2-DOF manipulator systems as nonlinear dynamical systems and mass-spring-damper (MSD) systems as linear systems.
\end{itemize}

The rest of the paper is structured as follows.
Section~\ref{sec:02} reviews PINNs for control \cite{AC24} and its MPC-based implementation \cite{NKFB22}. 
In Section~\ref{sec:03}, we propose an adaptive PID control led by the optimization problem with a PINNs-based dynamical model and show the implementation of the proposed PID control. 
In Sections~\ref{sec:05} and \ref{sec:06}, we demonstrate the effectiveness of the proposed control performance through simulations with two typical case studies of the 2-DOF manipulator systems and the MSD systems. 
Section~\ref{sec:07} concludes the paper.

\section{Preliminaries}
\label{sec:02}

\subsection{Original Control Problem}

In this paper, we consider a continuous-time dynamical control system during the time interval $\mathbb{T}:=[0,t_f]$, $t_f>0$, 
\begin{equation}
\dot{\boldsymbol{x}}(t) = \boldsymbol{f}(\boldsymbol{x}(t),\boldsymbol{u}(t)), \ t\in\mathbb{T}, \quad \boldsymbol{x}(0) = \boldsymbol{x}_0, 
\label{eq:ode_control}
\end{equation}
with the state $\boldsymbol{x}(t) \in \mathbb{X}\subseteq \mathbb{R}^n$ and the control input $\boldsymbol{u}(t) \in \mathbb{U} \subset \mathbb{R}^m$. 
We suppose that the dynamics (\ref{eq:ode_control}) represent a nominal (baseline) model, and that the function $\boldsymbol{f}$ is known and linear with respect to the control variable $\boldsymbol{u}$. 
Since the actual dynamics involve parametric uncertainties and unmodeled dynamics compared with the nominal model (\ref{eq:ode_control}), we identify the actual dynamics using PINNs. 
We also assume that the system function $\boldsymbol{f}$ is continuous and (locally) Lipschitz-continuous at $x\in\mathbb{X}$ and the initial value problem (\ref{eq:ode_control}) has the uniqueness of the (weak) solution for arbitrary control input $\boldsymbol{u} \in L^\infty(\mathbb{T}, \mathbb{U})$ under the compact subset property and convexity of $\mathbb{U}$ \cite[Thm.~54]{Sontag89}. 
Then, the state trajectory $\boldsymbol{x}(t)$, $t\in\mathbb{T}$, which is the solution to the dynamics (\ref{eq:ode_control}) with the initial state $\boldsymbol{x}(0)$ and the control input $\boldsymbol{u}(\tau)$ for~$\tau\in[0,t]$, satisfies the equation
\begin{equation}
\boldsymbol{x}(t) = \varphi(t,\boldsymbol{x}_0,\boldsymbol{u}), \ t\in\mathbb{T}, 
\label{eq:ode_control_flow}
\end{equation}
where $\varphi$ indicates the transition map corresponding to the dynamics (\ref{eq:ode_control}).

Given the reference state trajectory $\boldsymbol{x}^{\text{ref}}(t)$, $t\in\mathbb{T}$, over the duration $\mathbb{T}$, we consider the optimal control problem that minimizes the tracking error
\begin{equation}
    \boldsymbol{e}(t) = \boldsymbol{x}^{\text{ref}}(t) - \boldsymbol{x}(t)
    \label{eq:error}
\end{equation}
and the control input $\boldsymbol{u}$ approach to zero along the control system (\ref{eq:ode_control_flow}), i.e., 
\begin{equation}
\min_{\boldsymbol{u}(t)\in\mathbb{U}} 
\int_0^{t_f} J(\boldsymbol{e}(t), \boldsymbol{u}(t)) dt  + \Psi(\boldsymbol{e}(t_f))
\ \ \text{s.t.} \ \ (\ref{eq:ode_control_flow}), (\ref{eq:error}),
\tag{P1} \label{eq:def-OCP}
\end{equation}
where $J$ is the stage cost and $\Psi$ is the terminal cost. 
Assume that the stage cost $J$ satisfies $\nabla_{\boldsymbol{u}^2} J>0$ uniformly at $\boldsymbol{u}\in\mathbb{U}$.
The details of $J$ and $\Psi$ will be introduced in Section~\ref{sec:03}. 
The model-based optimal control problem (\ref{eq:def-OCP}) with the nominal model-based state transition (\ref{eq:ode_control_flow}) can be solved numerically or partially-analytically (e.g., \cite{Ohtsuka04,CasADi19}).
However, it is challenging to incorporate the actual dynamics, which include unmodeled dynamics and parametric uncertainty, into the optimal control problem (\ref{eq:def-OCP}).

\subsection{Physics-Informed Neural Networks for Control}

To overcome the issue, we next review the mathematical formulation of PINNs for control, which approximates the solution (\ref{eq:ode_control_flow}) using a data-driven approach. 
The standard PINNs \cite{MR19} is a method for approximating the transition map (\ref{eq:ode_control_flow}) of the continuous-time physically dynamical model, defined as an autonomous system without external control input, i.e., (\ref{eq:ode_control}) with $\boldsymbol{u}(t) \equiv \boldsymbol{0}$ for any time $t$. 
One of the typical issues in applying PINNs to the transition map of the dynamical control system, i.e., (\ref{eq:ode_control_flow}), is to prepare a large amount of training data to obtain a suitable PINNs model.
To be concrete, it is necessary to prepare the training data, considering not only the large number of candidates in the initial state but also the infinite dimensionality of the control $\boldsymbol{u}(t)$ in continuous time.

To overcome the issue, this paper uses an approach proposed in \cite{NKFB22,AC24}. 
The assumption proposed in \cite{NKFB22,AC24} is that the control input is restricted to a sampled signal with ZOH.
To be concrete, we discretize the time interval $\mathbb{T}$ with a sampling time $\Delta T\,(\ll t_f)$, where $\mathbb{T}$ is divided into $T\,(=t_f/\Delta T)$ equal intervals, and set the time points $t_k = k\Delta T$, $k \in \mathcal{K} :=\{ 0, 1, \ldots, T-1\}$. 
Then, the control input with ZOH is given by 
\begin{equation}
\boldsymbol{u}(t) = \boldsymbol{u}_k\in \mathbb{U}, \ \  
t \in [k\Delta T, (k+1)\Delta T), \ k\in\mathcal{K},
\label{eq:zoh}
\end{equation}
and the corresponding state trajectory with (\ref{eq:ode_control_flow}) and (\ref{eq:zoh}) is rewritten by  
\begin{equation}
\boldsymbol{x}(t) = \varphi(t,\boldsymbol{x}_k,\boldsymbol{u}_k), \ \ 
t \in [k\Delta T, (k+1)\Delta T), \ k\in\mathcal{K},
\label{eq:discrete-ode_control}
\end{equation}
where $\boldsymbol{x_k} = \boldsymbol{x}(k\Delta T)$.

Let us now focus on the state trajectory at the $k$-th interval.
The superior idea proposed in \cite{NKFB22,AC24} is to use the PINNs framework to approximate the state trajectory during the short interval $\Delta T$.
In short, the original transition map $\varphi$ in (\ref{eq:discrete-ode_control}) is replaced by a neural network $\hat{\varphi}$ generated by PINNs with learnable parameters $\omega\in\mathbb{R}^{w}$, i.e., 
\begin{equation}
\boldsymbol{x}(t) = \hat{\varphi}(t,\boldsymbol{x}_k,\boldsymbol{u}_k,\omega), \ \ 
t \in [k\Delta T, (k+1)\Delta T), \ k\in\mathcal{K}.
\label{eq:pinn_control}
\end{equation}
The primary problem with using (\ref{eq:pinn_control}) is to ensure that the learned model $\hat{\varphi}$ can accurately predict the state trajectory over the entire interval $[k\Delta T, (k+1)\Delta T]$.
We assume that the activation function of the neural network $\hat{\varphi}$ uses a continuous and differentiable function such as $\tanh$ and sigmoid functions. Therefore, the function $\hat{\varphi}$ is differentiable at time $t$.

Let us define the two training datasets. 
One is the time-series measured dataset $\mathcal{D}_{\text{data}}=\{(t_i,\boldsymbol{x_{0,i}},\boldsymbol{x_{f,i}},\boldsymbol{u_i})\}$, $i\in\mathcal{L}_{\text{data}}=\{1,\ldots,N_{\text{data}}\}$, where $\boldsymbol{x_{0,i}}\in\mathbb{X}$ is the initial state included in the admissible state set $\mathbb{X}$, $\boldsymbol{x_{f,i}}$ is the final state at time $t_i\in(0,\Delta T+\epsilon]$, $\epsilon>0$, and $\boldsymbol{u_i}$ is the control input for the $i$-th data point.
The other is the dataset $\mathcal{D}_{\text{phys}}=\{(t_j,\boldsymbol{x_{j}},\boldsymbol{u_j})\}$, $j\in\mathcal{L}_{\text{phys}}=\{1,\ldots,N_{\text{phys}}\}$ uniformly and randomly self-selected over $[0,\Delta T+\epsilon]\times\mathbb{X}\times\mathbb{U}$ to evaluate the physics-based dynamical model. 
Note that the PINNs model is trained for a period of length $\Delta T+\epsilon$, which is longer than the sampling time $\Delta T$, to ensure the automatic differentiation at the sampling time $\Delta T$ used in the optimization problem, which will be mentioned in Section~\ref{sec:03}. 
Then, the loss function for training the parameters $\omega$ in the PINNs model $\hat{\varphi}$ from the collected training dataset $\mathcal{D}_{\text{data}}$ and $\mathcal{D}_{\text{phys}}$ is defined as follows:
\begin{equation}
\min_{\omega} L(\omega), \quad L(\omega) := L_{\text{data}}(\omega) + \lambda L_{\text{phys}}(\omega) 
\label{eq:pinns-loss}
\end{equation}
with a hyper-parameter $\lambda > 0$ that balances the data-based loss and physics-based loss terms, where
the data imbalance term $L_{\text{data}}$ and the physically dynamical model loss term $L_{\text{phys}}$ are respectively defined by 
\begin{eqnarray*}
&& L_{\text{data}}(\omega) = \frac{1}{N_{\text{data}}} \sum_{i=1}^{N_{\text{data}}} \| \hat{\varphi}(t_i, \boldsymbol{x}_{0,i},\boldsymbol{u}_{i},\omega) - \boldsymbol{x}_{f,i} \|^2, \\
&& L_{\text{phys}}(\omega) = \frac{1}{N_{\text{phys}}} \sum_{j=1}^{N_{\text{phys}}} \| \varPhi(t_j, \boldsymbol{x}_{j},\boldsymbol{u}_{j},\omega) \|^2,
\end{eqnarray*}
where the residual of the physically dynamical model $\varPhi$ is defined as the error between the known nominal model and the time-differentiable trained dynamics, i.e., 
\begin{eqnarray}
\lefteqn{\varPhi(t_j, \boldsymbol{x}_{j},\boldsymbol{u}_{j},\omega)} && \nonumber \\
&&= \frac{\partial}{\partial t}\hat{\varphi}(t_j,\boldsymbol{x}_j,\boldsymbol{u}_j,\omega) - \boldsymbol{f}(\hat{\varphi}(t_j,\boldsymbol{x}_j,\boldsymbol{u}_j,\omega),\boldsymbol{u}_j).
    \label{def:eq-F}
\end{eqnarray}
Note that the norm $\|\cdot\|$ is the standard Euclidean norm.
The primary feature of PINNs is to consider the physics-based loss $L_{\text{phys}}(\omega)$ with (\ref{def:eq-F}). 
In particular, the function $\varPhi$ can be computed efficiently by using the automatic differentiation of the neural network.
Thus, if the physics-based loss is zero, i.e., $L_{\text{phys}}(\omega)=0$, then we obtain the same neural network model $\hat{\varphi}$ as the original state transition map, that is $\hat{\varphi}(t, \boldsymbol{x}_k,\boldsymbol{u}_k,\omega) \equiv \varphi(t, \boldsymbol{x}_k,\boldsymbol{u}_k)$.

\begin{figure}[!t]
\centerline{\includegraphics[bb=0 0 355 246,width=0.7\columnwidth]{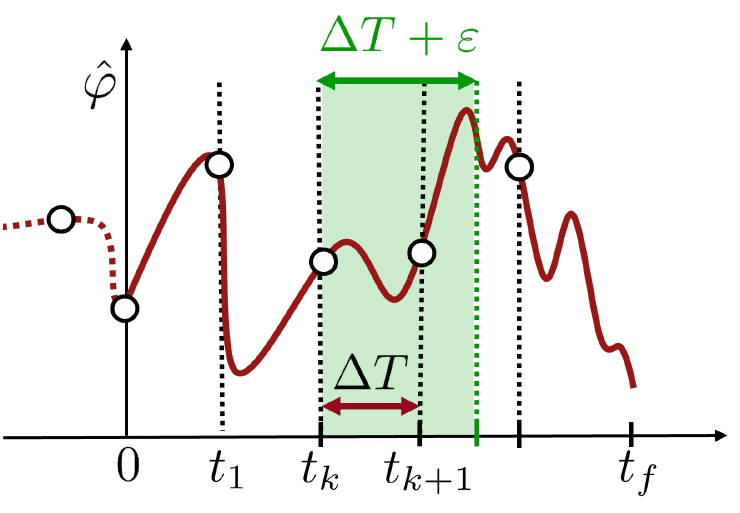}}
\caption{Sketch of PINNs-based model predictive control.}
\label{fig:3-time-steps}
\end{figure}

\begin{figure}[!t]
\centerline{\includegraphics[bb=0 0 964 436,width=\columnwidth]{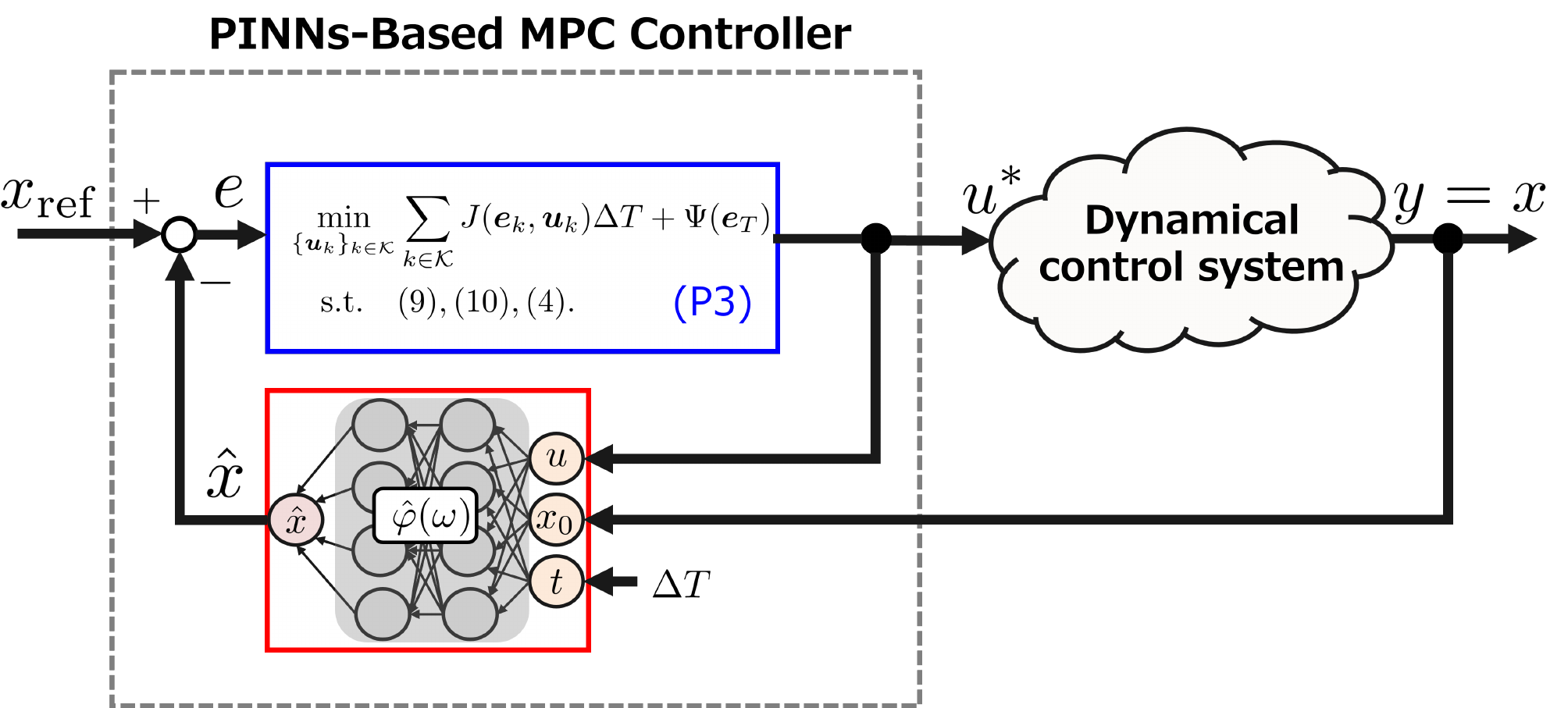}}
\caption{Block diagram of PINNs-based MPC.}
\label{fig:Block-diagram-pinns-mpc}
\end{figure}

We finally formulate the optimal control problem (\ref{eq:def-OCP}) based on PINNs model with (\ref{eq:pinn_control}), which is the solution of the optimization problem (\ref{eq:pinns-loss}). 
From the ZOH control input given by (\ref{eq:zoh}), we have the PINNs-based optimal control problem:
\begin{equation}
\min_{\boldsymbol{u}(t)\in\mathbb{U}} 
\int_0^{t_f} \! J(\boldsymbol{e}(t), \boldsymbol{u}(t)) dt + \Psi(\boldsymbol{e}(t_f))
\ \ \text{s.t.} \ \ (\ref{eq:pinn_control}), (\ref{eq:error}), (\ref{eq:zoh}).
\tag{P2} \label{eq:def-PINN-OCP}
\end{equation}
From the limited training data, the high-accuracy PINNs-based model (\ref{eq:pinn_control}) of the nonlinear systems (\ref{eq:ode_control}) is constrained in terms of the allowable time horizon $\Delta T+\epsilon$. 
In other words, the PINNs model (\ref{eq:pinn_control}) can be ensured only for short periods of time compared with the long-term prediction horizon $t_f$.
Therefore, the promising approach of the long-term prediction is to use the model predictive control (MPC) using recurrent self-loop prediction with ZOH control (\ref{eq:zoh}), i.e., 
\begin{equation}
\boldsymbol{x}_{k+1} = \hat{\varphi}(\Delta T,\boldsymbol{x}_k,\boldsymbol{u}_k,\omega), \quad k \in\mathcal{K}.
\label{eq:selfloop}
\end{equation}
Therefore, given the initial state $\boldsymbol{x}(0)=\boldsymbol{x}_0$ and the tracking error 
\begin{equation}
\boldsymbol{e}_k = \boldsymbol{x}^{\text{ref}}_k - \boldsymbol{x}_k
\label{eq:error_disc}
\end{equation}
with $\boldsymbol{x}^{\text{ref}}_k:=\boldsymbol{x}^{\text{ref}}(k\Delta T)$, the optimal control problem (\ref{eq:def-PINN-OCP}) is rewritten by PINNs and MPC-based optimal control problem \cite{NKFB22,AC24}:  
\begin{equation}
\min_{\{\boldsymbol{u}_k\}_{k\in\mathcal{K}}} \sum_{k\in\mathcal{K}} J(\boldsymbol{e}_k, \boldsymbol{u}_k) \Delta T + \Psi(\boldsymbol{e}_{T})
\ \text{s.t.} \ 
(\ref{eq:selfloop}), (\ref{eq:error_disc}), (\ref{eq:zoh}).
\tag{P3} \label{prob:MPC}
\end{equation}
The sketch of (\ref{prob:MPC}) is shown in Fig.~\ref{fig:3-time-steps}. 
The optimal control obtained in the optimization process is repeated while updating the initial conditions to the latest information.  put sequence $\{\boldsymbol{u}_k^*\}_{k\in\mathcal{K}}$ obtained from (\ref{prob:MPC}), the first control input $\boldsymbol{u}_0^*$ is applied to the real system (\ref{eq:ode_control}) until the duration $\Delta T$. 
The optimization process is then repeated at the next time step, using the new measured state of the system as the initial condition.
The block diagram of the implementation of the PINNs-based MPC is shown in Fig.~\ref{fig:Block-diagram-pinns-mpc}.

\section{PINNs-based Adaptive PID Control}
\label{sec:03}

In this section, we propose a PINNs-based optimization method for implementing adaptive PID control, which remains widely utilized in industrial applications.

\subsection{PID Gain Optimization Problem}

Given the tracking error $\boldsymbol{e}(t)$ defined in (\ref{eq:error}), we consider the nonlinear dynamics (\ref{eq:ode_control}) or the state trajectory (\ref{eq:ode_control_flow}) with the time-varying PID control law \cite{AH06}:
\begin{equation}
\boldsymbol{u}(t) = \boldsymbol{K}^p(t) \boldsymbol{e}(t) + \boldsymbol{K}^i(t) \int_0^t \boldsymbol{e}(\tau) d\tau + \boldsymbol{K}^d(t) \dot{\boldsymbol{e}}(t),
\label{eq:pid_control}
\end{equation}
where $\boldsymbol{F}(t):=[\boldsymbol{K}^p(t), \boldsymbol{K}^i(t), \boldsymbol{K}^d(t)]$ is the time-varying PID feedback gain vector satisfying $\boldsymbol{K}^p(t) \in \mathbb{R}^{m \times n}$, $\boldsymbol{K}^i(t) \in \mathbb{R}^{m \times n}$, and $\boldsymbol{K}^d(t) \in \mathbb{R}^{m \times n}$. 
The admissible convex set of the gain vector satisfying $\boldsymbol{u}(t)\in\mathbb{U}$, which is defined in (\ref{eq:pid_control}), is denoted as $\mathcal{F}(t,\boldsymbol{e}) \subset \mathbb{R}^{m \times 3n}$. 
Afterwards, $\mathcal{F}(t,\boldsymbol{e})$ is simply rewritten by $\mathcal{F}$.

From the discussion in Section~\ref{sec:02}, the state trajectory (\ref{eq:ode_control_flow}) can be replaced with a PINNs-based prediction model $\hat{\varphi}$ in (\ref{eq:pinn_control}) and (\ref{eq:selfloop}). 
Then, to stabilize the nonlinear system (\ref{eq:ode_control}) over a wide state range, a time-varying PID gain optimization of the controller (\ref{eq:pid_control}) given as the solution of the optimization problem (\ref{eq:def-OCP}) is provided by the continuous-time formulation
\begin{equation}
\min_{\boldsymbol{F}(t)\in \mathcal{F}} \int_0^{t_f} J(\boldsymbol{e}(t), \boldsymbol{u}(t); \boldsymbol{F}(t)) dt 
\ \ \text{s.t.} \ \ 
(\ref{eq:pinn_control}), (\ref{eq:error}), (\ref{eq:pid_control}), 
\tag{P4} \label{op-AdaptivePID-conti-PINNs}
\end{equation}
Referring to (\ref{op-AdaptivePID-conti-PINNs}), the MPC based optimal control problem (\ref{prob:MPC}) with ZOH PID control 
\begin{equation}
\boldsymbol{F}(t) = \boldsymbol{F}_k :=[\boldsymbol{K}^p_k, \boldsymbol{K}^i_k, \boldsymbol{K}^d_k], \ t\in[k\Delta T, (k+1)\Delta T), \ k\in\mathcal{K}
\nonumber
\end{equation}
can be formulated as the discrete-time PID gain optimization: 
\begin{align}
\min_{\substack{\{\boldsymbol{F}_k\}_{k\in\mathcal{K}},\\ \boldsymbol{F}_k\in\mathcal{F}}} 
& \sum_{k\in\mathcal{K}} J(\boldsymbol{e}_k, \boldsymbol{u}_k; \boldsymbol{F}_k) \Delta T + \Psi(\boldsymbol{e}_{T})
\tag{P5} \label{op-AdaptivePID-discrete-PINNs}
\\
\text{s.t.} \quad & 
(\ref{eq:selfloop}), (\ref{eq:error_disc}), \ 
\boldsymbol{u}_k = \boldsymbol{F}_k \boldsymbol{E}_k, \ k\in\mathcal{K}, 
\nonumber
\end{align}
where $\boldsymbol{E}_k:=[(\boldsymbol{e}^{\text{prop}}_k)^\top, (\boldsymbol{e}^{\text{int}}_k)^\top, (\boldsymbol{e}^{\text{deri}}_k)^\top]^\top$, $k\in\mathcal{K}$,  
\begin{subequations}
\begin{align*}
\boldsymbol{e}^{\text{prop}}_{k} &= \boldsymbol{e}_{k} = \boldsymbol{x}^{\text{ref}}_{k}-\boldsymbol{x}_{k}, \ k\in\mathcal{K}, 
\\
\boldsymbol{e}^{\text{int}}_{k+1} &= \boldsymbol{e}^{\text{int}}_{k} + \!\int_{0}^{\Delta T} \!\!\!\!\!\left\{ \boldsymbol{x}^{\text{ref}}_{k} - \hat{\varphi}(\tau,\boldsymbol{x}_{k},\boldsymbol{u}_{k},\omega) \right\}\! d\tau, \ k\in\mathcal{K}, \!\!
\\ 
\boldsymbol{e}^{\text{deri}}_{k+1} &= \frac{\boldsymbol{e}_{k+1}-\boldsymbol{e}_{k}}{\Delta T}, \ k\in\mathcal{K}, 
\\
\boldsymbol{e}^{\text{int}}_{0} &=0, \ \boldsymbol{e}^{\text{deri}}_{0}=\frac{\boldsymbol{x}^{\text{ref}}_0-\boldsymbol{x}^{\text{ref}}_{\text{init}}}{\Delta T}-\frac{\partial}{\partial t}\hat{\varphi}(0,\boldsymbol{x}_0,\boldsymbol{u}_0,\omega), 
\end{align*}
\end{subequations}%
with a suitable initial reference value $\boldsymbol{x}^{\text{ref}}_{\text{init}}$.

The optimal PID control gain problem (\ref{op-AdaptivePID-conti-PINNs}) with the nominal model of the original nonlinear dynamics (\ref{eq:ode_control_flow}) is generally complex to solve analytically.
Meanwhile, the proposed optimization problem (\ref{op-AdaptivePID-discrete-PINNs}) utilizes an efficient gradient computation through the automatic differentiation of PINNs, obtaining a numerical solution for data-driven optimal control with unmodeled dynamics.

From empirical evidence, it is not easy to obtain suitable PID gains while minimizing the objective function (\ref{op-AdaptivePID-discrete-PINNs}). 
To prevent overfitting and ensure the stability of the learned PID gains, a regularization term $\Theta(\boldsymbol{F}_k)$ is introduced. 
Therefore, the optimization problem (\ref{op-AdaptivePID-discrete-PINNs}) is reformulated as
\begin{align}
\min_{\substack{\{\boldsymbol{F}_k\}_{k\in\mathcal{K}},\\ \boldsymbol{F}_k\in\mathcal{F}}} 
& \sum_{k\in\mathcal{K}} \left\{ J(\boldsymbol{e}_k, \boldsymbol{u}_k; \boldsymbol{F}_k) \Delta T + \mu \Theta( \boldsymbol{F}_k ) \right\} + \Psi(\boldsymbol{e}_{T})
\tag{P6} \label{prob:proposed_PID}
\\
\text{s.t.} \quad & 
(\ref{eq:selfloop}), (\ref{eq:error_disc}), \ 
\boldsymbol{u}_k = \boldsymbol{F}_k \boldsymbol{E}_k, \ k\in\mathcal{K}, 
\nonumber
\end{align}
with hyper-parameter $\mu > 0$.

\begin{remark}
From the perspective of industrial applications, a large number of datasets exist that consist of specific initial conditions, control inputs, and corresponding output results in fundamental companies.
In the case of closed-loop system design, such as in VRFT \cite{CLS02} and FRIT \cite{SKF04}, adapting to changes caused by system nonlinearities requires retraining during the control design phase.
Meanwhile, as the proposed method can decouple optimal control from precise (open-loop) system modeling, it is practical for real-world applications with complex and high-dimensional dynamical systems.
As the main contribution of this paper is to propose a novel data-driven approach for adaptive PID control using PINNs, industrial applications involving real data will be discussed in a separate paper. 
\end{remark}

\begin{remark}
Appropriate selection of the admissible gain set $\mathcal{F}$ yields a controller that guarantees the internal stability of the systems. 
Compared to the nominal model (\ref{eq:ode_control}), a PINNs-based model calibrated to real physical data can accommodate unmodeled dynamics and uncertainties more effectively. 
Consequently, the proposed approach is efficient when integrated with real-world data and is promising from a Sim2Real perspective. 
Meanwhile, it is challenging to obtain a PINNs-based, theoretically guaranteed, stabilized controller. 
A partial discussion regarding the guarantee of internal stability will be discussed in Section~\ref{sec:V-C}.
Validation on real-data applications is left for future work.
\end{remark}

\subsection{Implementation}
\label{sec:03_implement}

In this subsection, we introduce the methodology to implement the proposed adaptive PID control (\ref{prob:proposed_PID}). 
The typical system models (\ref{eq:ode_control}) used for validation will be discussed in Sections~\ref{sec:06} and \ref{sec:07}.

We first construct the prediction model (\ref{eq:pinn_control}) using PINNs to design the adaptive PID controller.
The activation function of PINNs employs the $\tanh$ function to guarantee the differentiability of the neural network.
The training dataset consists of $N_{data}$ and $N_{phys}$ samples selected randomly and uniformly from $\Delta T+\epsilon$, the admissible state set $\mathbb{X}$, and the control input space $\mathbb{U}$, using Latin Hypercube Sampling (LHS) to efficiently and comprehensively cover the state-input space $\mathbb{X}\times\mathbb{U}$ with a limited number of samples.
We set the hyper-parameter $\lambda=1.0$ in the loss function (\ref{eq:pinns-loss}) to equivalently balance the contributions of the data and physics terms based on empirical evidence.
The optimization method of $\omega$ in (\ref{eq:pinns-loss}) uses Adam\cite{KB14} or/and L-BFGS method \cite{LN89}, which will be discussed in the later sections.

Once a highly accurate PINNs model, $\hat{\varphi}$, is obtained, we next utilize it for the adaptive PID control design using gradient-based optimization with automatic differentiation of PINNs.
The cost functions in (\ref{prob:proposed_PID}) are given by quadratic-form representation
\begin{subequations} \label{eq:lqr}
\begin{align}
J(\boldsymbol{e}_k,\boldsymbol{u}_k;\boldsymbol{F}_k) &= \frac{1}{2} \qty( \boldsymbol{e}_{k}^\top Q_k \boldsymbol{e}_{k} + \boldsymbol{u}_{k}^\top R_k \boldsymbol{u}_{k} ), \\
\Theta(\boldsymbol{F}_k) &= \|\boldsymbol{F}_k\|^2, \label{lqr_cost_theta} %\\
\end{align}
\begin{align}
\Psi(\boldsymbol{e}_T) &= \frac{1}{2} \boldsymbol{e}_{T}^\top Q_T \boldsymbol{e}_{T},
\end{align}
\end{subequations}%
with the positive-semidefinite weight matrices of states $Q_k\in \mathbb{R}^{n\times n}$, $Q_T\in \mathbb{R}^{n\times n}$, and the positive-definite weight matrix of control input $R_k\in \mathbb{R}^{m\times m}$. 
The hyper-parameter in (\ref{prob:proposed_PID}) is set as $\mu=1$ and the initial reference value is set as $\boldsymbol{x}^{\text{ref}}_{\text{init}}\equiv 0$. 
Then, the proposed gradient-based optimization to solve the adaptive PID gain optimization problem (\ref{prob:proposed_PID}) is as follows:

To solve the adaptive PID gain optimization problem (\ref{prob:proposed_PID}), we employ a segment-wise optimization strategy. 
The entire prediction horizon $t_f$ is divided into $T = t_f/\Delta T$ equal intervals, and independent optimization is performed for each interval. 
Since the prediction model is constructed using PINNs, gradients of the objective function with respect to the gains can be efficiently computed via automatic differentiation of PINNs. 
The PINNs-based approach enables us to leverage various optimizers available in deep learning libraries for updating the gains.

In this paper, we employ the Adam optimizer \cite{KB14}, which is one of the reliable solvers for non-convex optimization problems. 
Let us denote the stage cost function at time step $k$ by $J_k(\boldsymbol{F}_k) := J(\boldsymbol{e}_{k}, \boldsymbol{u}_k;\boldsymbol{F}_k) \Delta T + \mu \Theta(\boldsymbol{F}_k)$. 
Then, we can obtain the gradient of the cost function $\nabla_{\boldsymbol{F}_k} J_k(\boldsymbol{F}_k)$ with respect to the gain $\boldsymbol{F}_k$. 
Under the initial parameters $\boldsymbol{m}^{(0)}=0$ and $\boldsymbol{v}^{(0)}=0$, the Adam update rule for the gain $\boldsymbol{F}_k$ at iteration $\iota\,(\geq 1)$ is given as follows:
\begin{subequations} \label{eq:adam_update}
\begin{align*}
\boldsymbol{m}^{(\iota)} &= \beta_1 \boldsymbol{m}^{(\iota-1)} + (1-\beta_1) \nabla_{\boldsymbol{F}_k} J_k, \\
\boldsymbol{v}^{(\iota)} &= \beta_2 \boldsymbol{v}^{(\iota-1)} + (1-\beta_2) (\nabla_{\boldsymbol{F}_k} J_k)^2, \\
\boldsymbol{F}_k^{(\iota)} &= \boldsymbol{F}_k^{(\iota-1)} - \alpha \frac{\boldsymbol{m}^{(\iota)}/(1-\beta_1^\iota)}{\sqrt{\boldsymbol{v}^{(\iota)}/(1-\beta_2^\iota)} + \epsilon_{\mathrm{Adam}}}, 
\end{align*}
\end{subequations}%
where $\beta_1,\beta_2\in[0,1)$, $\alpha>0$ and $\epsilon_{\mathrm{Adam}}>0$. 
The variables $\boldsymbol{m}^{(\iota)}$ and $\boldsymbol{v}^{(\iota)}$ indicate the first and second moment estimates, respectively. 
The terms $1/(1-\beta_1^\iota)$ and $1/(1-\beta_2^\iota)$ indicate the learning rate with the bias-correction term.
The update rule is repeated until $|\boldsymbol{F}_k^{(\iota)}-\boldsymbol{F}_k^{(\iota-1)}|$ achieves a sufficient small value. 
This segment-wise optimization approach enables the real-time implementation of time-varying PID gains, considering system nonlinear dynamics constraints, input constraints, and gain constraints within the feasible set $\mathcal{F}$.
The proposed adaptive PID control algorithm is summarized in Algorithm~\ref{alg:pid}.
In this paper, we set the learning rate to $\alpha = 1 \times 10^{-2}$ and use the default parameters $\beta_1 = 0.9$, $\beta_2 = 0.999$, and $\epsilon_{\mathrm{Adam}} = 1 \times 10^{-7}$.

\begin{algorithm}[!t]
    \caption{PINNs-based Adaptive PID Control Algorithm}
    \label{alg:pid}
    \begin{algorithmic}[1]
        \Require{Initial state $\boldsymbol{x}_0$, reference trajectory $\boldsymbol{x}_{\text{ref}}$, trained PINNs model $\hat{\varphi}$, sampling time $\Delta T$, terminal prediction time $t_f$, and prediction steps $T \gets t_f/\Delta T$}
        \Ensure{Optimized gain sequences $\{\boldsymbol{F}_k\}_{k\in\mathcal{K}}$}
        \For{$k \gets 0$ to $N-1$}
            \State \textbf{Initialize:} $\boldsymbol{F}_k$ randomly from feasible set $\mathcal{F}$
            \Repeat
                \State Compute control input: $\boldsymbol{u}_k = \boldsymbol{F}_k \boldsymbol{E}_k$
                \State Predict: $\boldsymbol{x}_{k} \gets \hat{\varphi}(\Delta T, \boldsymbol{x}_{k}, \boldsymbol{u}_k, \omega)$
                \State Compute cost: $J_k \gets J(\boldsymbol{e}_{k}, \boldsymbol{u}_k;\boldsymbol{F}_k) \Delta T + \mu \Theta(\boldsymbol{F}_k)$
                \State Update gains: $\boldsymbol{F}_k \gets \text{Adam}(\boldsymbol{F}_k, \nabla_{\boldsymbol{F}_k} J_k)$ 
                \State Project $\boldsymbol{F}_k$ onto feasible set $\mathcal{F}$
            \Until{convergence or maximum iterations reached}
            \State Apply $\boldsymbol{u}_k$ to system for duration $\Delta T$
            \State \textbf{Update:} current state: $\boldsymbol{x}_0 \gets$ measured system state
            \State \textbf{Update:} reference trajectory: $\boldsymbol{x}_{\text{ref}}$
        \EndFor
    \end{algorithmic}
\end{algorithm}

To implement the above methodology through simulations discussed in Sections~\ref{sec:06} and \ref{sec:07}, we utilize the virtual machine computer Proxmox Virtual Environment (Proxmox VE), which is recognized as one of the leading open-source virtualization platforms.%
\footnote{The detailed information of Proxmox VE can be found at \url{https://www.proxmox.com/en/products/proxmox-virtual-} \url{environment/overview}}
As for the hardware environment, we allocated 10 cores from the host machine's Intel Core i9-12900 processor (24 cores) to the virtual machine, along with 48GB of memory from a total of 128GB, and configured it to utilize the NVIDIA RTX 3090 through GPU passthrough directly. 
For the software environment, we conducted experiments on Linux and adopted TensorFlow 2.12.0 as the deep learning framework.
All code was written in Python. 
To ensure reproducibility, all random number seed values are fixed at a common value.

Under the above settings, we evaluate the effectiveness of the proposed control methodology through simulations of two system models in Sections~\ref{sec:05} and \ref{sec:06}.

\begin{figure}[!t]
\centering
\includegraphics[bb=0 0 386 320,width=0.6\linewidth]{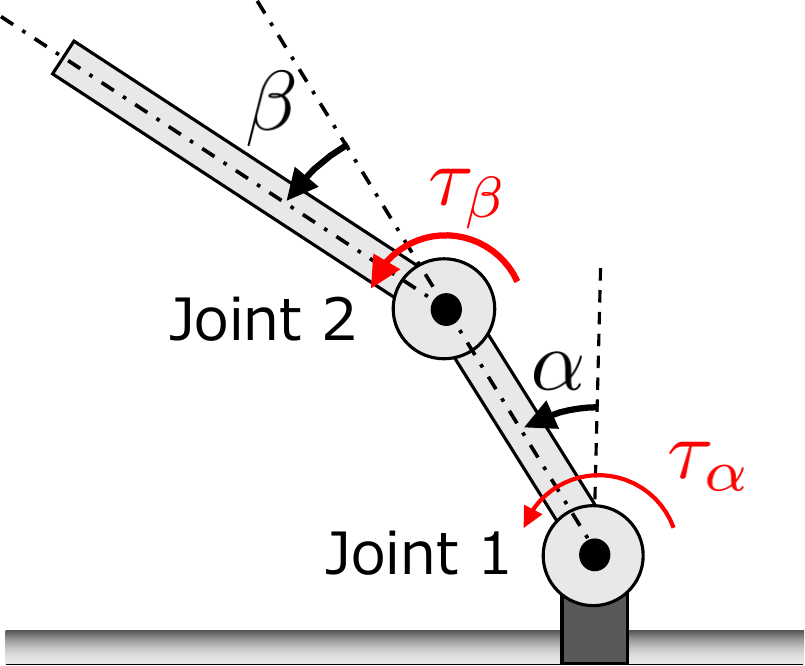}
\caption{A 2-DOF manipulator system.}
\label{fig:2DOF-model}
\end{figure}

\section{Case Study 1: 2-DOF Robot Manipulator}
\label{sec:05}

We first evaluate the effectiveness of the proposed PINNs-based adaptive PID control methodology, as shown in Algorithm~\ref{alg:pid}, using a typical nonlinear system: a 2-DOF robot manipulator system \cite{NKFB22}, as depicted in Fig.~\ref{fig:2DOF-model}. 
In particular, we will verify the effectiveness through both the regulation problem at the equilibrium point and the servo problem. 

The motion dynamics of the 2-DOF robot manipulator with two joint angles $\alpha$ and $\beta$ are generally written by 
\begin{equation}
\boldsymbol{D}(\boldsymbol{q}) \ddot{\boldsymbol{q}} + \boldsymbol{C}(\boldsymbol{q}, \dot{\boldsymbol{q}}) \dot{\boldsymbol{q}} + \boldsymbol{g}(\boldsymbol{q}) = \boldsymbol{\tau}
\label{eqn:motion}
\end{equation}
with the generalized coordinates $\boldsymbol{q}=[\alpha, \beta]^\top \in \mathbb{R}^2$ and the torque variables at the joints $\boldsymbol{\tau}=[\tau_{\alpha},\tau_{\beta}]^\top \in \mathbb{R}^2$\cite{SHV20}. 
The control input variables, consisting of the motor currents at the joints, are given by  $\boldsymbol{\tau}=\boldsymbol{B} \boldsymbol{u}={\rm diag}(b_\alpha,b_\beta)[u_{\alpha},u_{\beta}]^\top \in \mathbb{R}^2$, where ${\rm diag}(\cdot)$ denotes the operator that converts a vector to a diagonal matrix.
Let us denote by $\boldsymbol{D}(\boldsymbol{q}) \in \mathbb{R}^{2 \times 2}$ the invertible inertia matrix, by $\boldsymbol{C}(\boldsymbol{q},\dot{\boldsymbol{q}}) \in \mathbb{R}^{2\times 2}$ the matrix corresponding to the centrifugal, coriolis and gyroscopic forces, and by $\boldsymbol{g}(\boldsymbol{q}) \in \mathbb{R}^{2}$ the gravity vector, respectively.
See \cite{NKFB22,FSSE20} for the detailed values of the parameters. 
From (\ref{eqn:motion}), we obtain the state dynamics with the state variables $\state=(\boldsymbol{q},\boldsymbol{\qdot})^\top\in \mathbb{R}^4$ as follows:
\begin{equation}
\dv{}{t}\mqty[\boldsymbol{q} \\ \boldsymbol{\qdot}] 
= 
\mqty[ \boldsymbol{\qdot} \\ 
-\boldsymbol{D}^{-1}(\boldsymbol{q})( \boldsymbol{C}(\boldsymbol{q}, \boldsymbol{\qdot})\boldsymbol{\qdot} +\boldsymbol{g}(\boldsymbol{q}) ) ] 
+ \mqty[ O \\ \boldsymbol{D}^{-1}(\boldsymbol{q})\boldsymbol{B}] \boldsymbol{u}.
\label{eq:plant_state_space}
\end{equation}
The dynamics (\ref{eq:plant_state_space}) are control-affine systems and satisfy the mathematical conditions on (\ref{eq:ode_control}).

\subsection{Properties on steady-state PID control}
\label{sec:2dof_theory}

Let us now consider the steady-state properties of the PID control law (\ref{eq:pid_control}) for the robot manipulator system (\ref{eqn:motion}) in the context of the regularized optimal control problem (\ref{prob:proposed_PID}) with (\ref{eq:lqr}).

The tracking error $\boldsymbol{e}_\infty$ at the steady state converges to zero, i.e., $\boldsymbol{e}_\infty = \lim_{t \to \infty} \boldsymbol{e}(t) = \boldsymbol{0}$, which implies $\boldsymbol{q} = \boldsymbol{q}^{\text{ref}}$ and $\dot{\boldsymbol{q}} = \boldsymbol{0}$. 
We suppose the PID gains during the prediction period $\mathbb{T}$ at the steady state are fixed, i.e., $\boldsymbol{F}(t) \equiv \boldsymbol{F}_{\infty} := [\boldsymbol{K}^p_{\infty}, \boldsymbol{K}^i_{\infty}, \boldsymbol{K}^d_{\infty}]\in\mathcal{F}$ for all $t\in\mathbb{T}$. 
Then, the stage cost function $L_\infty$ with (\ref{eq:lqr}) becomes
\begin{equation}
\!\!\! L_\infty = \frac{1}{2}\boldsymbol{u}_\infty^\top R \boldsymbol{u}_\infty + \mu \left( \|\boldsymbol{K}^p_\infty\|^2 + \|\boldsymbol{K}^i_\infty\|^2 + \|\boldsymbol{K}^d_\infty\|^2 \right)
\label{eq:steady_cost}
\end{equation}
and the corresponding PID control law (\ref{eq:pid_control}) is reduced to
\begin{equation}
\boldsymbol{u}_\infty = \boldsymbol{K}^i_\infty \boldsymbol{\xi}_\infty, \quad 
\boldsymbol{\xi}_\infty = \lim_{t \to \infty} \int_0^t \boldsymbol{e}(\tau) d\tau.
\label{eq:steady_control}
\end{equation}
The steady-state integral error $\boldsymbol{\xi}_\infty$ converges to a constant value.
From (\ref{eq:steady_cost}) and (\ref{eq:steady_control}), the optimality conditions for minimizing (\ref{eq:steady_cost}) yield $\boldsymbol{K}^p_\infty = \boldsymbol{0}$ and $\boldsymbol{K}^d_\infty = \boldsymbol{0}$.
As the steady-state robot manipulator dynamics (\ref{eqn:motion}) achieve
\begin{equation}
\boldsymbol{g}(\boldsymbol{q}^{\text{ref}}) = \boldsymbol{B} \boldsymbol{K}^i_\infty \boldsymbol{\xi}_\infty, 
\label{eq:gravity_constraint}
\end{equation}
the steady-state integral gain minimizing (\ref{eq:steady_cost}) satisfies
\begin{align}
\boldsymbol{K}^i_{\infty} &= \begin{cases}
    \boldsymbol{0} & \text{if } \boldsymbol{g}(\boldsymbol{q}^{\text{ref}}) = \boldsymbol{0} \\
    \boldsymbol{K}^{i*}(\boldsymbol{q}^{\text{ref}})\ (\neq \boldsymbol{0}) & \text{if } \boldsymbol{g}(\boldsymbol{q}^{\text{ref}}) \neq \boldsymbol{0}
\end{cases}, \label{eq:ki_converge}
\end{align}
where $\boldsymbol{K}^{i*}(\boldsymbol{q}^{\text{ref}})$ is the minimum-norm solution satisfying the constraint (\ref{eq:gravity_constraint}) for the reference state $\boldsymbol{q}^{\text{ref}}$.
The condition (\ref{eq:ki_converge}) indicates the two cases:
\begin{itemize}
\item If $\boldsymbol{g}(\boldsymbol{q}^{\text{ref}}) = \boldsymbol{0}$ (e.g., vertical configuration where gravity effects vanish), the constraint (\ref{eq:gravity_constraint}) becomes $\boldsymbol{K}^i_\infty \boldsymbol{\xi}_\infty = \boldsymbol{0}$. The regularization term $\mu\|\boldsymbol{K}^i_\infty\|^2$ is minimized when $\boldsymbol{K}^i_\infty = \boldsymbol{0}$.

\item If $\boldsymbol{g}(\boldsymbol{q}^{\text{ref}}) \neq \boldsymbol{0}$ (general configuration with gravity effects), the constraint (\ref{eq:gravity_constraint}) must be satisfied with $\boldsymbol{K}^i_\infty \neq \boldsymbol{0}$. The optimal value of $\boldsymbol{K}^{i}$ minimizes the combined cost of control effort and regularization while satisfying the gravity compensation requirement.
\end{itemize}

From the perspective of the transient state, it is expected that
\begin{equation*}
\lim_{t\to\infty} \boldsymbol{F}(t) = \boldsymbol{F}_{\infty}
\end{equation*}
will hold since the regularization term $\mu\Theta(\boldsymbol{F}_k)$ in (\ref{prob:proposed_PID}) penalizes non-zero gains that do not contribute to the control performance and automatically eliminates unnecessary control gains at steady state.
This result suggests that the proportional and derivative gains, which only contribute during transient responses, naturally decay to zero as the system reaches equilibrium, and that only the integral gain persists to provide the necessary steady-state compensation for gravitational forces. 
Therefore, during the transient phase, the time-varying gains $\boldsymbol{F}(t)$ contribute to achieving fast and stable convergence to the reference state. 
Moreover, when the system approaches a steady state, it is expected that the time-varying PID controller will approach an integral controller.
We will verify the hypothesis through the following simulations.

\begin{figure}[!t]
\centering
\begin{minipage}[t]{0.48\linewidth}
\centering
\includegraphics[clip,bb=6 12 1004 625,width=\columnwidth]{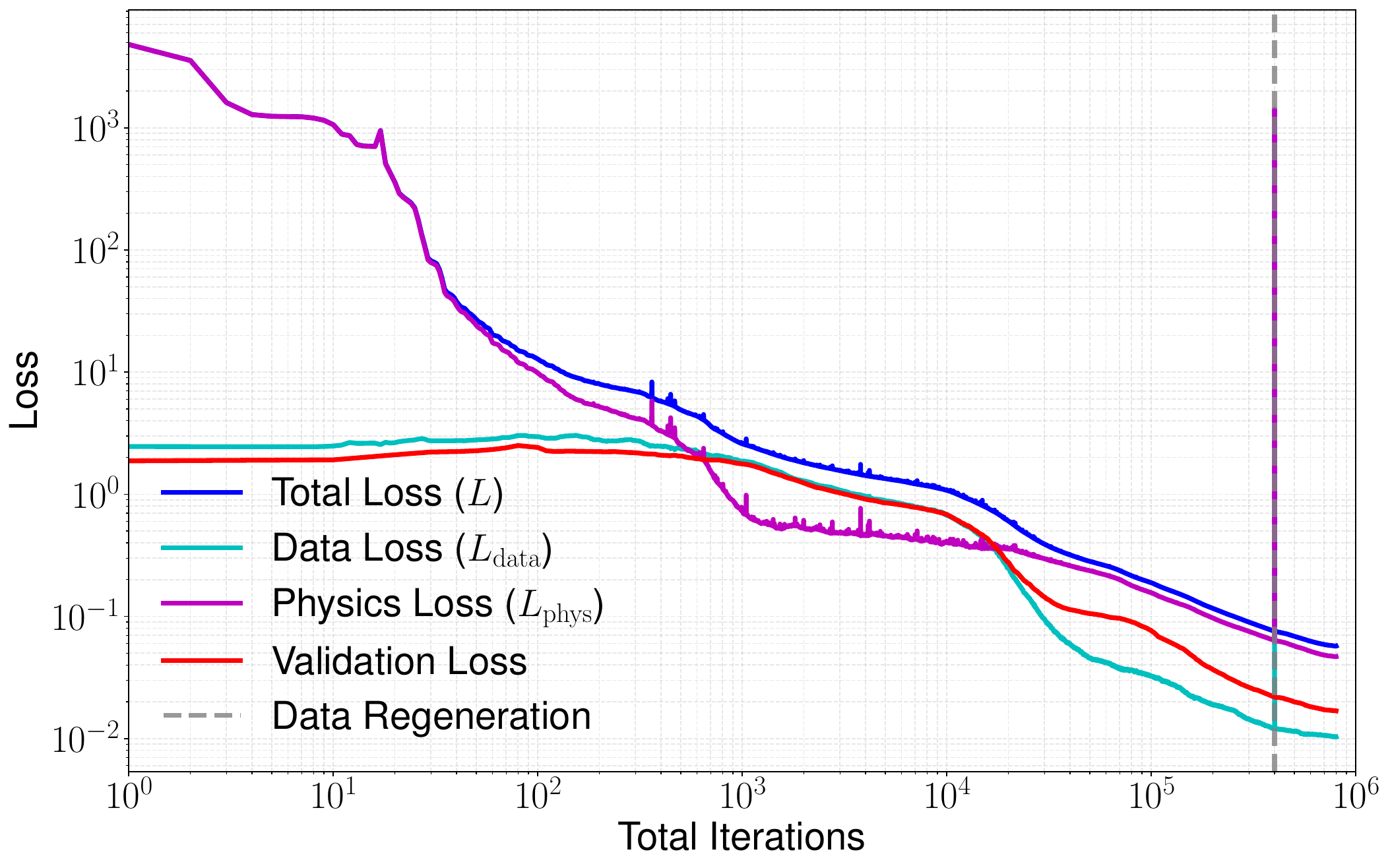}
\subcaption{Evolution of loss functions during training.}
\label{fig:result_2DOF_model_loss}
\end{minipage}
\hspace{0.01\linewidth}
\begin{minipage}[t]{0.48\linewidth}
\centering
\includegraphics[clip,bb=6 7 1098 668,width=\columnwidth]{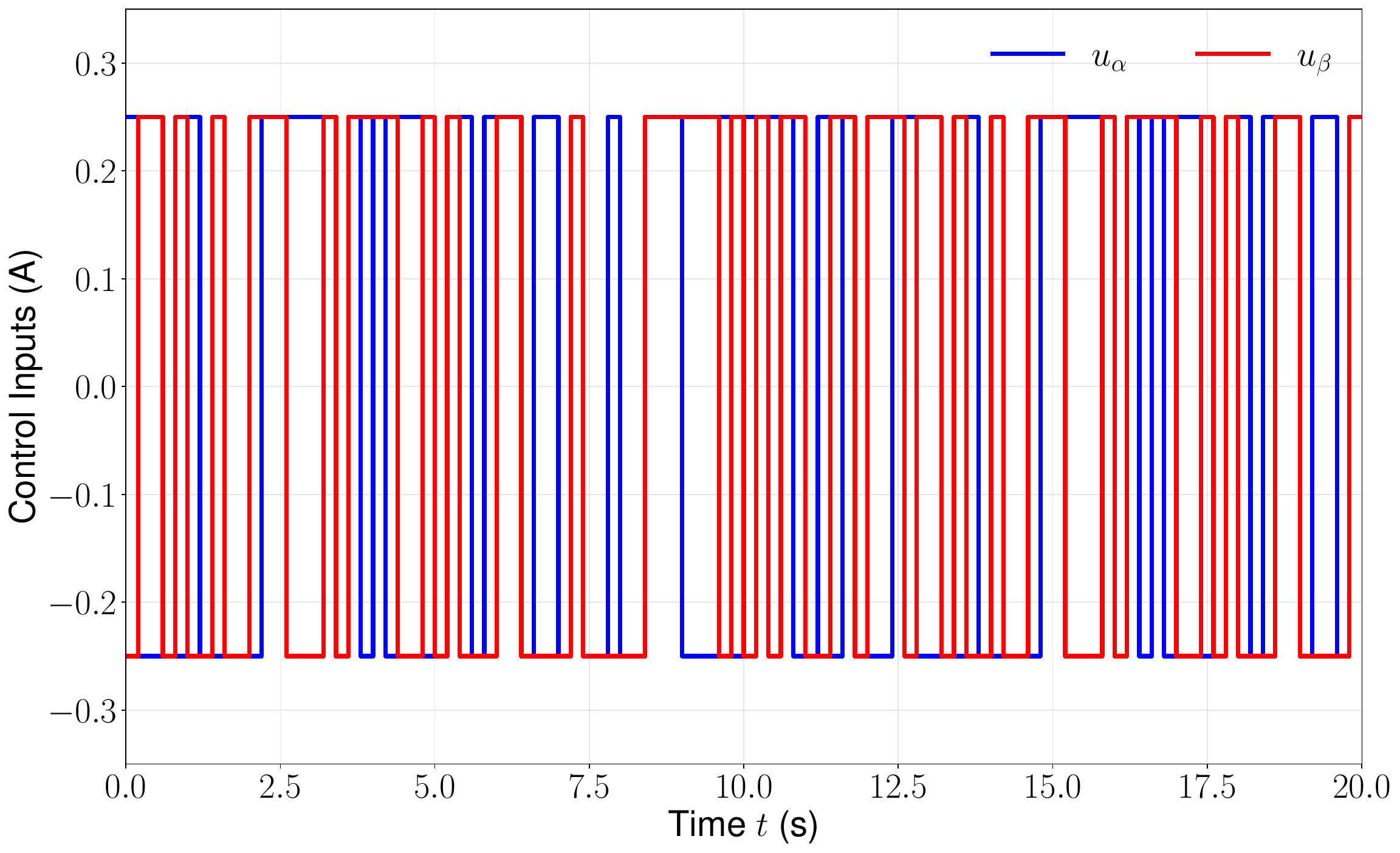}
\subcaption{Control input signals for the test dataset.}
\label{fig:result_2DOF_model_input}
\end{minipage}
\\ \smallskip
\begin{minipage}[t]{0.48\linewidth}
\centering
\includegraphics[clip,bb=7 7 1107 676,width=\columnwidth]{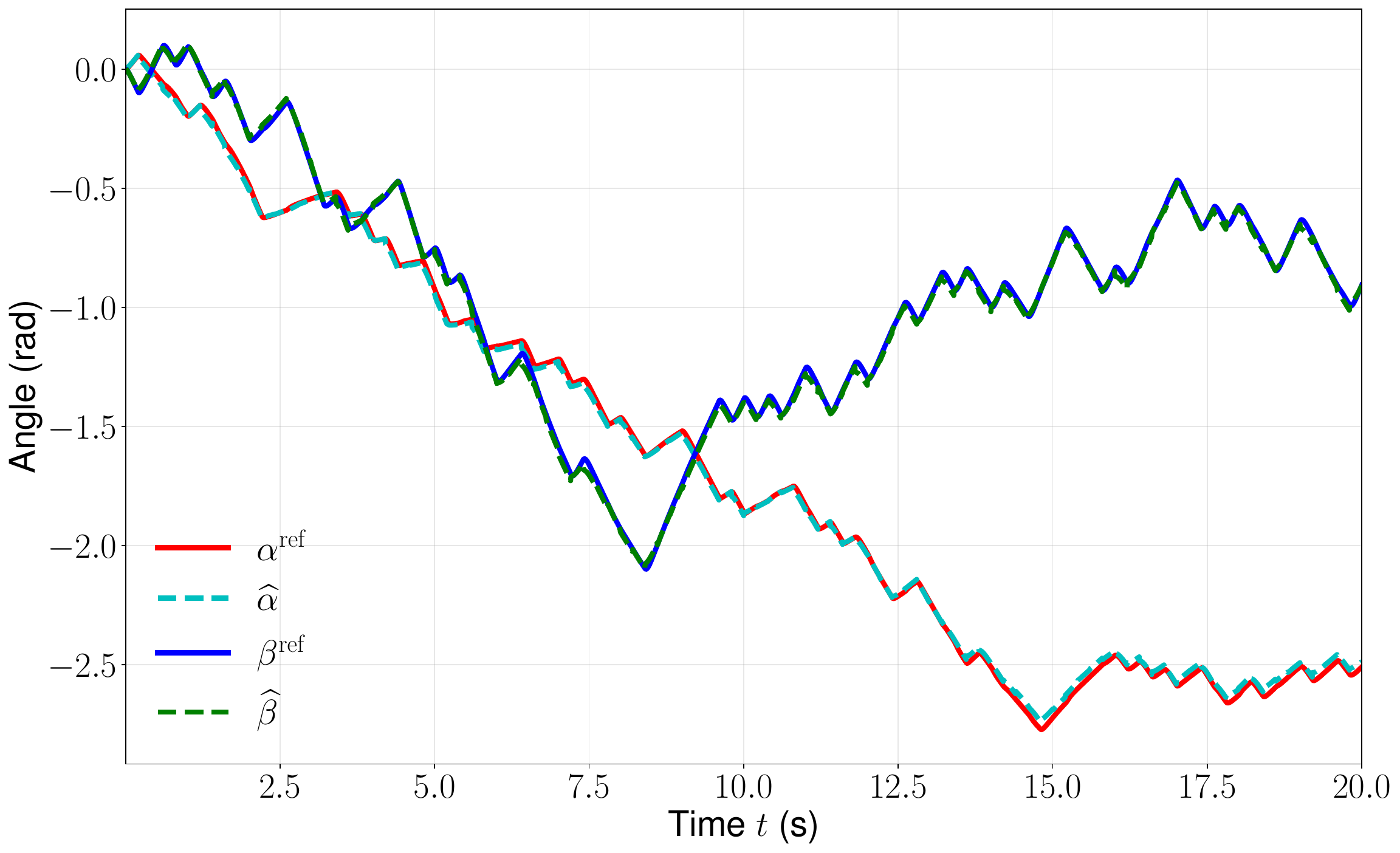}
\subcaption{Time evolutions of joint angles with the input trajectory.}
\label{fig:result_2DOF_model_angle}
\end{minipage}
\hspace{0.01\linewidth}
\begin{minipage}[t]{0.48\linewidth}
\centering
\includegraphics[clip,bb=6 7 1111 679,width=\columnwidth]{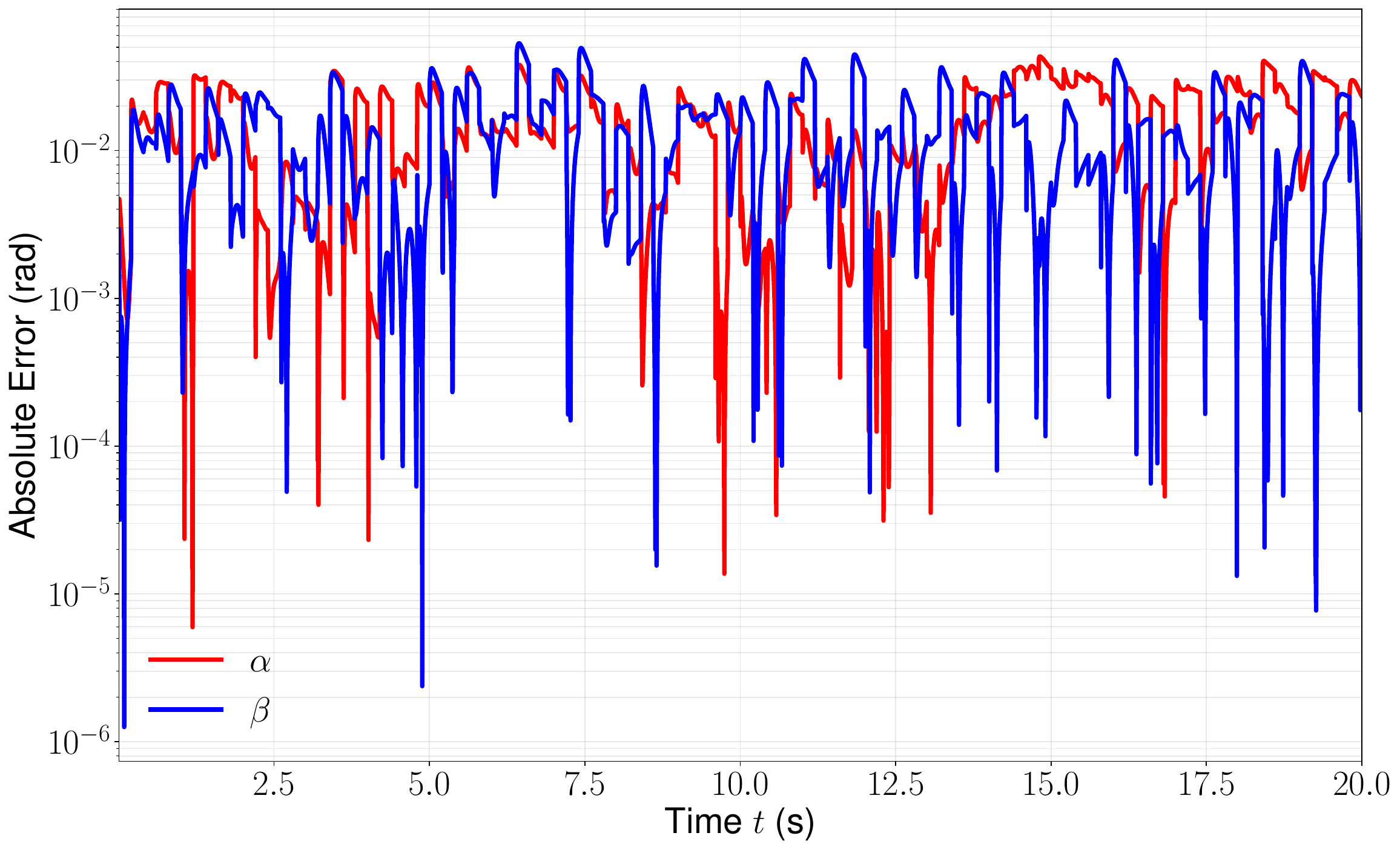}
\subcaption{Time evolutions of prediction absolute errors for both joints.}
\label{fig:result_2DOF_model_MAE}
\end{minipage}
\caption{Validation of the trained PINNs model for the 2-DOF manipulator system.}
\label{fig:result_2DOF_model}
\end{figure}

\subsection{Evaluation on PINNs-based model}
\label{sec:05-PINNmodel}

Following \cite{NKFB22}, we use the predicted state trajectory model (\ref{eq:pinn_control}) composed of a forward propagation-type deep neural network with $7$-dimensions at the input layer, $4$-hidden layers, which is based on $64$-units at each layer, and $4$-dimensions at the output layer. 
We consider the optimization problem (\ref{prob:proposed_PID}) with the prediction horizon $T=20$ based on the recurrent self-loop PINNs prediction of the sampling period $\Delta T=0.200$\,s.
The training datasets, $\mathcal{D}_{\text{data}}$ and $\mathcal{D}_{\text{phys}}$, are composed of $N_{\text{data}}=2\times 10^4$ and $N_{\text{phys}}=1\times 10^5$ samples, respectively. These samples are generated from the admissible state set $\mathbb{X} = [-\pi,\pi]^2 \times [-2.5,2.5]^2$ and the control input space $\mathbb{U} = [-0.5,0.5]^2$, with $\Delta T+\epsilon= 0.250$\,s.
We use an open-data published in \cite{NKFB22} as the validation dataset, which contains admissible initial states, control inputs, and the corresponding accurate state trajectories of the 2-DOF manipulator system~(\ref{eqn:motion}). 

By using the above dataset, we train the PINNs model (\ref{eq:pinn_control}) for a total of $8\times 10^5$ iterations. 
To improve the model accuracy, the training dataset is regenerated after $4\times 10^5$ iterations. 
The validation is performed every $10$ iterations by evaluating the Mean Squared Error (MSE) between the predicted states from the PINNs model and the ground-truth states on the validation dataset.
The optimization method of $\omega$ in (\ref{eq:pinns-loss}) uses L-BFGS method \cite{LN89}.

The results of the trained PINNs model are shown in Fig.~\ref{fig:result_2DOF_model}.
From Fig.~\ref{fig:result_2DOF_model_loss} showing the evolution of the loss function (\ref{eq:pinns-loss}) during the training process, the learning is in a steady decline.
The terminal value of the loss function $L(\omega)$ for the trained PINNs model is about $10^{-1}$, which is dominated by the physically dynamical model loss term $L_{\text{phys}}(\omega)$.
The validation loss score, evaluated as prediction error (MSE), decreases consistently throughout training, confirming that the model generalizes well without overfitting.
We next evaluate the trained PINNs model for the test dataset, which consists of state trajectory $(\alpha^{\text{ref}}, \beta^{\text{ref}})$ of the dynamical system (\ref{eqn:motion}) obtained with the control inputs presented in Fig.~\ref{fig:result_2DOF_model_input}.
To track the reference trajectory $(\alpha^{\text{ref}}, \beta^{\text{ref}})$ (solid lines), the predicted state trajectory $(\hat{\alpha}, \hat{\beta})$ for a sample input trajectory using the obtained PINNs model (dashed lines) is shown in Fig.~\ref{fig:result_2DOF_model_angle}. 
The absolute error between the reference trajectory and the predicted trajectory is shown in Fig.~\ref{fig:result_2DOF_model_MAE}.
The trained PINNs model achieves mean absolute errors (MAE) of $1.56\times10^{-2}$ and $1.39\times10^{-2}$ for the angular trajectories $\alpha$ and $\beta$, respectively. 
These values indicate an accuracy level of approximately $10^{-2}$.
The evidence of the figures suggests that the obtained PINNs model has not overfitted and has achieved sufficient generalization performance for the 2DOF manipulator control with nonlinear dynamics.

\subsection{Application to Regulation Problem}
\label{sec:2dof_regulation}

We next verify the effectiveness of the proposed method for the regulation problem, specifically the stabilization control problem at the vertical unstable equilibrium point $\boldsymbol{x}^{\text{ref}} = \boldsymbol{0}$. 
We set the state $\state(0) = \qty[-2.0, 1.5, 0.0, 0.0]^\top$ at time $t=0$, and control sets $\mathbb{U} =[-0.48, 0.48]^2$, $\mathcal{F}=[0.0, 3.0]^2\times[-3.0, 3.0]^2\times[0.0, 3.0]^2$. 
Note that the system with the initial state must swing up to reach the inverted equilibrium point.
We conduct the stabilization control at the equilibrium point during $40.0$\,s while implementing the MPC-based adaptive PID gain optimization (\ref{prob:proposed_PID}) with $\Delta T=0.200$\,s, $t_f=40.0$\,s, and the weight matrices of (\ref{eq:lqr}) given by $Q = {\rm diag}(100,100,0.01,0.01)$ and $R = 0.01I_2$, where $I_2$ denotes the 2-dimensional identity matrix.
The optimizer employed to solve the problem (\ref{prob:proposed_PID}) uses Adam \cite{KB14} with a learning rate of $10^{-2}$, and the maximum iteration to solve the problem is set as $1.6\times 10^4$. 
For comparison, we consider a conventional control design with fixed PID gains guaranteeing the stability property for the linearized system at the equilibrium, i.e., $\boldsymbol{F}_k\equiv [\boldsymbol{K}^p_k,\boldsymbol{K}^i_k,\boldsymbol{K}^d_k] = [2.0,2.0,0.3,0.3,0.2,0.2]$ at any time $k$.

\begin{figure}[!t]
\centering
\begin{minipage}[t]{0.48\linewidth}
\centering
    \includegraphics[clip,bb= 7 7 1004 626,width=\columnwidth]{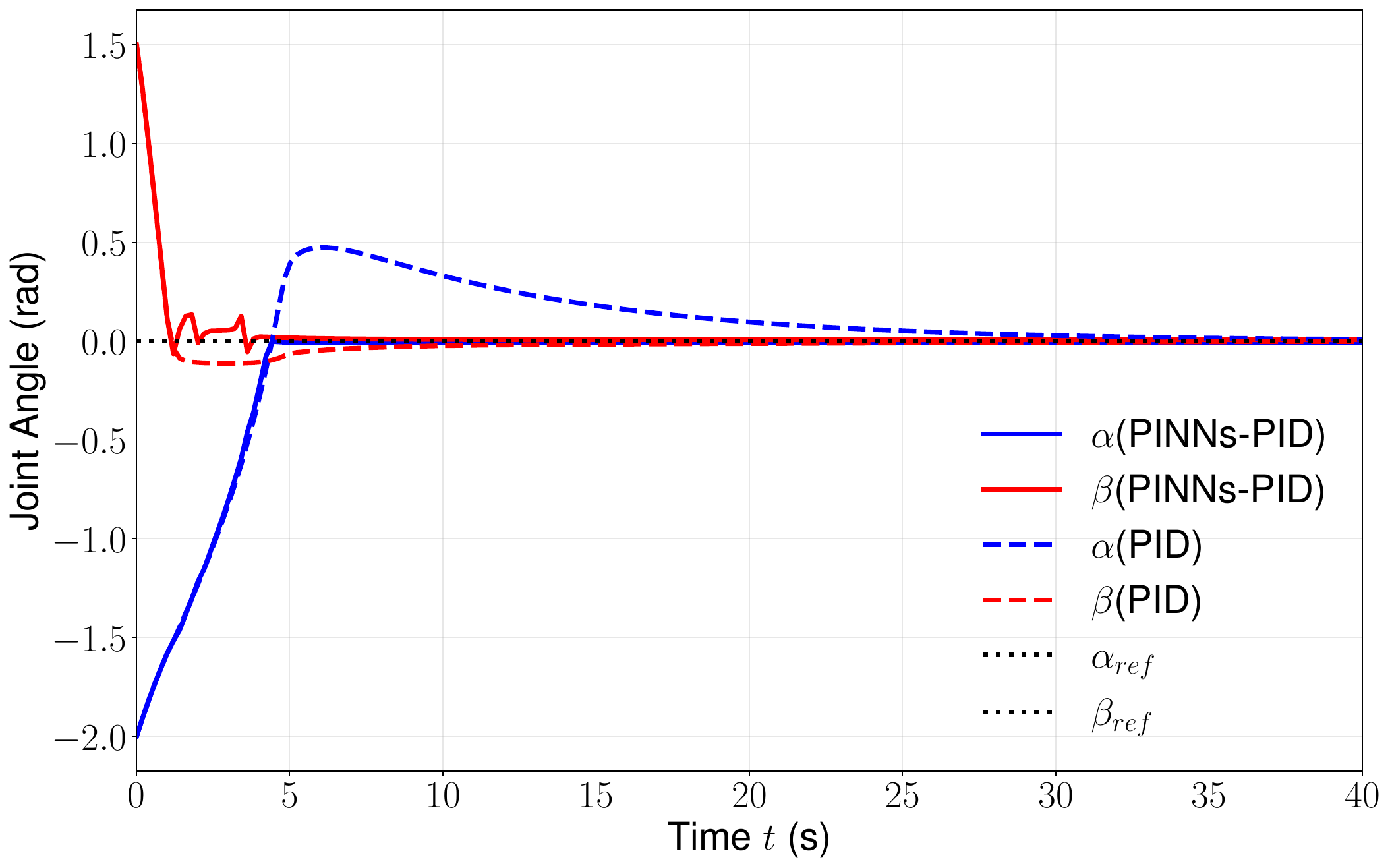}
    \subcaption{Joint angle responses of the two-link manipulator.}
    \label{fig:equilibrium_angles}
\end{minipage}
\hspace{0.01\linewidth}
\begin{minipage}[t]{0.48\linewidth}
\centering
    \includegraphics[clip,bb=7 7 999 626,width=\columnwidth]{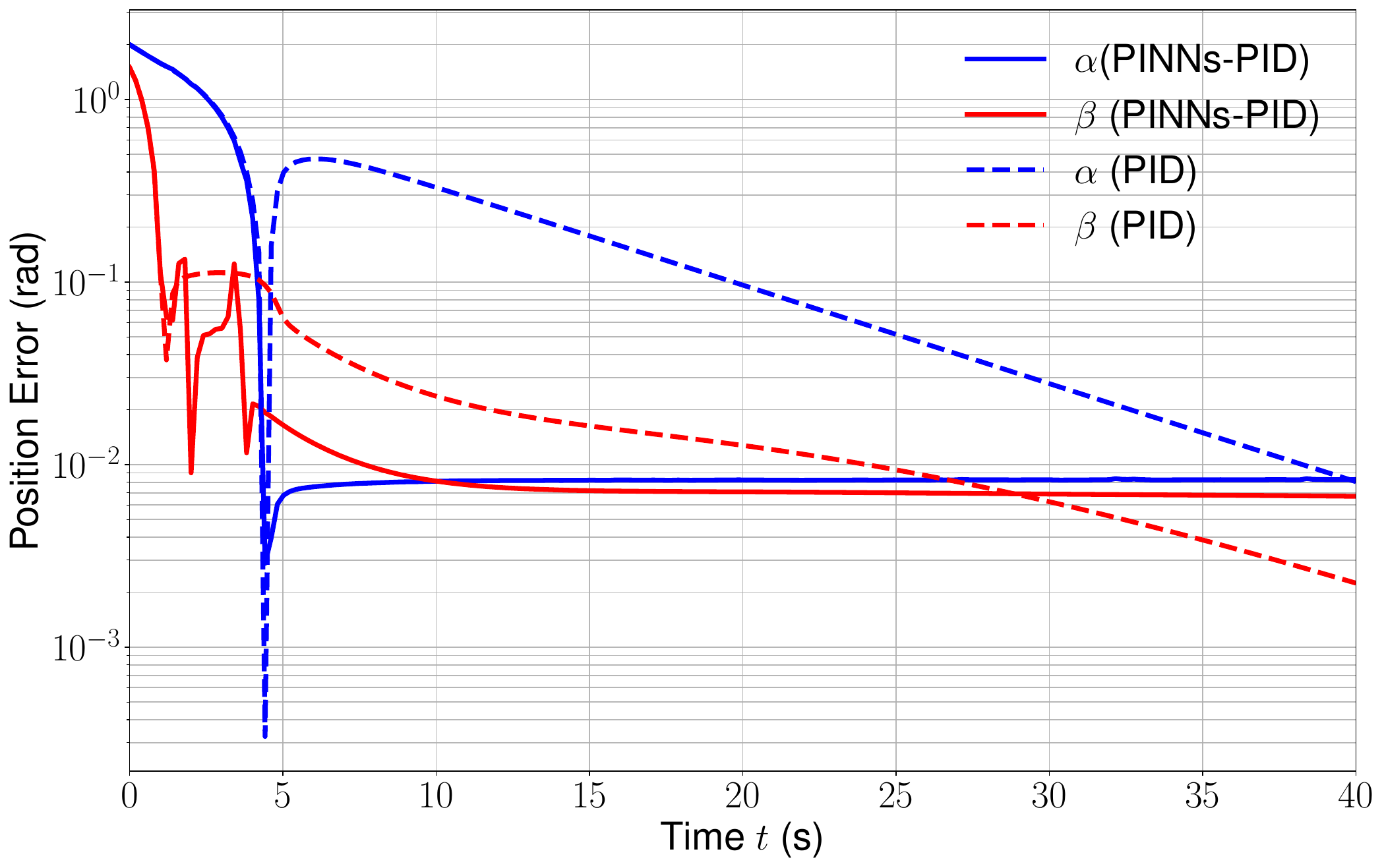}
    \subcaption{Position tracking errors on logarithmic scale.}
    \label{fig:equilibrium_error}
\end{minipage}
\\ \smallskip
\begin{minipage}[t]{0.48\linewidth}
\centering
    \includegraphics[clip,bb=6 7 1004 626,width=\columnwidth]{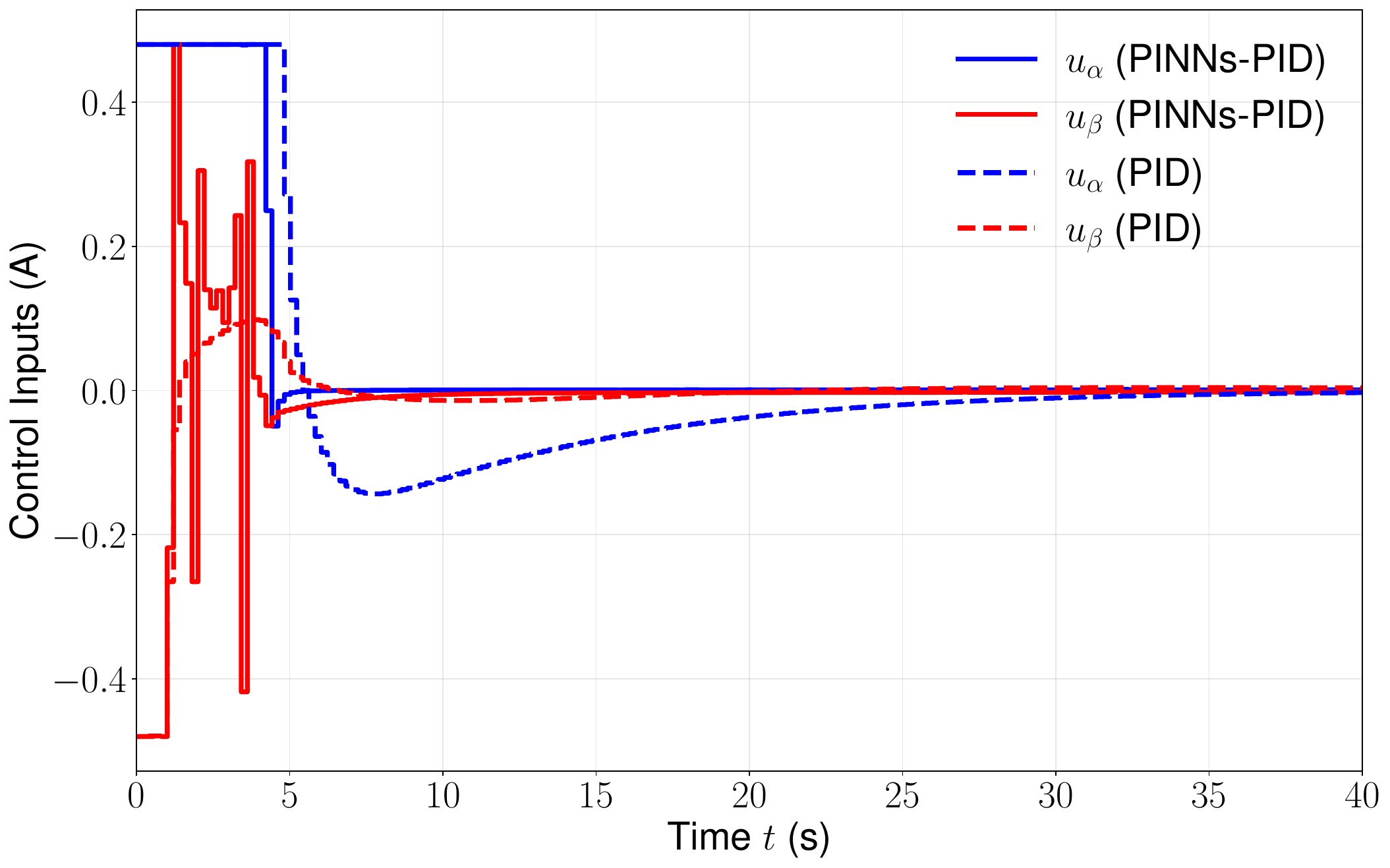}
    \subcaption{Control input signals during stabilization.}
    \label{fig:equilibrium_input}
\end{minipage}
\hspace{0.01\linewidth}
\begin{minipage}[t]{0.48\linewidth}
\centering
    \includegraphics[clip,bb= 6 7 982 626,width=\columnwidth]{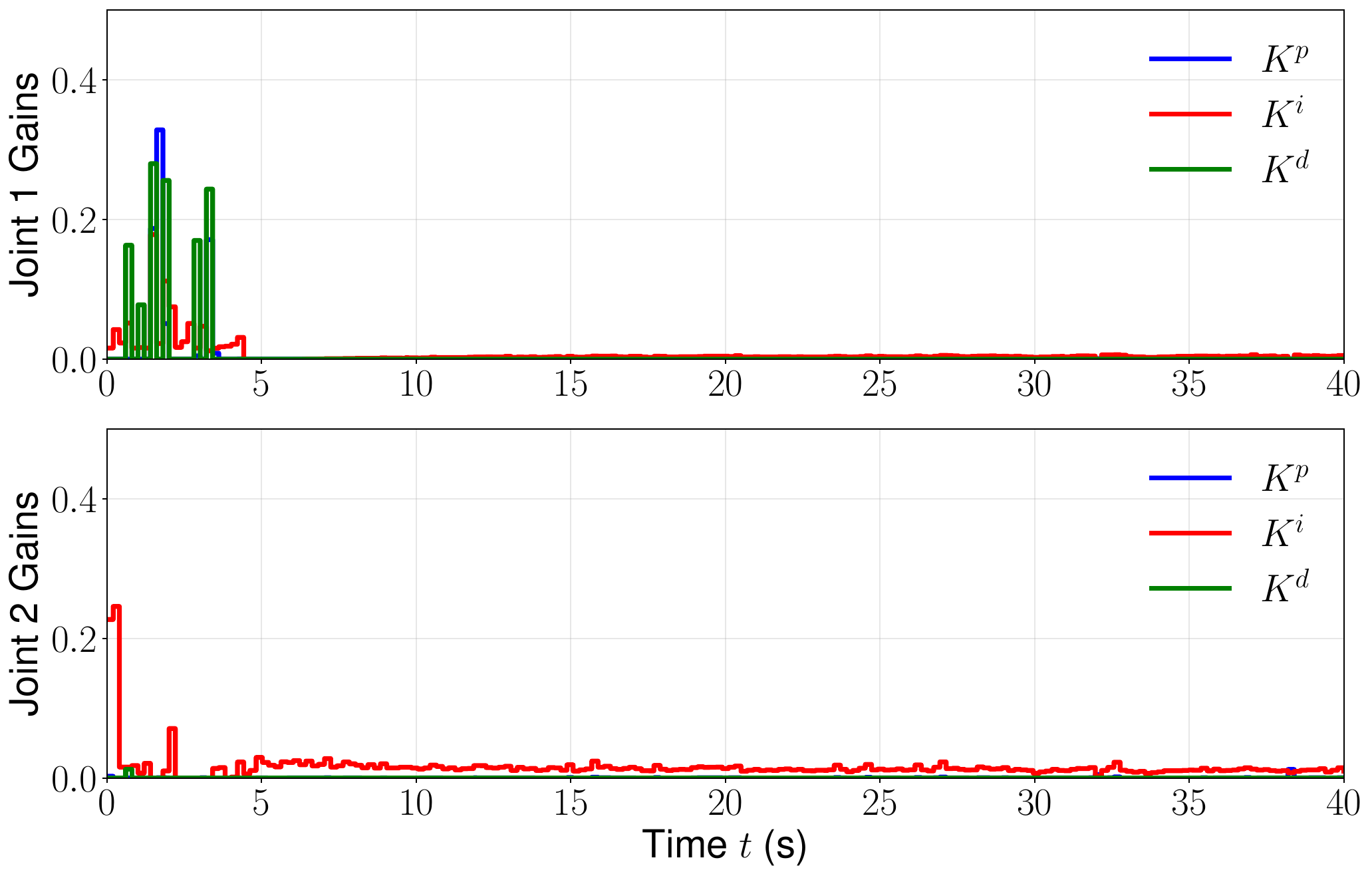}
    \subcaption{Time evolution of the proposed PID gains.}
    \label{fig:equilibrium_gains}
\end{minipage}
\caption{Regulation problem results comparing the proposed adaptive PID controller with fixed-gain PID for the 2-DOF manipulator.}
\label{fig:result_2DOF_regulation}
\end{figure}

The results for both controllers are shown in Fig.~\ref{fig:result_2DOF_regulation}. 
From the time evolutions of the angles $(\alpha,\beta)$ shown in Fig.~\ref{fig:equilibrium_angles} and the corresponding tracking error shown in Fig.~\ref{fig:equilibrium_error}, we see that the convergence property of the proposed method is superior to the fixed PID method. 
Actually, when the settling time for each angle is defined as the time required to reach and maintain a value within $2$\% of the reference value, then the fixed PID method requires $38.39$\,s and $36.18$\,s, respectively. 
In contrast, the settling times for the proposed method are $5.43$\,s and $11.26$\,s, respectively. 
The saturation of the tracking error at approximately $10^{-2}$ for the proposed method can be attributed to the limitations in prediction accuracy of the PINNs model. 
As shown in Fig.~\ref{fig:result_2DOF_model_MAE}, our test results reveal that the PINNs model achieves mean absolute errors (MAE) of $1.56\times10^{-2}$ and $1.39\times10^{-2}$ for the predicted state trajectories $\hat{\alpha}$ and $\hat{\beta}$, respectively. 
Since the proposed MPC-based optimization framework depends on these PINNs predictions, this accuracy bound imposes a fundamental limitation on the achievable tracking performance. 
Nevertheless, this error magnitude remains well within acceptable tolerances for practical control applications.

The time evolutions of the control input $\boldsymbol{u}(t)=[u_{\alpha}(t), u_{\beta}(t)]^\top$ and the corresponding proposed PID gains $\boldsymbol{F}_k$ are shown in Fig.~\ref{fig:equilibrium_input} and Fig.~\ref{fig:equilibrium_gains}, respectively.
The fixed PID controller exhibits relatively low-frequency behavior during the transient state.
In other words, the convergence to the steady-state of $\alpha$ is relatively poor because the initial response saturates at the upper input constraint. 
Conversely, the proposed method improves this situation by increasing the gain of the control input $u_{\beta}$ according to the size of the deviation and by finely adjusting the input within the constraint range.
From the time evolutions of the proposed PID gains shown in Fig.~\ref{fig:equilibrium_gains}, we observe different characteristics for transient and steady-state conditions.
In the transient state, the proportional and differential gains are dominant, whereas in the steady state, the integral gain plays a significant role in stabilizing the system.
The numerical results of the steady-state are equivalent to the tendency of the theoretical analysis mentioned in Section~\ref{sec:2dof_theory}.
The phenomenon of a temporary increase in $\boldsymbol{K}^p_k$ and $\boldsymbol{K}^d_k$ during the transient response can be interpreted as the result of dynamically optimizing the trade-off between the convergence property and the stabilization property.

\begin{figure}[!t]
\centering
\begin{minipage}[t]{0.48\linewidth}
\centering
    \includegraphics[bb=7 7 1004 626,width=\columnwidth]{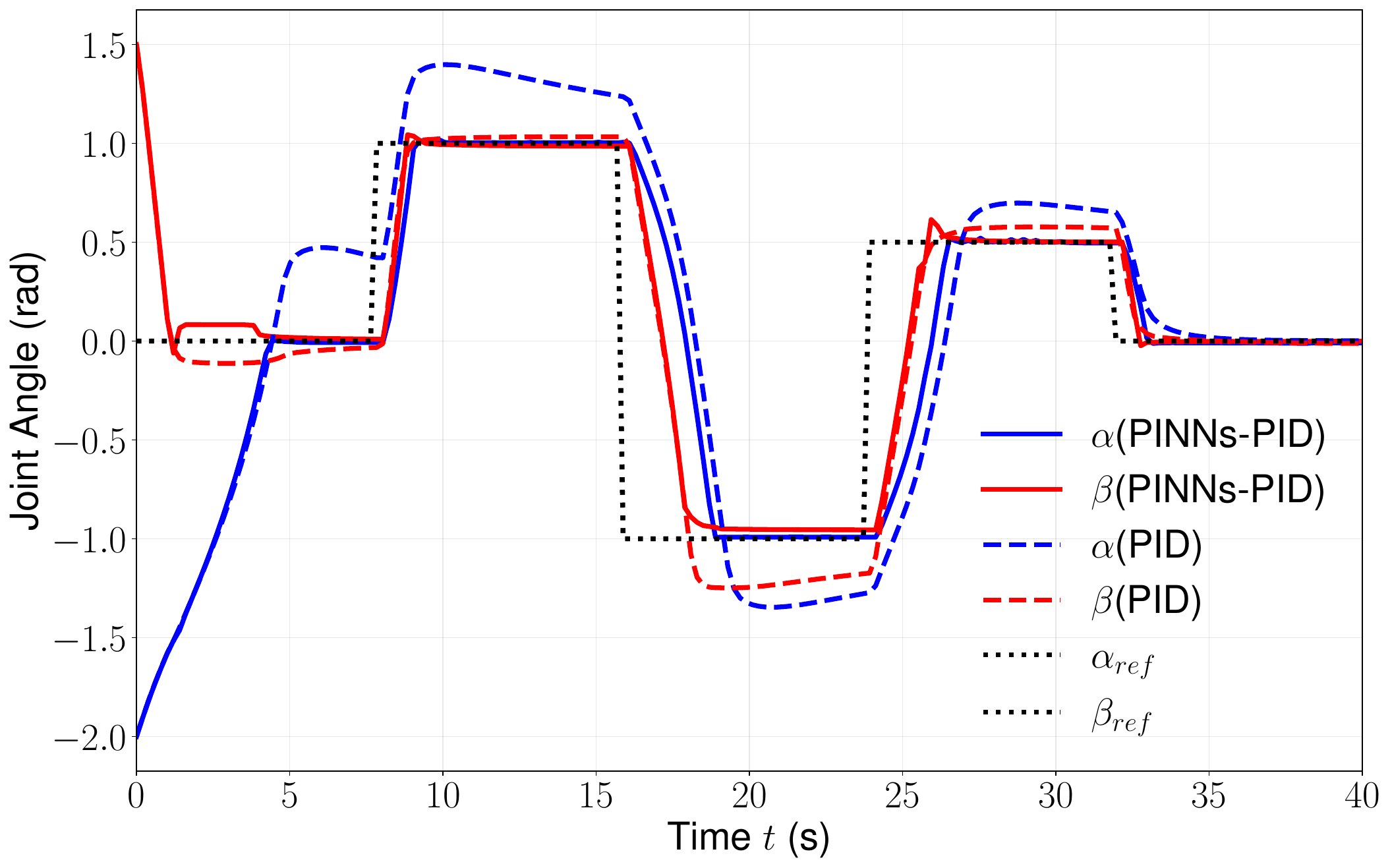}
    \subcaption{Joint angles for stepwise reference signals.}
    \label{fig:step_angles}
\end{minipage}
\hspace{0.01\linewidth}
\begin{minipage}[t]{0.48\linewidth}
\centering
    \includegraphics[bb=7 7 999 626,width=\columnwidth]{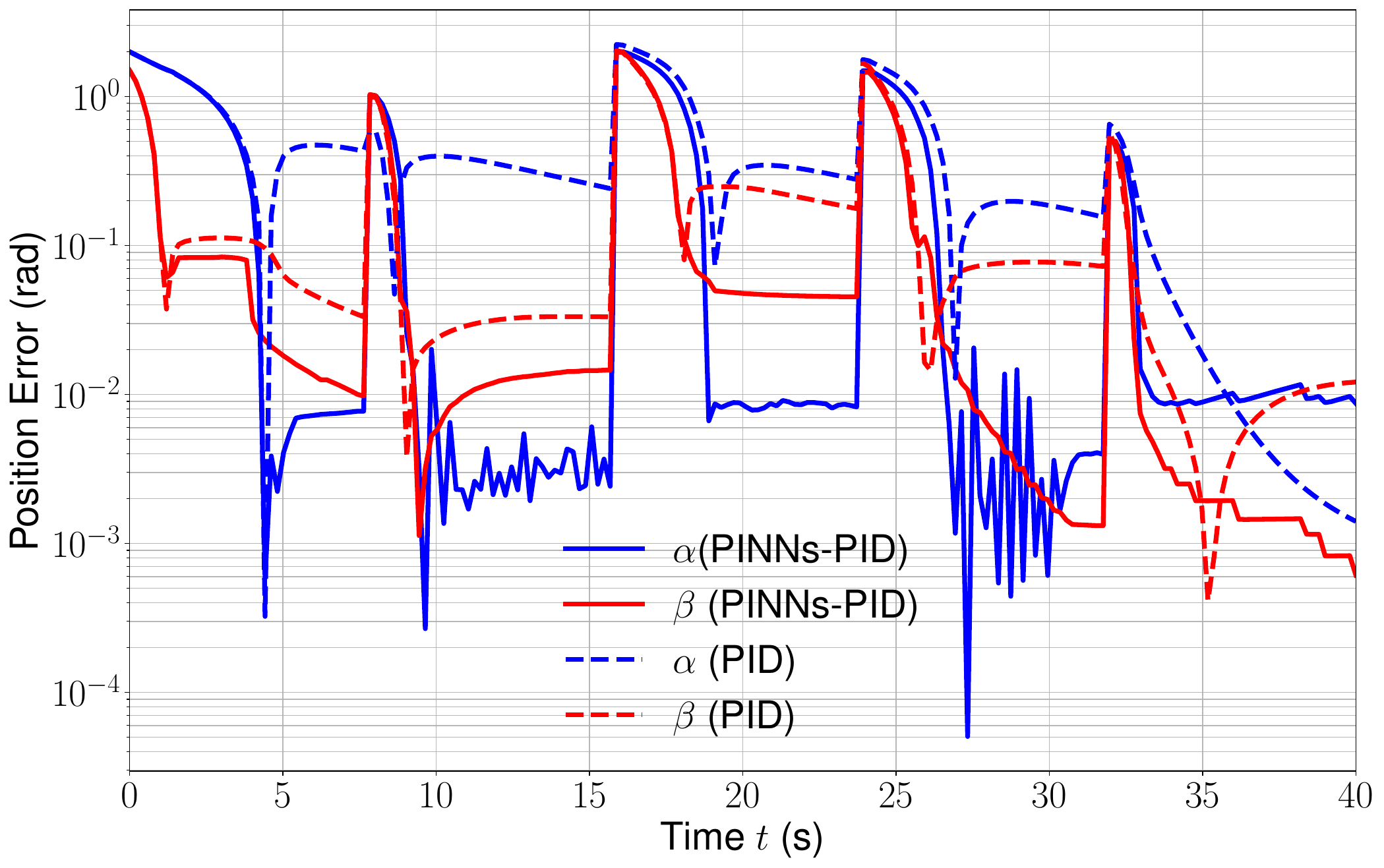}
    \subcaption{Angle position tracking errors.}
    \label{fig:step_error}
\end{minipage}
\\ \smallskip
\begin{minipage}[t]{0.48\linewidth}
\centering
    \includegraphics[bb= 6 7 1004 626,width=\columnwidth]{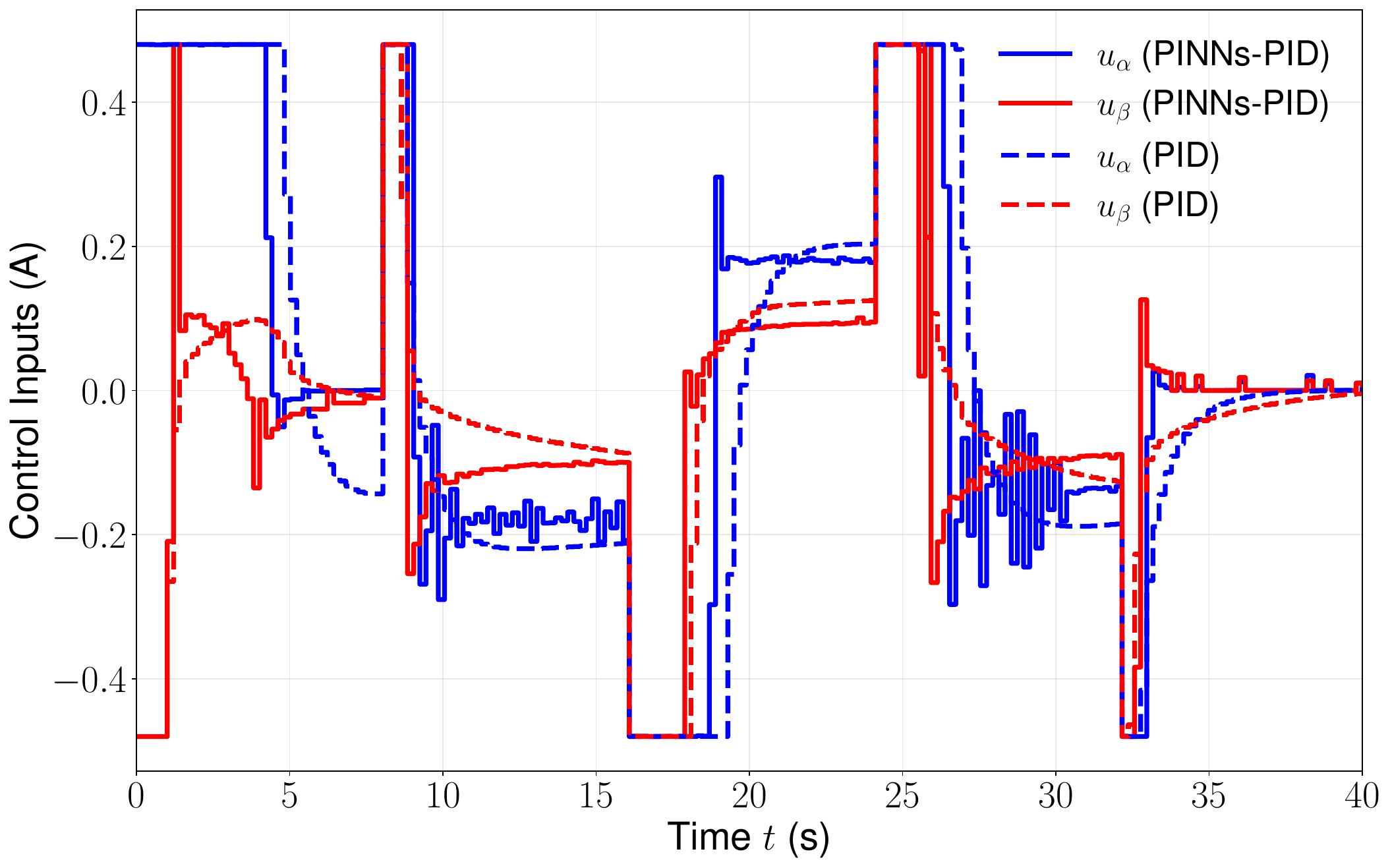}
    \subcaption{Control input signals.}
    \label{fig:step_input}
\end{minipage}
\hspace{0.01\linewidth}
\begin{minipage}[t]{0.48\linewidth}
\centering
    \includegraphics[bb=6 7 1017 631,width=\columnwidth]{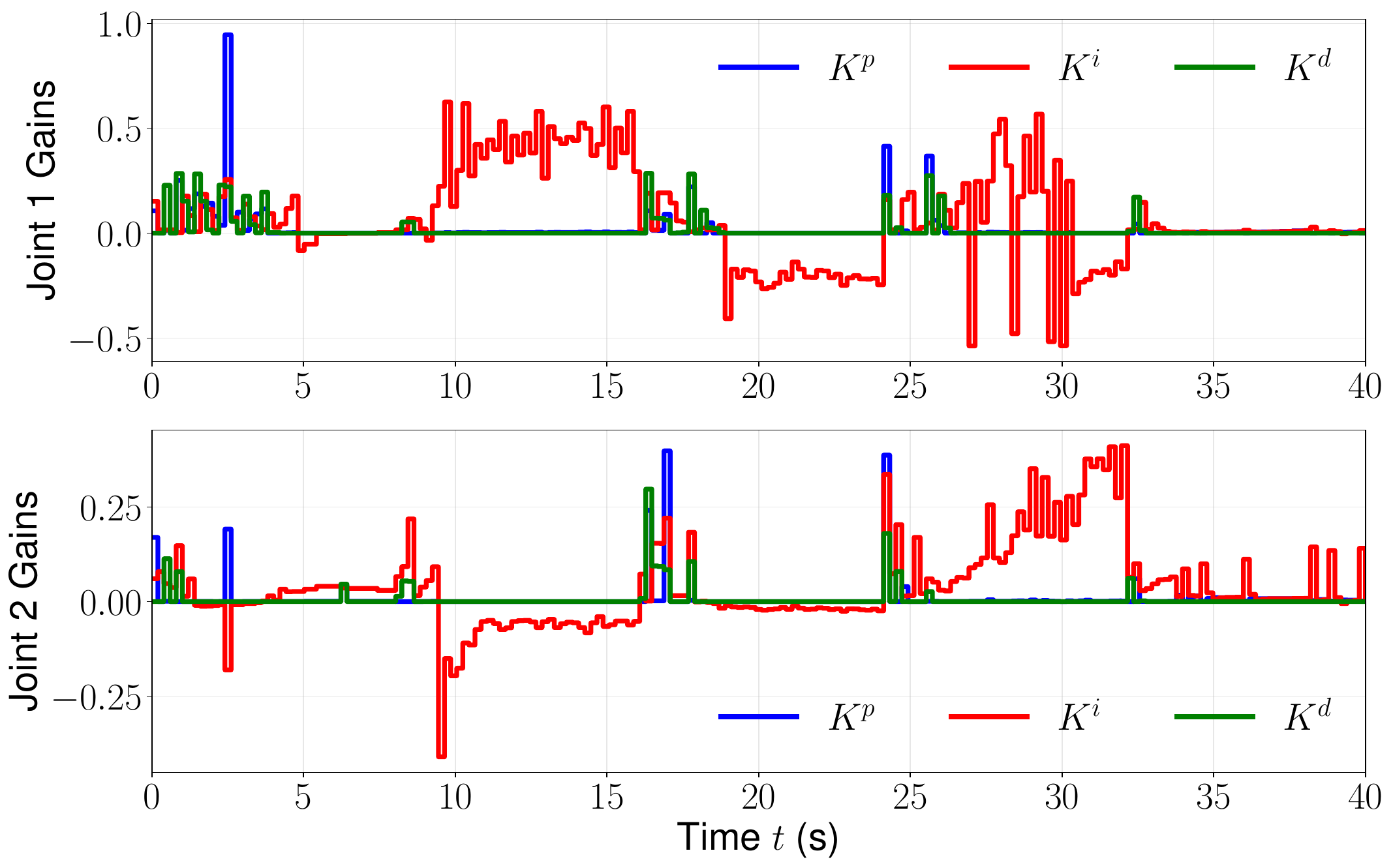}
    \subcaption{PID gains by the proposed online adaptive PID method.}
    \label{fig:step_gains}
\end{minipage}
\caption{Servo tracking performance of the proposed adaptive PID controller for stepwise reference changes in the 2-DOF manipulator.}
\label{fig:result_2DOF_servo}
\end{figure}

\subsection{Application to Servo Problem}
\label{sec:2dof_servo}
Finally, we evaluate the performance for the servo problem, which involves the tracking control problem with sequentially stepwise changes in target values.
The reference trajectory $\boldsymbol{x}^{\text{ref}}=[\alpha^{\text{ref}}, \beta^{\text{ref}},0,0]^\top$ is defined as a series of step changes in joint angles, as shown in Fig~\ref{fig:step_angles}.
The reference trajectory of the angles $[\alpha^{\text{ref}}, \beta^{\text{ref}}]^\top$ rad transitions to different target angles at specific time intervals; $[1.0, 0.0]^\top$ at $t=8$\,s, $[-1.0, 0.0]^\top$ at $16$\,s, $[0.5, 0.0]^\top$ at $24$\,s and $[0, 0]^\top$ at $32$\,s.
We emphasize that the reference trajectory is significantly affected by gravity and cannot be stabilized without control input. 
The other experimental conditions and the fixed PID method are the same as those for the regulation problem discussed in Section~\ref{sec:2dof_regulation}.

Then, the results are shown in Fig.~\ref{fig:result_2DOF_servo}. 
We can observe from the time evolutions of the angles shown in Fig.~\ref{fig:step_angles} and the error in Fig.~\ref{fig:step_error} that the fixed PID method (dashed lines) remains the steady-state tracking error at the reference except for the unstable inverted equilibrium point (dotted lines) due to the state-dependent gravity term $\boldsymbol{g}(\boldsymbol{q})$, meanwhile the proposed method (solid lines) achieves better tracking performance and convergence property on a scale of about $10^{2}$ or less during almost all the simulation time. 
In particular, focusing on the state trajectories from $8$\,s to $24$\,s, we observe that the fixed PID method required a maximum overshoot of about $1.3$ rad and a settling time of more than $5$\,s and maintains a persistent error of approximately $0.2$\,rad. 
Meanwhile, the proposed PID controller produces very little overshoot and reduces the error to below $0.02$\,rad within $2$\,s.
Similar trends were observed in the responses to the other steps.
Therefore, one of the essential advantages of the proposed method is to be capable of quickly and adaptively leading the nonlinear system to an arbitrarily admissible equilibrium state with non-zero inputs such as the gravity compensation term, given only the desired reference state trajectory by using the optimization (\ref{prob:proposed_PID}) based on the automatic differentiability of the PINNs-based prediction model.

We next discuss the control input shown in Fig.~\ref{fig:step_input} and the corresponding PID gain shown in Fig.~\ref{fig:step_gains}. 
We see from the figures that the gains of the proposed PID method are automatically and high-frequently adjusted based on the online state information of the nonlinear system and the corresponding error between the real-time state and the stepwise reference trajectory. 
The numerical results of the dominant reference-dependent integral gain in steady state are equivalent to the tendency observed in the theoretical analysis mentioned in Section~\ref{sec:2dof_theory}.
Note that, as the proposed method can effectively handle the uncertainties and variations in the system dynamics by leveraging the data-driven nature of the PINNs-based prediction model and computing the optimization quickly using only data and automatic differentiation, the proposed method is decisively different from the conventional model-based approach.

In summary, these simulation results demonstrate the superior adaptability and tracking performance of the proposed method under nonlinear dynamics and input constraints.

\begin{remark}
With the computational resources used in this study, the computation time required for learning exceeds $12$ hours, and it is difficult to apply the method to higher-dimensional states than those mentioned above due to a computational hang. 
Since the contribution of this paper is to propose an adaptive PID method based on PINNs, the applicability to higher-dimensional states is left as an issue for future research.
\end{remark}

\begin{figure}[!t]
\centering
\includegraphics[bb=0 0 559 272,width=0.7\linewidth]{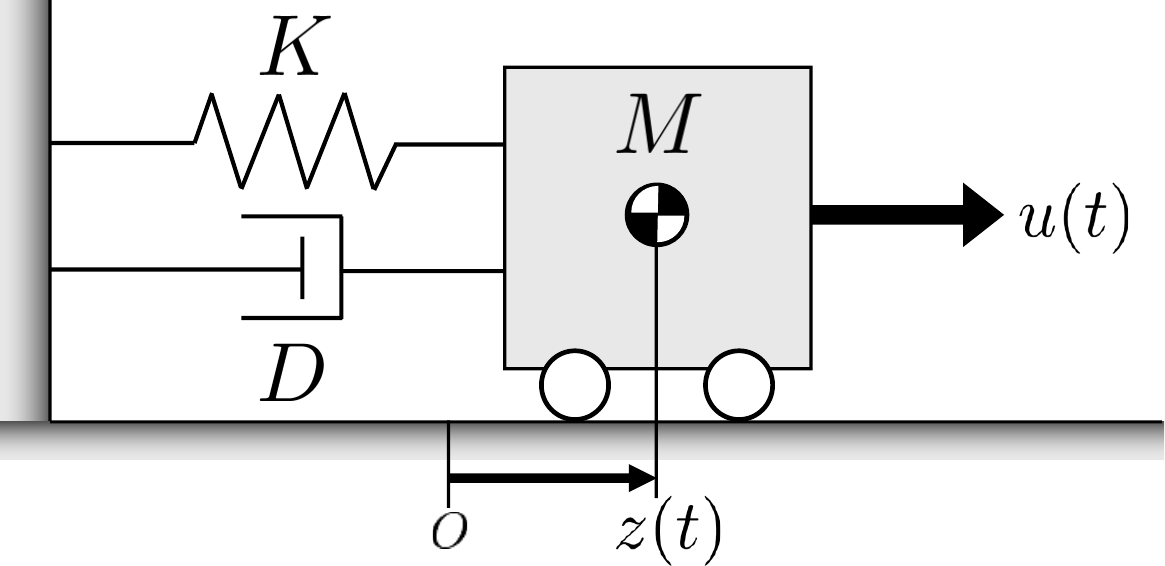}
\caption{A mass-spring-damper system.}
\label{fig:MSD-model}
\end{figure}

\section{Case Study 2: Mass-Spring-Damper System}
\label{sec:06}

To evaluate the stability margin performance of the proposed adaptive PID control, this section examines a mass-spring-damper (MSD) model, a typical example of a linear system, as shown in Fig.~\ref{fig:MSD-model}. 
This model has extensive applications in control system design methods \cite{AH06,AH95}.

Given the position $z\in\mathbb{R}$ from the equilibrium point under the unforced dynamics and the (force) control $u\in\mathbb{R}$, the mechanical motion dynamics can be given by 
\begin{equation}
M\ddot{z} + D\dot{z} + Kz = u \label{eq:MSD_sys}
\end{equation}
with the mass $M\,(>0)$, the damping factor $D\,(>0)$ and the spring constant $K\,(>0)$. 
The motion dynamics (\ref{eq:MSD_sys}) can be rewritten by the state dynamics with the state $\boldsymbol{x}:=[z,\dot{z}]^\top$: 
\begin{equation}
\frac{d}{dt}\begin{bmatrix}z \\ \dot{z}\end{bmatrix} = \begin{bmatrix}0 & 1 \\ -(K/M) & -(D/M) \end{bmatrix}\begin{bmatrix}z \\ \dot{z}\end{bmatrix} + \begin{bmatrix}0 \\ 1/M \end{bmatrix}u.
\label{eq:MSD_sys2}
\end{equation}
In this simulation, we use $M=1.0$\,kg, $K=1.0$\,N/m, $D=0.5$\,Ns/m,
$\mathbb{X}:=\{(z,\dot{z})\,|\, -2.0 \leq z \leq 2.0, -1.0 \leq \dot{z} \leq 1.0\}$, and $\mathbb{U}:=\{u\,|\,-1.0 \leq u \leq 1.0\}$.
Therefore, the system is a second-order vibration system with a damping ratio $\zeta = D/(2\sqrt{MK}) = 0.25$ and a natural angular frequency $\omega_n = \sqrt{K/M} = 1.0$ rad/s.

\begin{figure}[!t]
\centering
\begin{minipage}[t]{0.48\linewidth}
\centering
    \includegraphics[bb=6 12 988 625,width=\columnwidth]{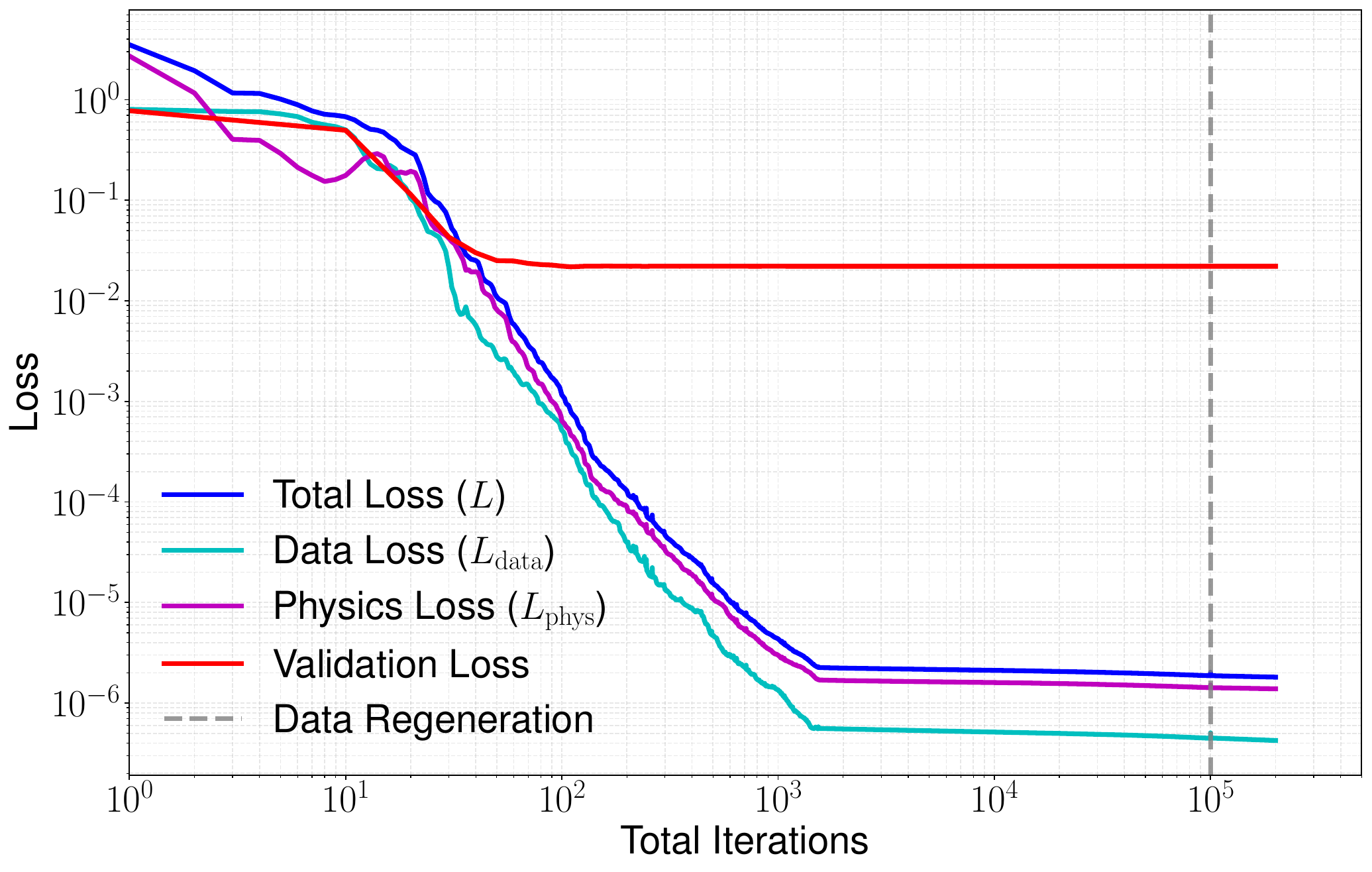}
    \subcaption{Evolution of loss functions during training.}
    \label{fig:loss_curves_spring_mass}
\end{minipage}
\hspace{0.01\linewidth}
\begin{minipage}[t]{0.48\linewidth}
\centering
    \includegraphics[bb=6 7 1100 666,width=\columnwidth]{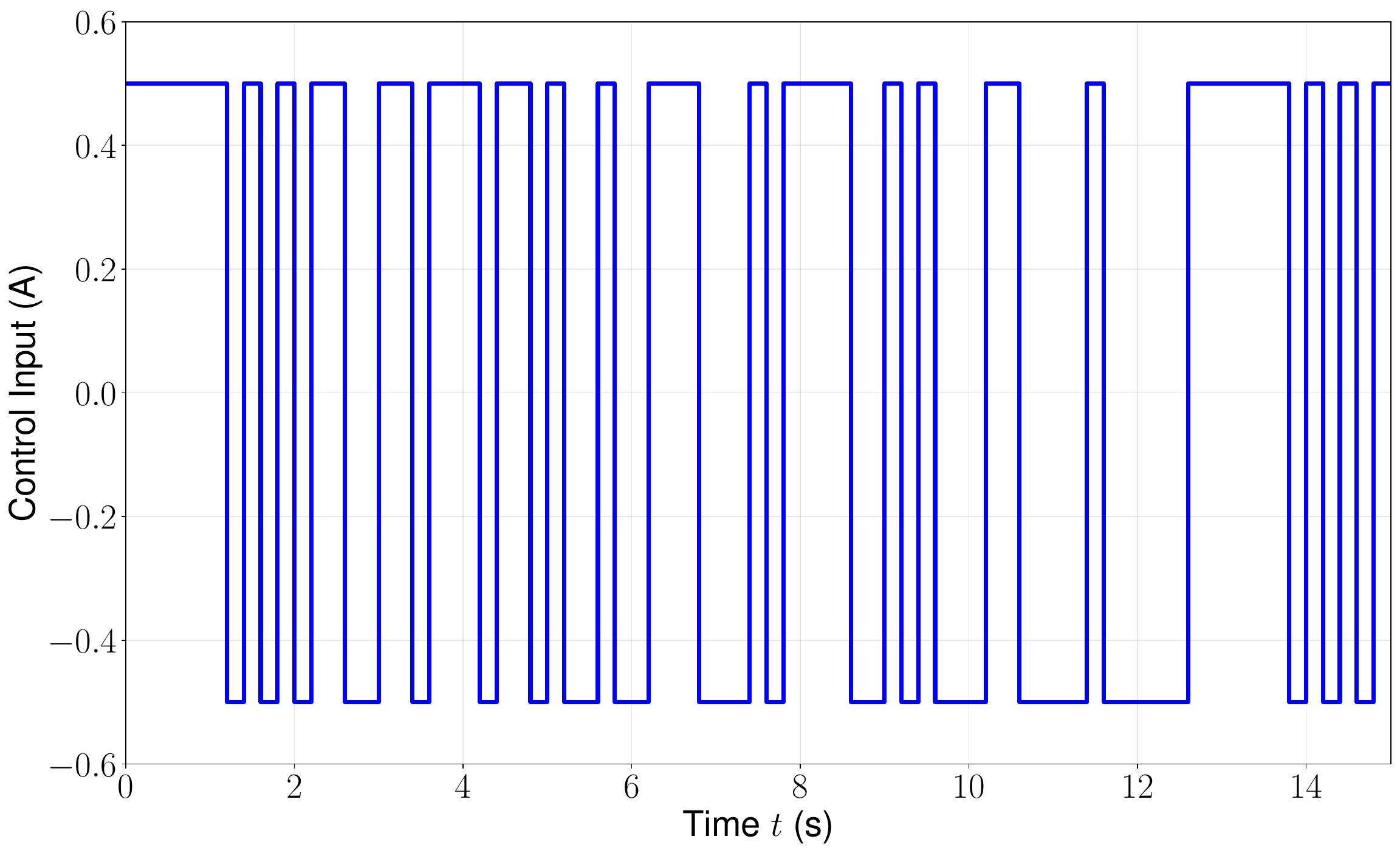}
    \subcaption{Control input signal for the test dataset.}
    \label{fig:MSD_sys_PINN_inputs}
\end{minipage}
\\ \smallskip
\begin{minipage}[t]{0.48\linewidth}
\centering
    \includegraphics[bb=7 7 1100 668,width=\columnwidth]{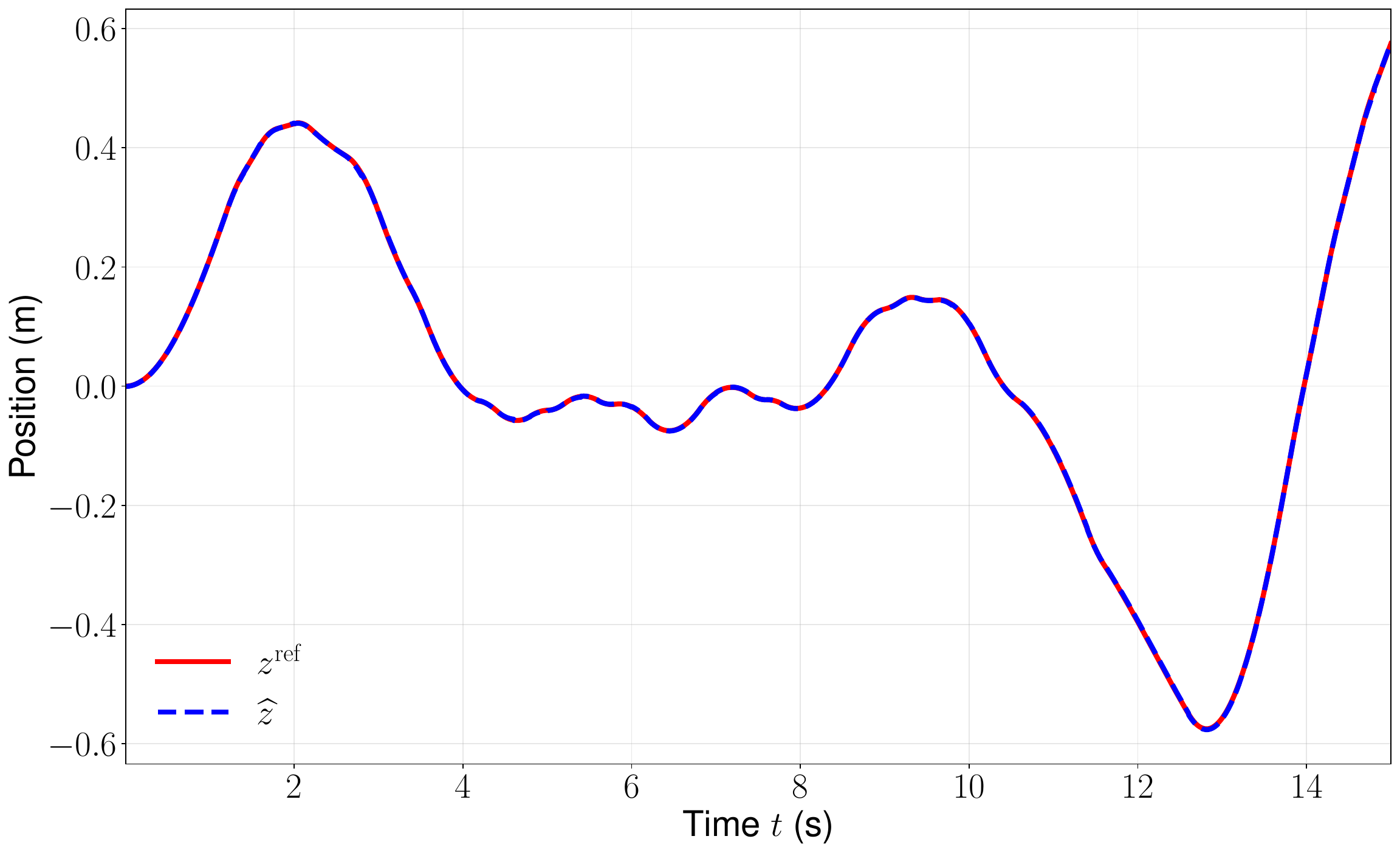}
    \subcaption{Time evolutions of the state with the input trajectory.}
    \label{fig:MSD_sys_PINN_state}
\end{minipage}
\hspace{0.01\linewidth}
\begin{minipage}[t]{0.48\linewidth}
\centering
    \includegraphics[bb= 7 7 1091 668,width=\columnwidth]{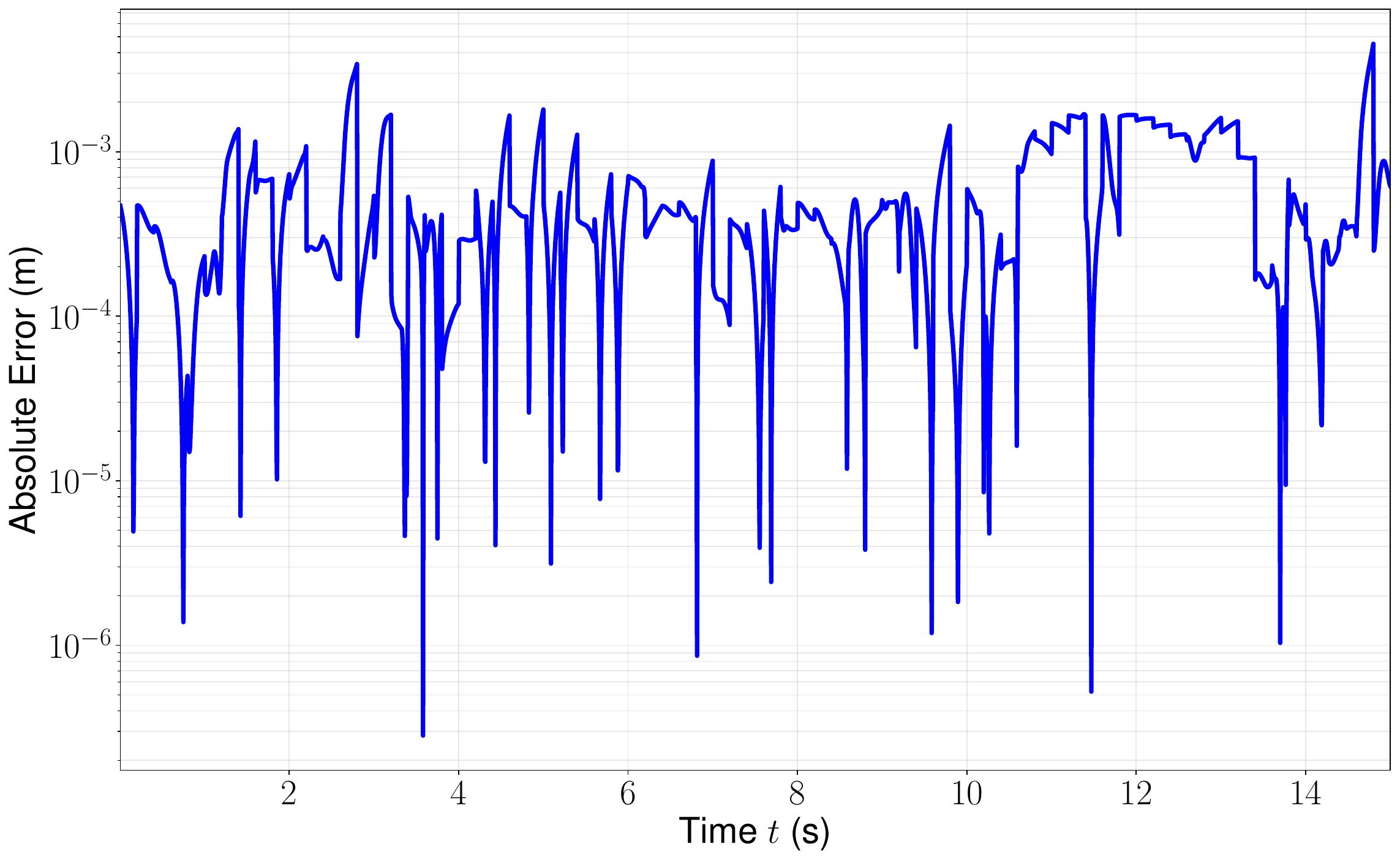}
    \subcaption{Time evolutions of prediction absolute error for the state.}
    \label{fig:MSD_sys_PINN_MAE}
\end{minipage}
\caption{Validation of the trained PINNs model for the MSD system.}
\label{fig:MSD_sys_PINN}
\end{figure}

\subsection{Evaluation on PINNs-based model}
\label{sec:MSD_PINN_eval}

The training dataset to obtain the model (\ref{eq:pinn_control}) is generated by the same method shown in Sections~\ref{sec:03_implement} and \ref{sec:05-PINNmodel}. 
We consider the optimization problem (\ref{prob:proposed_PID}) with the prediction horizon $T=4.0$ based on the recurrent self-loop PINNs prediction of the sampling period $\Delta T=0.200$\,s.
The training datasets, $\mathcal{D}_{\text{data}}$ and $\mathcal{D}_{\text{phys}}$, are composed of $N_{\text{data}}=2\times 10^4$ and $N_{\text{phys}}=1\times 10^5$ samples, respectively. These samples are generated with $\Delta T+\epsilon= 0.250$\,s.

We use the predict state trajectory model (\ref{eq:pinn_control}) composed by a forward propagation-type deep neural network with $4$-dimensions ($k,z_{k},\dot{z}_{k},u_{k}$) at the input layer, $3$-hidden layers, which is based on $32$-units at each layer, and $2$-dimensions ($z_{k+1},\dot{z}_{k+1}$) at the output layer. 
The PINNs model training spans a total of $2\times 10^5$ iterations.
The optimization of $\omega$ in (\ref{eq:pinns-loss}) employs the L-BFGS method~\cite{LN89}.
To enhance model accuracy and avoid overfitting to specific data patterns, the training dataset is regenerated at $1\times 10^5$ iteration intervals.
From the loss function evolution during the training process in Fig.~\ref{fig:loss_curves_spring_mass}, the learning exhibits steady convergence reaching a stable minimum after approximately $2\times 10^3$ iterations.
The terminal value of the loss function $L(\omega)$ for the trained PINNs model is approximately $7 \times 10^{-5}$, with the physics imbalance loss term $L_{\text{phys}}(\omega)$ contributing the dominant component.

The validation dataset contains admissible initial states, control inputs, and the corresponding accurate state trajectories of the MSD system (\ref{eq:MSD_sys2}) over $40$\,s.
The test dataset consists of state trajectories of the dynamical system (\ref{eq:MSD_sys2}) obtained using the fourth-order Runge-Kutta method (RK4) with the control inputs presented in Fig.~\ref{fig:MSD_sys_PINN_inputs}.
To track the reference trajectory $z^{\text{ref}}$ (solid lines), the predicted state trajectory $\hat{z}$ for the test input trajectory using the obtained PINNs model (dashed lines) is shown in Fig.~\ref{fig:MSD_sys_PINN_state}.
The absolute error between the reference trajectory and the predicted trajectory is shown in Fig.~\ref{fig:MSD_sys_PINN_MAE}.
The trained PINNs model achieves a mean absolute error (MAE) of $5.70\times 10^{-4}$ for the position trajectory.
The evidence of the figures suggests that the obtained PINNs model has not overfitted and has achieved sufficient generalization performance for the MSD system control.

\begin{figure}[!t]
\centering
\begin{minipage}[t]{0.48\linewidth}
\centering
    \includegraphics[clip,bb=7 7 1017 634,width=\columnwidth]{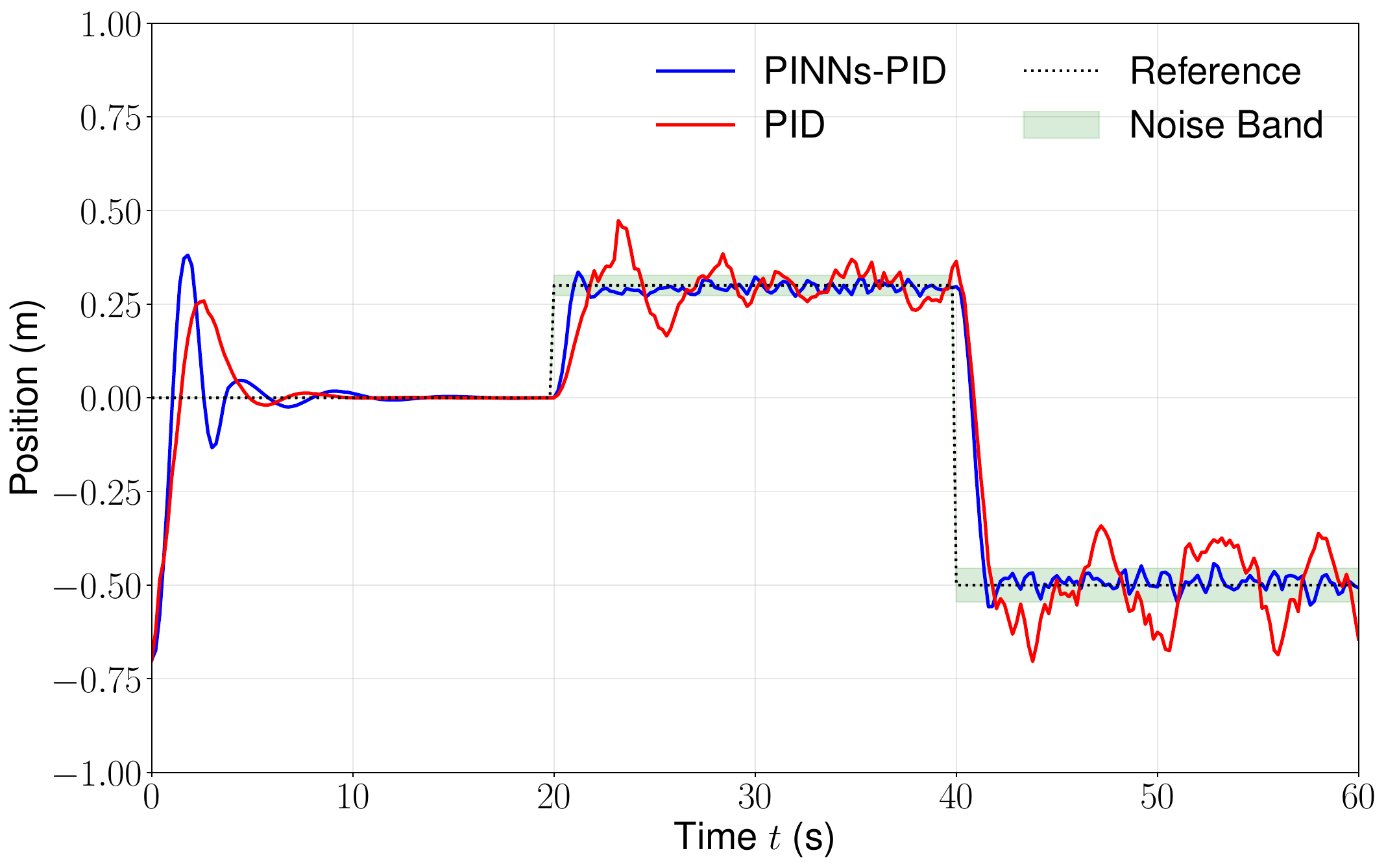}
    \subcaption{Position responses for stepwise reference signals.}
    \label{fig:smd_position_comparison}
\end{minipage}
\hspace{0.01\linewidth}
\begin{minipage}[t]{0.48\linewidth}
\centering
    \includegraphics[clip,bb=7 7 999 626,width=\columnwidth]{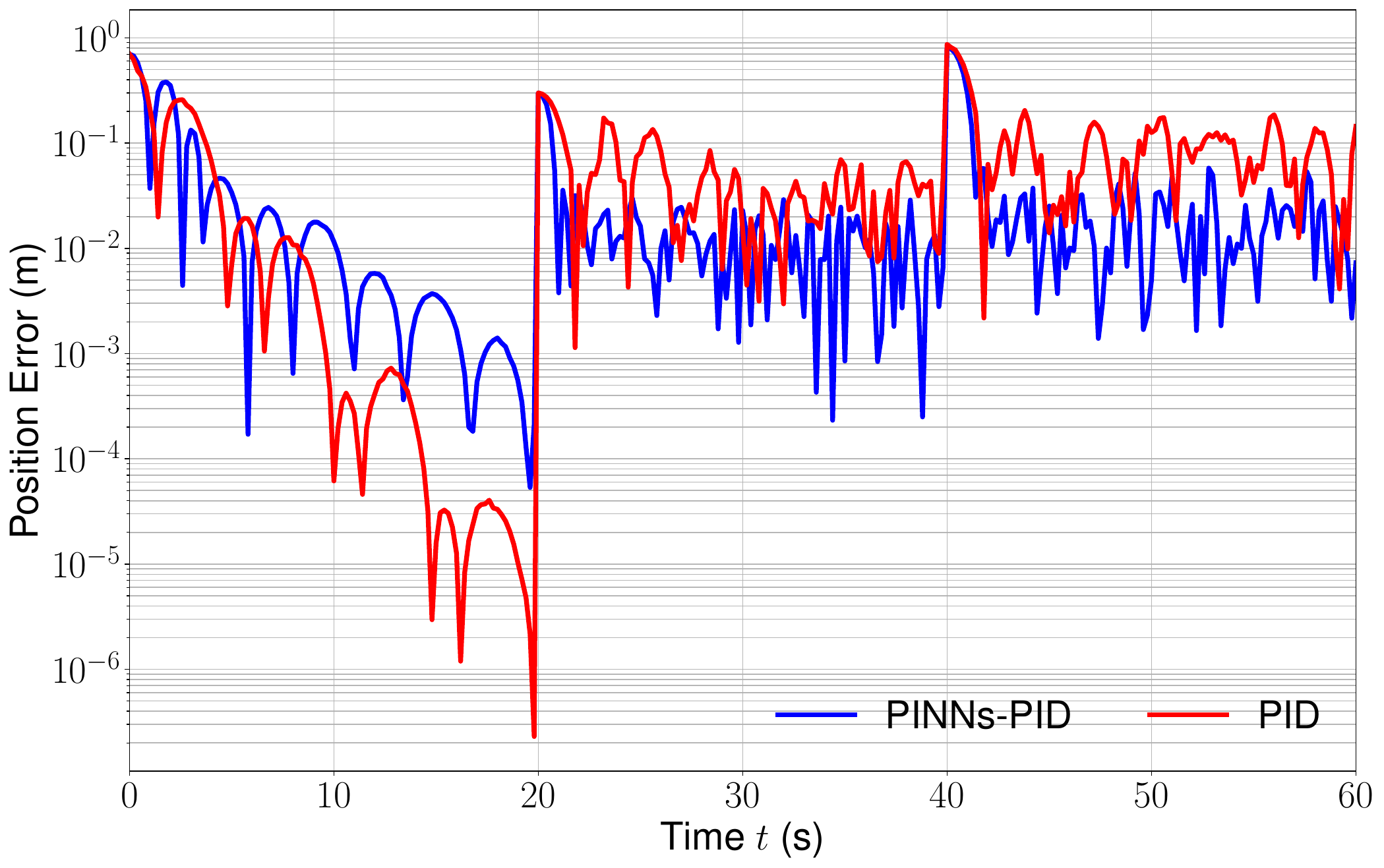}
    \subcaption{Position tracking errors under measurement noise.}
    \label{fig:smd_tracking_error_comparison}
\end{minipage}
\\ \smallskip
\begin{minipage}[t]{0.48\linewidth}
\centering
    \includegraphics[clip,bb=6 7 1017 626,width=\columnwidth]{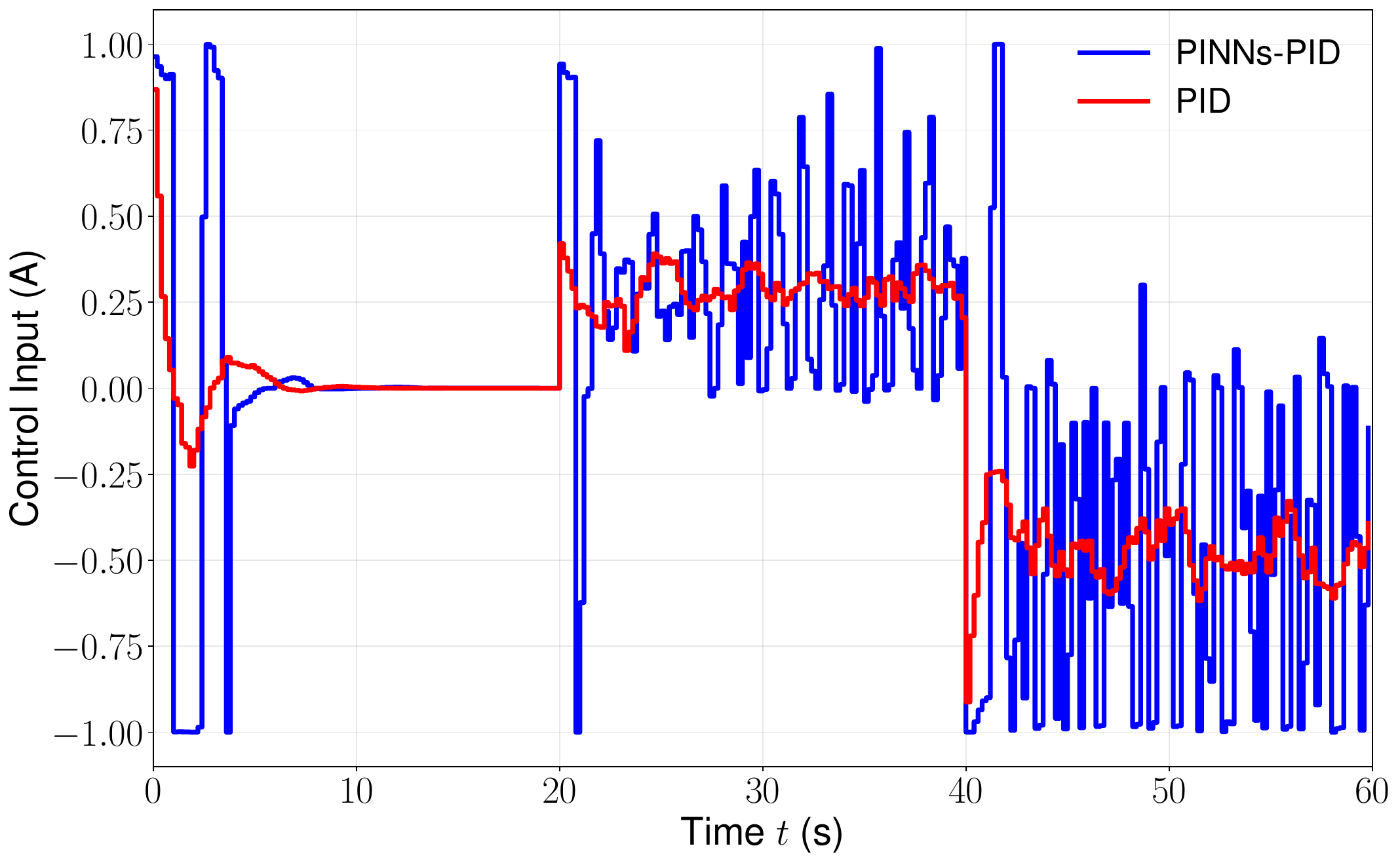}
    \subcaption{Control input signals during tracking.}
    \label{fig:smd_control_input_comparison}
\end{minipage}
\hspace{0.01\linewidth}
\begin{minipage}[t]{0.48\linewidth}
\centering
    \includegraphics[clip,bb=6 7 962 631,width=\columnwidth]{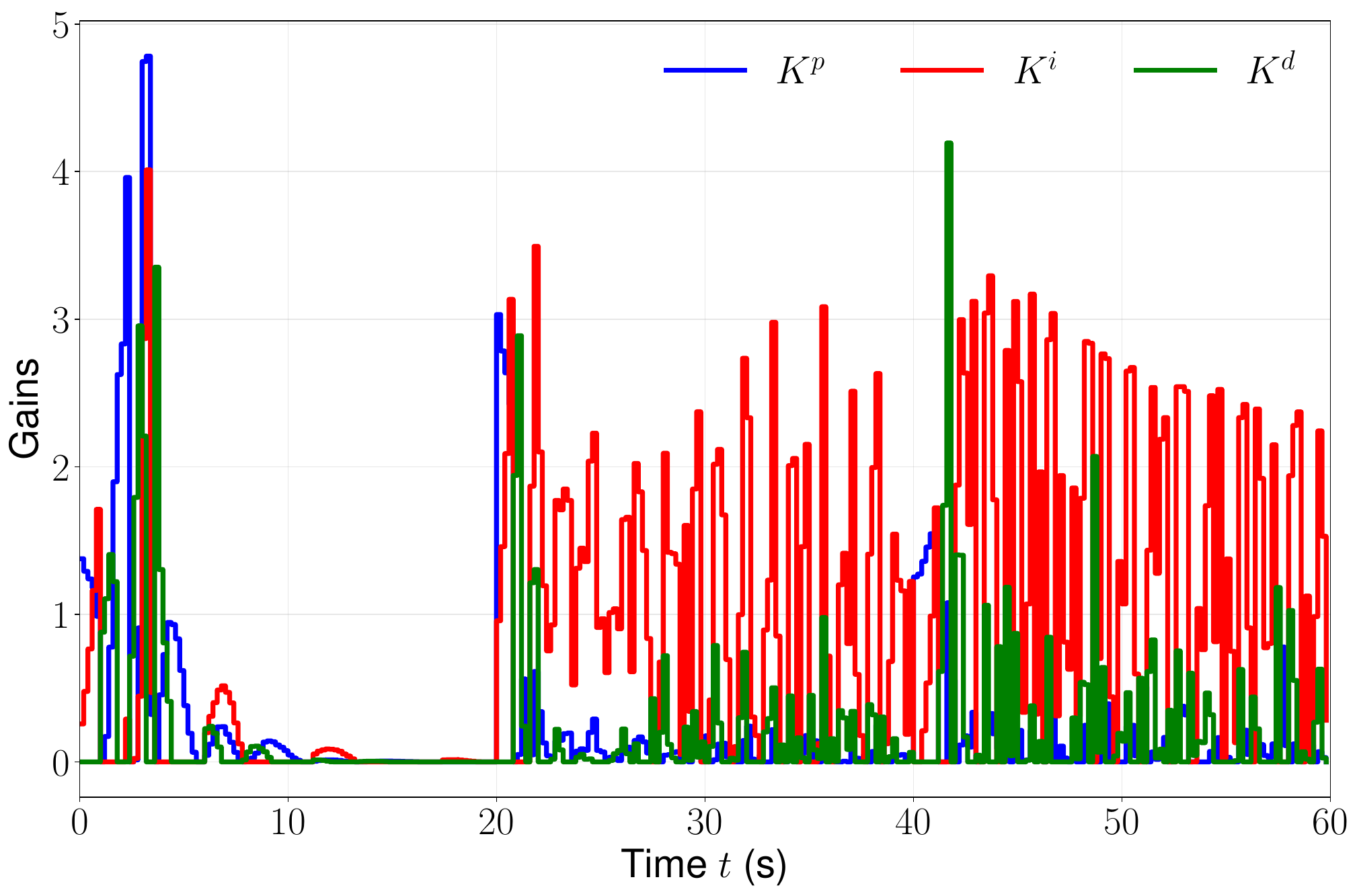}
    \subcaption{Time evolution of proposed adaptive PID gains.}
    \label{fig:smd_gains_evolution}
\end{minipage}
\caption{Comparative servo tracking performance for the MSD system under 3\% measurement noise.}
\label{fig:result_MSD_servo}
\end{figure}

\subsection{Application to Servo Problem}
\label{sec:MSD_servo}

We next evaluate the performance of the proposed method for the servo problem of the dynamical system (\ref{eq:MSD_sys2}), similar to Section~\ref{sec:2dof_servo}. 

Note that in this servo problem, while the PINNs model was trained without noise consideration, we introduce 3\% multiplicative Gaussian sensor noise in the dynamical system (\ref{eq:MSD_sys2}) to test the robustness of the control performance under measurement uncertainty.
The reference trajectory $\boldsymbol{x}^{\text{ref}}=[z^{\text{ref}}, 0]^\top$ is defined as a series of step changes in position, as shown in Fig.~\ref{fig:smd_position_comparison}. 
The position reference trajectory $z^{\text{ref}}$ transitions to different target positions at specific time intervals: $0.3$ at $t=20$\,s, $-0.5$ at $40$\,s.
These transitions test the system's ability to track both positive and negative step changes in the presence of measurement noise.
We set the initial state $\boldsymbol{x}(0) = [-0.7, 0.0]^\top$, and the control sets $\mathbb{U} =[-1.0, 1.0]$, $\mathcal{F}=[0.0, 5.0]\times[0.0, 5.0]\times[0.0, 5.0]$.
The adaptive PID control is implemented during $60.0$\,s using the MPC-based adaptive PID gain optimization (\ref{prob:proposed_PID}) with $\Delta T=0.200$\,s, $t_f=60.0$\,s, and the weight matrices of (\ref{eq:lqr}) given by $Q = {\rm diag}(1000, 1)$ and $R = 0.01$.
The optimizer to solve the problem (\ref{prob:proposed_PID}) uses Adam \cite{KB14} with a learning rate of $10^{-2}$, and the maximum iteration is set as $2.0\times 10^4$.
For comparison, we employ a conventional fixed-gain PID controller with $\boldsymbol{F}_k\equiv [K^p_k,K^i_k,K^d_k] = [1.2, 1.0, 1.2]$ for all time steps $k$. 
These gains are tuned empirically to achieve acceptable tracking performance while maintaining stability under the input constraints.

The simulation results for both controllers are shown in Fig.~\ref{fig:result_MSD_servo}. 
From the position time evolution shown in Fig.~\ref{fig:smd_position_comparison} and the tracking error depicted in Fig.~\ref{fig:smd_tracking_error_comparison}, significant performance differences are observed between the two methods. 
The fixed PID controller (red line) exhibits persistent oscillations and fails to converge to the reference positions. 
After each step change at $t=20$\,s and $40$\,s, the fixed PID response exhibits considerable overshoot, followed by sustained oscillations around the reference value, with tracking errors remaining at the order of $10^{-1}$\,m throughout the simulation.
In contrast, the proposed adaptive PID method (blue line) achieves rapid convergence to each reference position with minimal overshoot. 
The tracking error of the proposed method decreases to the order of $10^{-2}$ to $10^{-3}$\,m, effectively reaching the noise band imposed by the 3\% multiplicative measurement uncertainty.

Next, we discuss the control input shown in Fig.~\ref{fig:smd_control_input_comparison} and the corresponding PID gains shown in Fig.~\ref{fig:smd_gains_evolution}. 
The proportional gain $K^p$ exhibits peaks during transient periods immediately following reference changes, providing rapid error correction.
The integral gain $K^i$ gradually increases during steady-state periods to eliminate residual errors while maintaining stability.
The derivative gain $K^d$ shows moderate variations, contributing to damping during transitions.
This coordinated gain adaptation enables the controller to switch between rapid transient response and stable steady-state regulation based on the current operating conditions.

\begin{figure}[!t]
\centering
\begin{minipage}[t]{0.48\linewidth}
\centering
    \includegraphics[clip,bb=6 12 1100 589,width=\columnwidth]{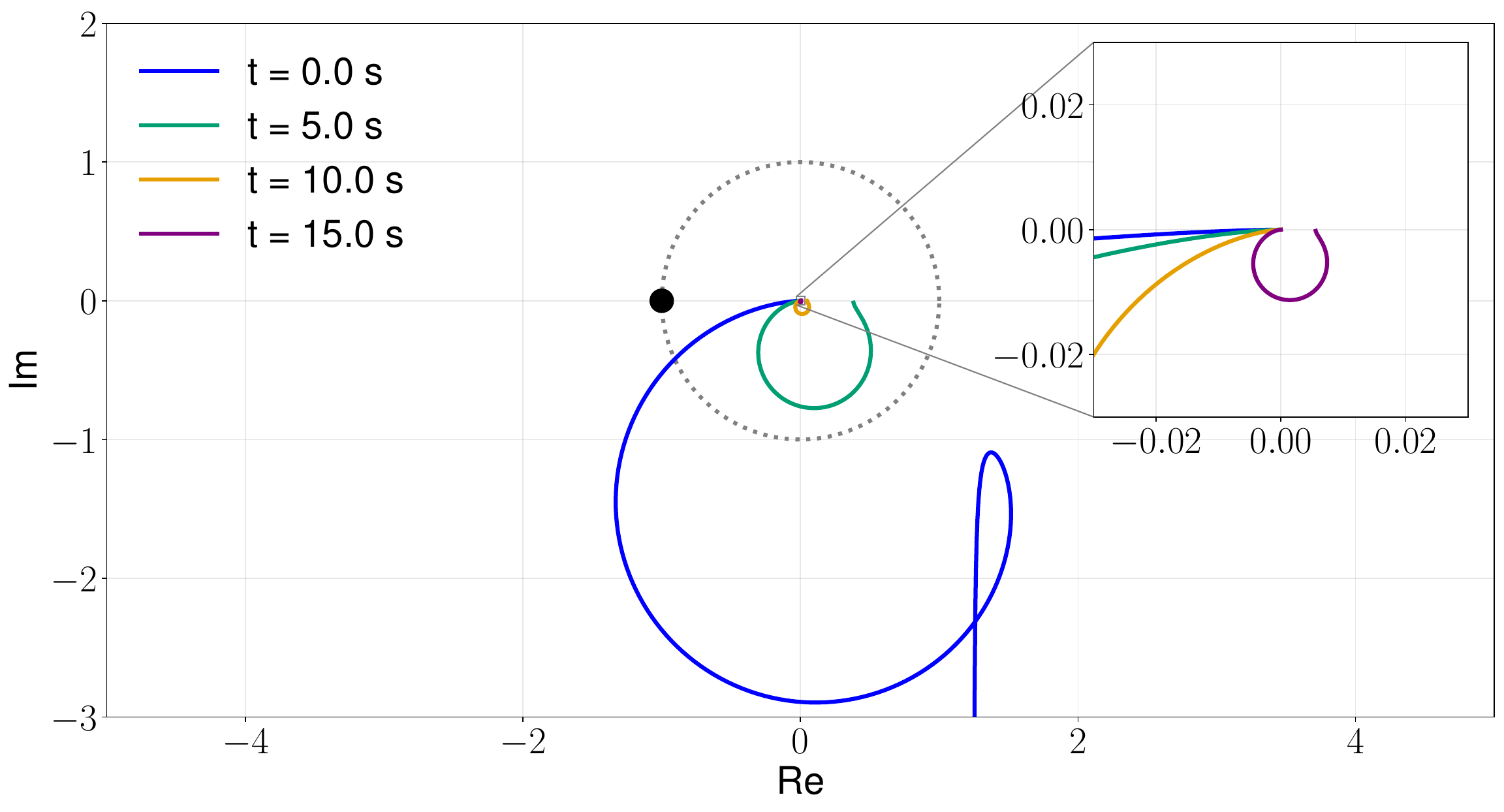}
    \subcaption{Nyquist plot for $[0, 20]$\,s.}
    \label{fig:nyquist_0_20}
\end{minipage}
\hspace{0.01\linewidth}
\begin{minipage}[t]{0.48\linewidth}
\centering
    \includegraphics[clip,bb= 6 12 972 525,width=\columnwidth]{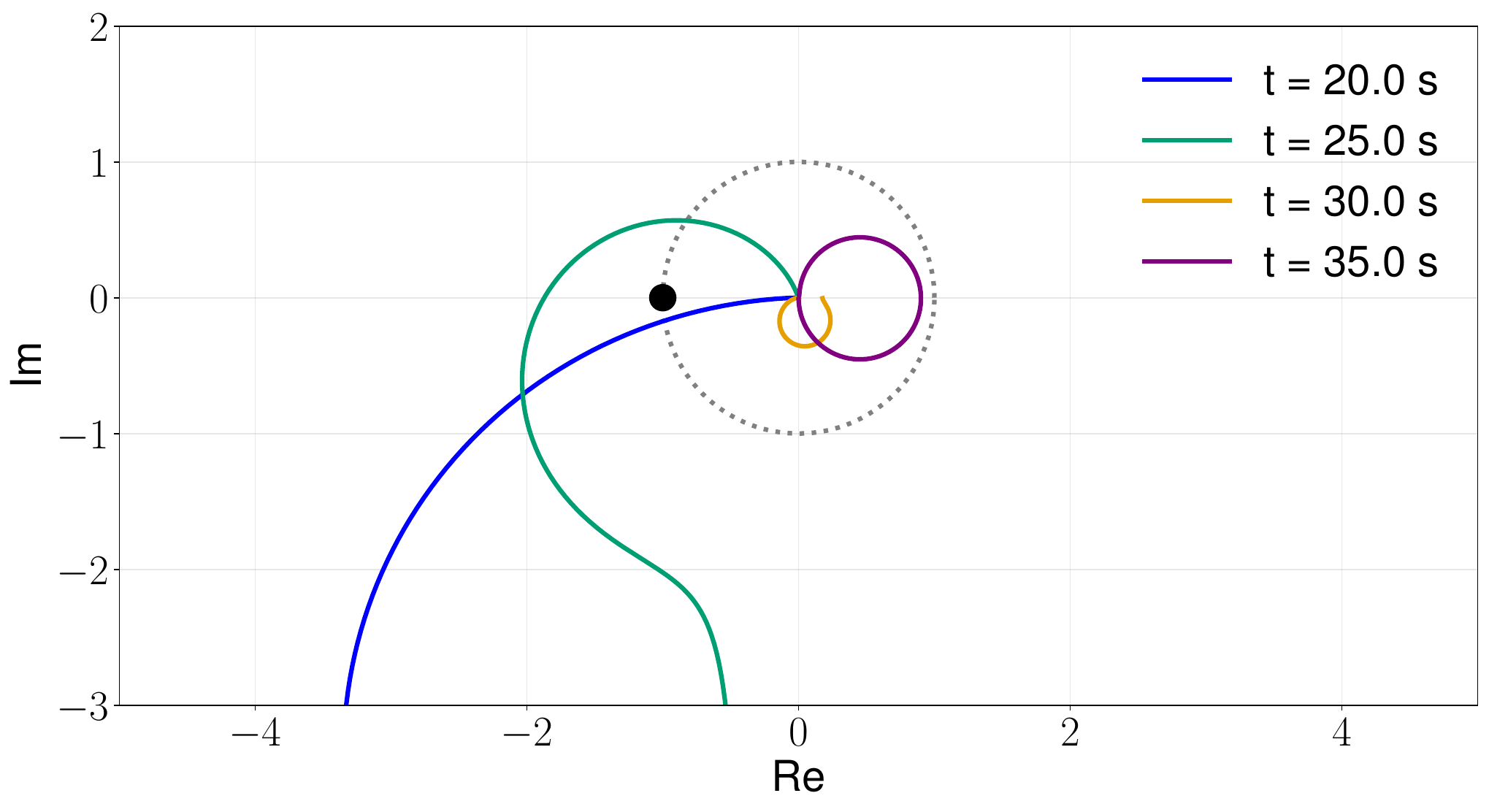}
    \subcaption{Nyquist plot for $[20, 40]$\,s.}
    \label{fig:nyquist_20_40}
\end{minipage}
\\ \smallskip
\begin{minipage}[t]{0.48\linewidth}
\centering
    \includegraphics[clip,bb=6 12 972 525,width=\columnwidth]{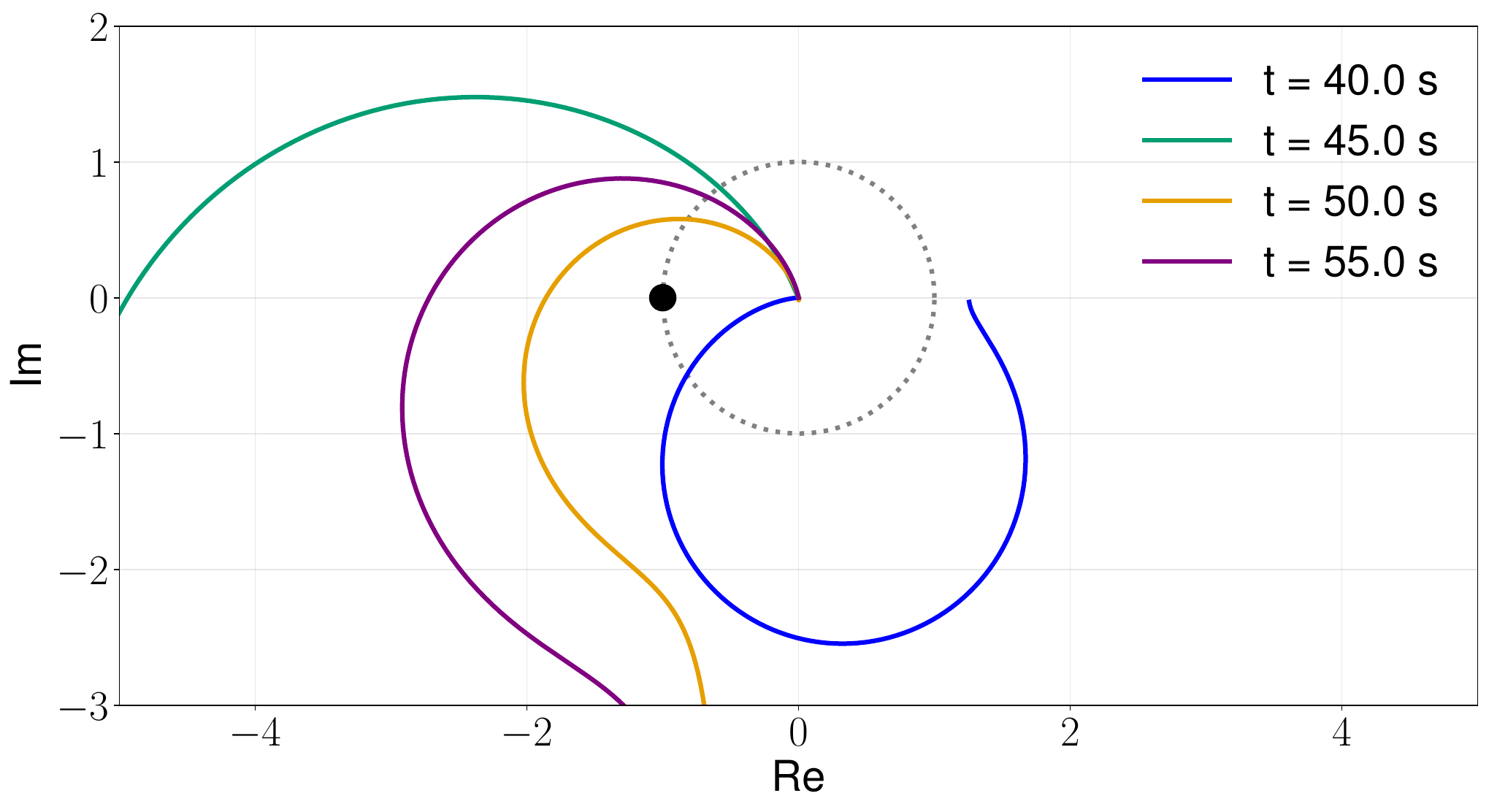}
    \subcaption{Nyquist plot for $[40, 60]$\,s.}
    \label{fig:nyquist_40_60}
\end{minipage}
\hspace{0.01\linewidth}
\begin{minipage}[t]{0.48\linewidth}
\centering
    \includegraphics[clip,bb=6 7 1017 626,width=\columnwidth]{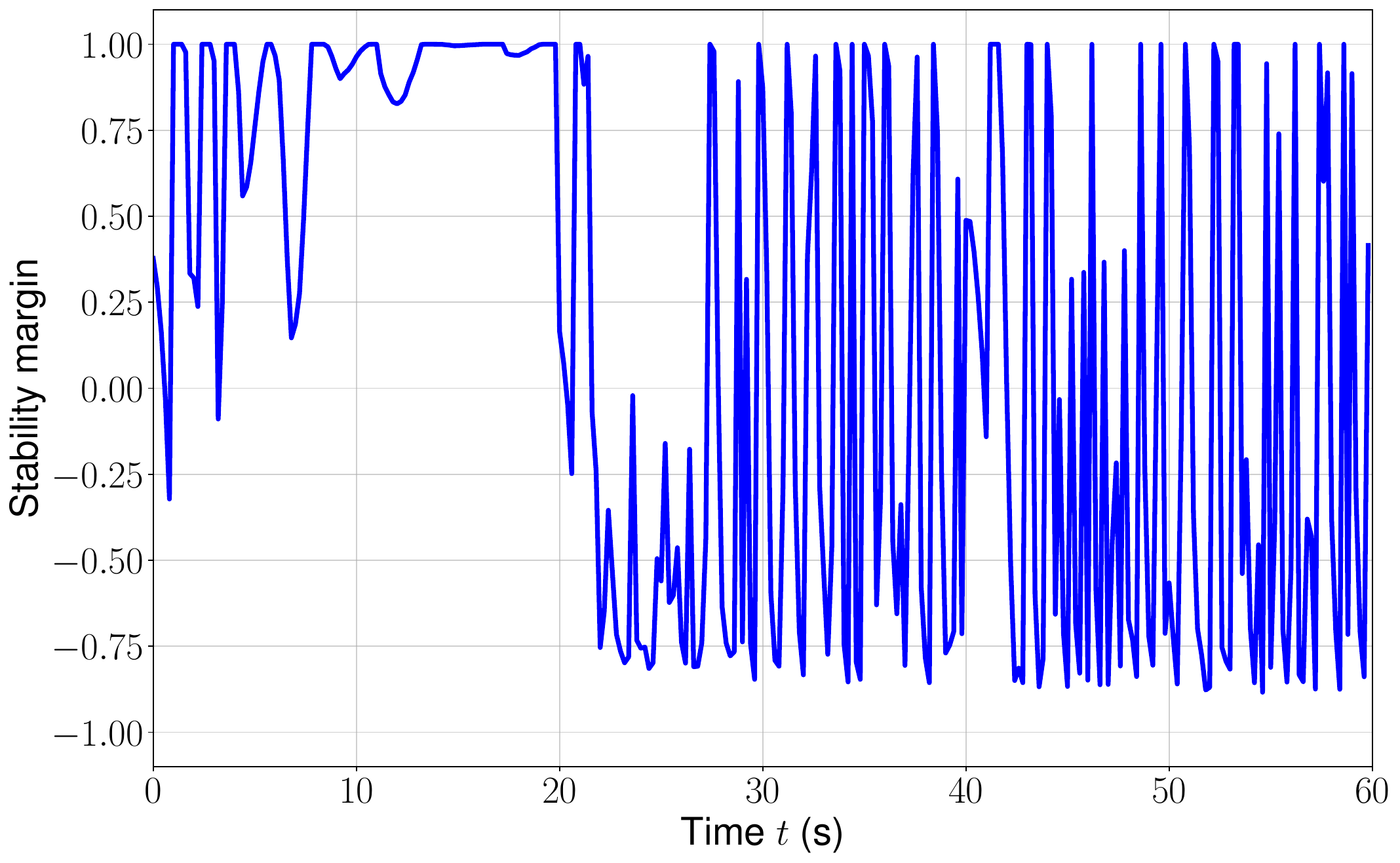}
    \subcaption{Temporal evolution of stability margins.}
    \label{fig:stability_evolution}
\end{minipage}
\caption{Stability analysis of the proposed adaptive PID controller without stability-guaranteed constraints: (a)--(c) Nyquist plots at different time intervals and (d) time-varying stability margins during servo control.}
\label{fig:stability_analysis}
\end{figure}

\begin{figure}[!t]
\centering
\includegraphics[bb=7 7 1098 668,width=0.7\linewidth]{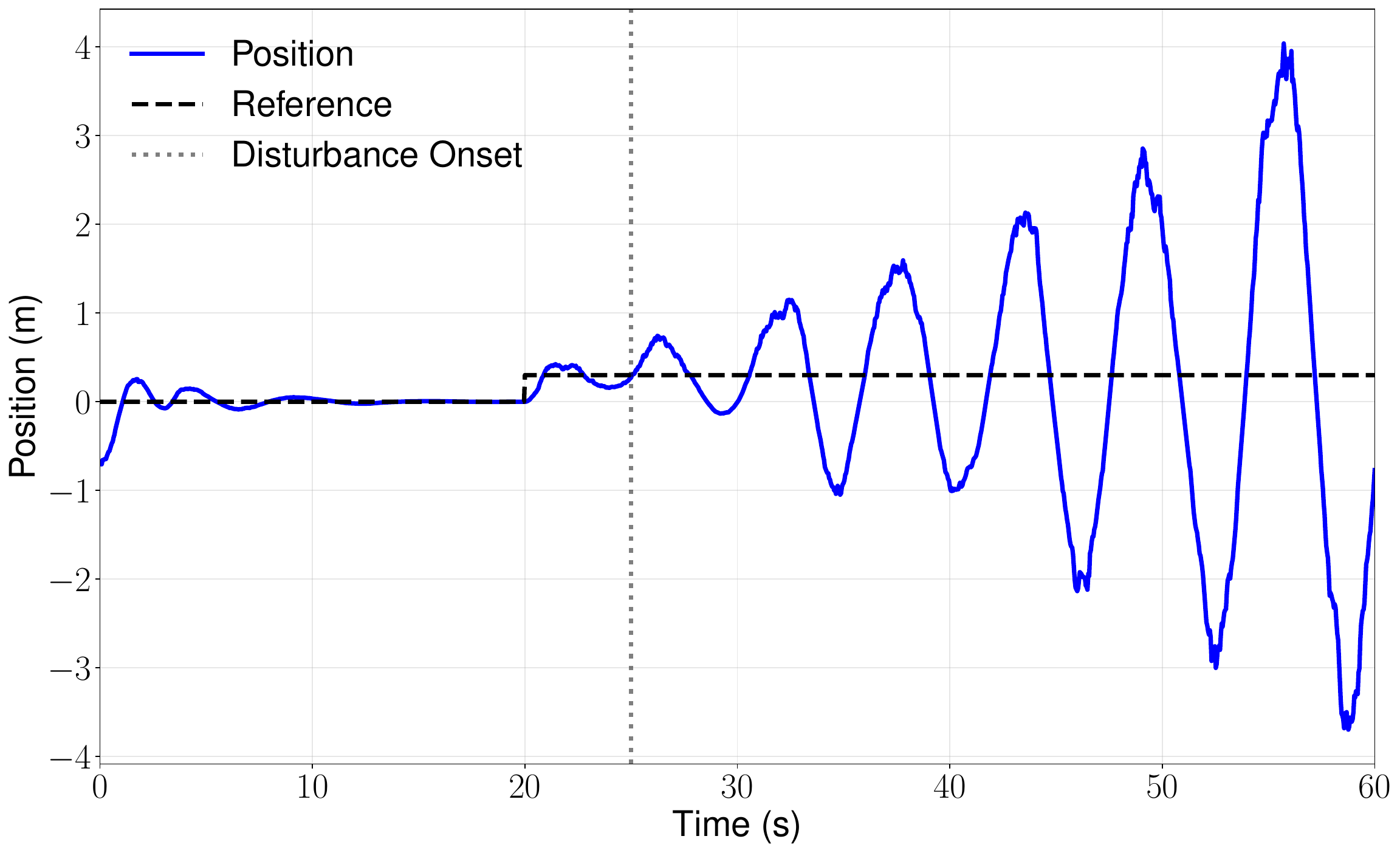}
\caption{System response with time-varying gains ($t \leq 25$\,s) followed by frozen gains and disturbance $d(t)=0.3$ ($t > 25$\,s), showing divergent behavior.}
\label{fig:disturbance_response}
\end{figure}

\subsection{Stability Analysis}
\label{sec:V-C}

In this subsection, we analyze the stability condition and stability margins of the proposed adaptive PID controller.
Due to the linear time-invariant system (\ref{eq:MSD_sys2}), we can evaluate stability using the open-loop transfer function at each discrete time step $k$.
Given the PID gains $\boldsymbol{F}_k = [K^p_k, K^i_k, K^d_k]$ at time $k$, the corresponding open-loop transfer function, assuming the gains are frozen at their instantaneous values, is
\begin{align}
L_k(s) &= \frac{1}{Ms^2 + Ds + K} \left(K^p_k + \frac{K^i_k}{s} + K^d_k s\right). 
\label{eq:open_loop_frozen}
\end{align}
Although this frozen-time analysis does not guarantee global stability for time-varying controlled systems, it provides valuable insights into the controller's instantaneous behavior.

To analyze the stability characteristics of the proposed adaptive PID controller, we examine the instantaneous open-loop transfer function (\ref{eq:open_loop_frozen}) throughout the servo control operation. 
The Nyquist plots at different time intervals during the simulation of Fig.~\ref{fig:result_MSD_servo} are shown in Figs.~\ref{fig:nyquist_0_20}--\ref{fig:nyquist_40_60}.
The temporal evolution of the stability margins, which is the signed distance of the shortest distance to the critical point $-1$, is depicted in Fig.~\ref{fig:stability_evolution}.
Note that the sign of the stability margin is positive when the system is internally stable, and negative when it is not.
As Fig.~\ref{fig:stability_evolution} shows, the optimized gain after $20$s does not guarantee the closed-loop stability. 
Note that the average gain crossover frequency for the unconstrained case is $1.15$\,rad/s, indicating relatively slow closed-loop dynamics.

To verify this instability, we conducted an additional simulation with a hybrid gain strategy: optimized time-varying gains were applied for $t \leq 25$\,s as in the original simulation, while for $t > 25$\,s, the gains were frozen at their $t=25$\,s values and a constant load disturbance $0.3$ was introduced to the control input $u(t)$. 
From the resulting time response shown in Fig.~\ref{fig:disturbance_response}, the system response diverges after the gain freezing at $t=25$\,s, confirming the instability of the frozen-gain configuration.

As illustrated in Fig.~\ref{fig:result_MSD_servo}, the utilization of the MPC appears to facilitate successful convergence to the reference trajectory. 
However, this stability analysis reveals that the optimization process can yield potentially destabilizing gains during certain operational phases, as shown in Figs.~\ref{fig:stability_evolution} and \ref{fig:disturbance_response}. 
This observation highlights the importance of explicitly incorporating stability-guaranteed constraints into the optimization formulation to ensure robust closed-loop performance across all operating conditions.

To address the stability issues identified above, we reformulate the servo problem from Section~\ref{sec:MSD_servo} by incorporating explicit stability-guaranteed constraints.
Applying the Routh-Hurwitz stability criterion to the open-loop transfer function (\ref{eq:open_loop_frozen}), the closed-loop stability conditions for the PID gains are derived as
\begin{equation}
g_k(\boldsymbol{F}_k) = (K^d_k + D)(K^p_k + K) - MK^i_k > 0, \label{eq:stability_condition-1}
\end{equation}
and $K^p_k, K^i_k, K^d_k \geq 0$.
As the positive gain condition is inherently satisfied through the feasible set $\mathcal{F}$, the inequality constraint (\ref{eq:stability_condition-1}) must be explicitly enforced to guarantee the internal stability of the system. 
One of the options is that the feasible set $\mathcal{F}$ is redefined to satisfy the inequality constraint (\ref{eq:stability_condition-1}).
However, since the PINNs-based model (\ref{eq:pinn_control}) is slightly different from the nominal model (\ref{eq:MSD_sys2}), not all values contained within a set satisfying the inequality constraint (\ref{eq:stability_condition-1}) can necessarily be stabilized. 
Moreover, from the primary purpose of using the PINNs-based model, it is necessary to realize a form even in nonlinear systems.
Therefore, to incorporate this stability condition into the optimal control problem framework (\ref{prob:proposed_PID}), we modify the regularization term to include a logarithmic barrier method, i.e., 
\begin{equation}
\Theta(\boldsymbol{F}_k) = \|\boldsymbol{F}_k\|^2 - \frac{1}{\rho}\log(g_k(\boldsymbol{F}_k)) \label{eq:modify_cost_theta}
\end{equation}
instead of (\ref{lqr_cost_theta}), where $\rho$ is the barrier parameter that decreases linearly from $10^4$ to $10^{-3}$ throughout the optimization iterations, progressively tightening the stability-guaranteed constraint while maintaining numerical tractability.

\begin{figure}[!t]
\centering
\begin{minipage}[t]{0.48\linewidth}
\centering
    \includegraphics[clip,bb=7 7 1017 634,width=\columnwidth]{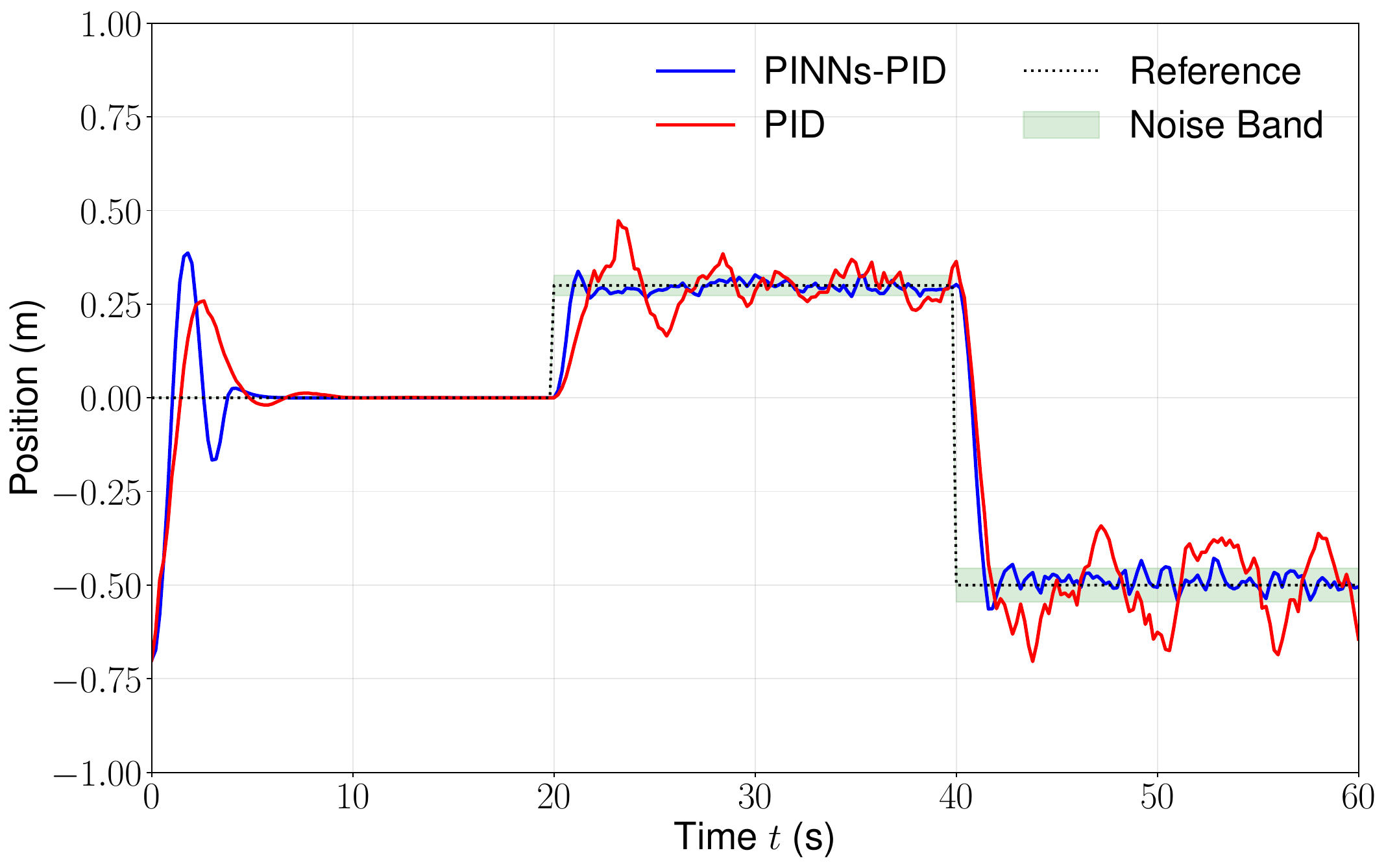}
    \subcaption{Position responses.}
    \label{fig:stable_position}
\end{minipage}
\hspace{0.01\linewidth}
\begin{minipage}[t]{0.48\linewidth}
\centering
    \includegraphics[clip,bb=7 7 999 626,width=\columnwidth]{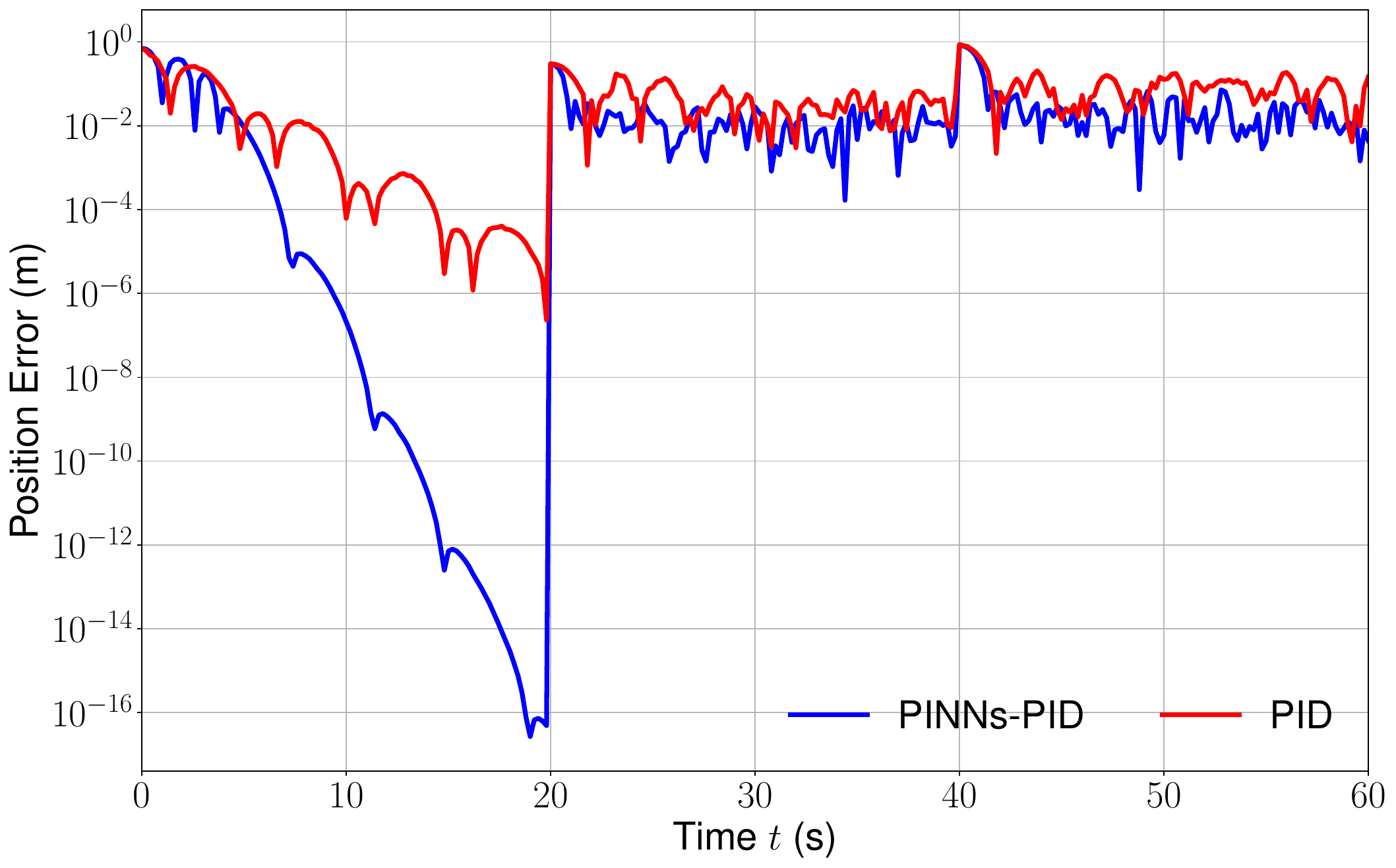}
    \subcaption{Tracking errors under measurement noise.}
    \label{fig:stable_tracking_error}
\end{minipage}
\\ \smallskip
\begin{minipage}[t]{0.48\linewidth}
\centering
    \includegraphics[clip,bb=6 7 1017 626,width=\columnwidth]{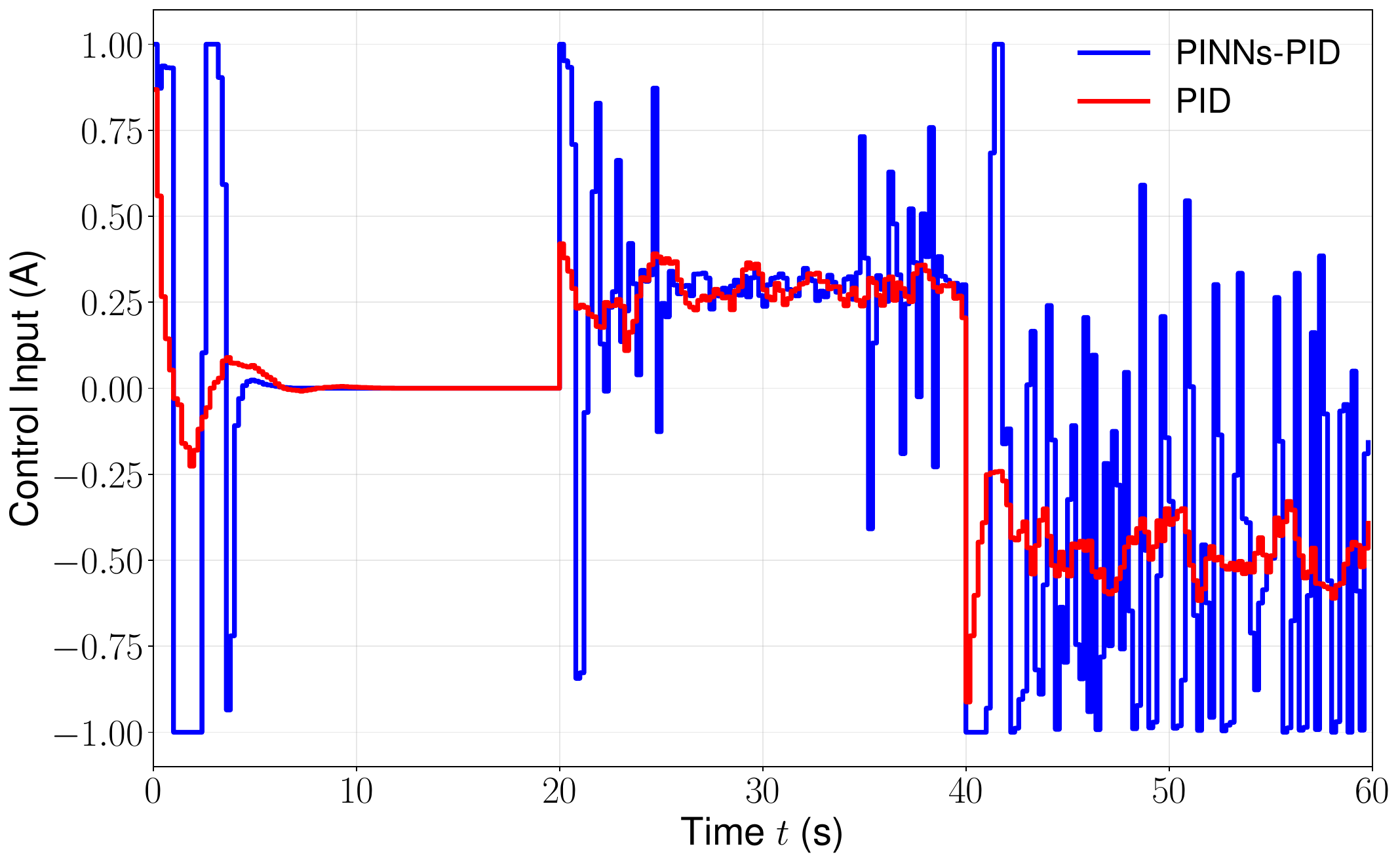}
    \subcaption{Control input signals.}
    \label{fig:stable_control_input}
\end{minipage}
\hspace{0.01\linewidth}
\begin{minipage}[t]{0.48\linewidth}
\centering
    \includegraphics[clip,bb=6 7 962 631,width=\columnwidth]{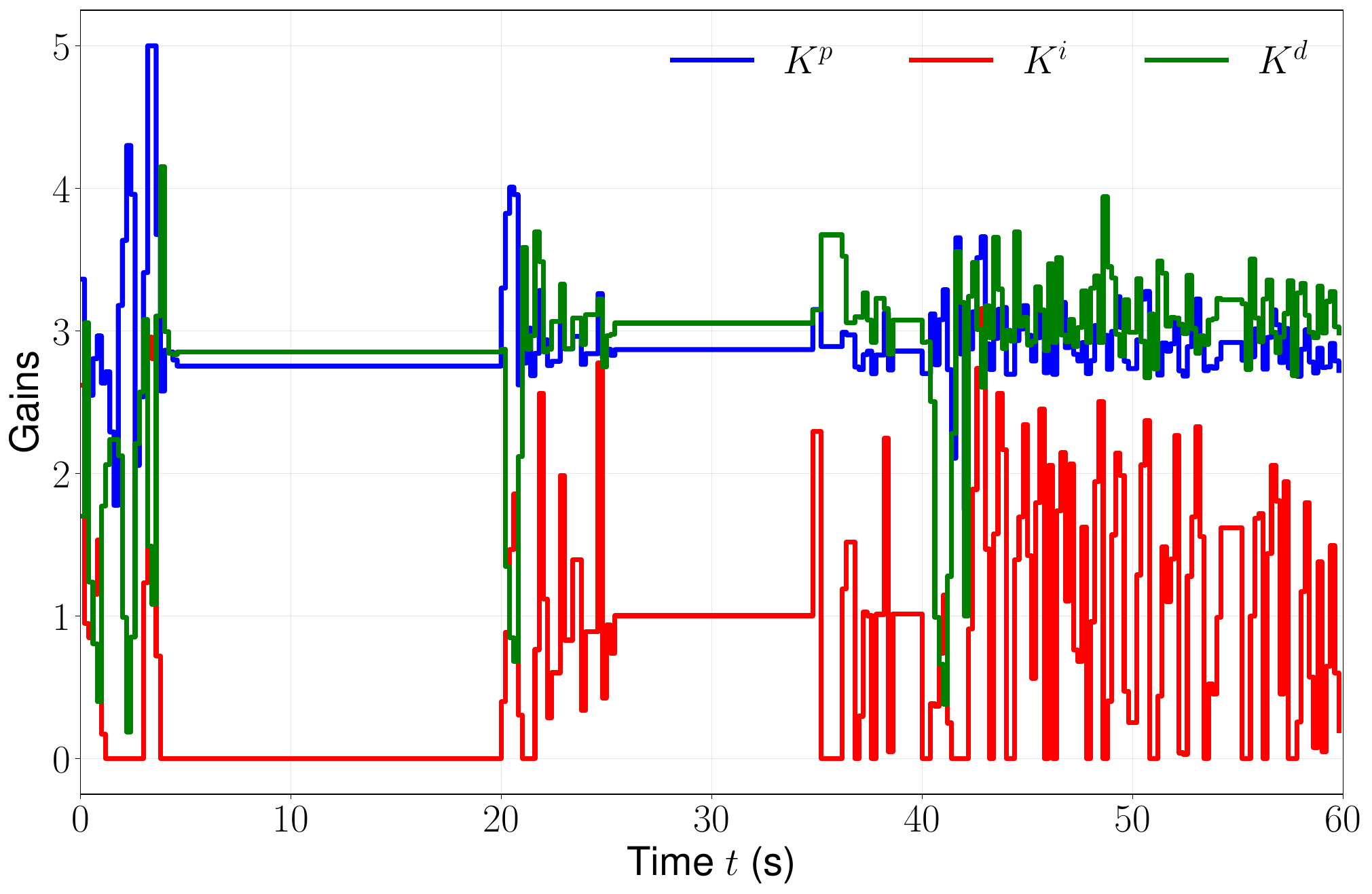}
    \subcaption{Time evolution of adaptive PID gains.}
    \label{fig:stable_gains}
\end{minipage}
\caption{Performance of the stability-guaranteed adaptive PID controller for the MSD servo problem.}
\label{fig:stable_servo_results}
\end{figure}

\begin{figure}[!t]
\centering
\begin{minipage}[t]{0.48\linewidth}
\centering
    \includegraphics[clip,bb=6 12 977 573,width=\columnwidth]{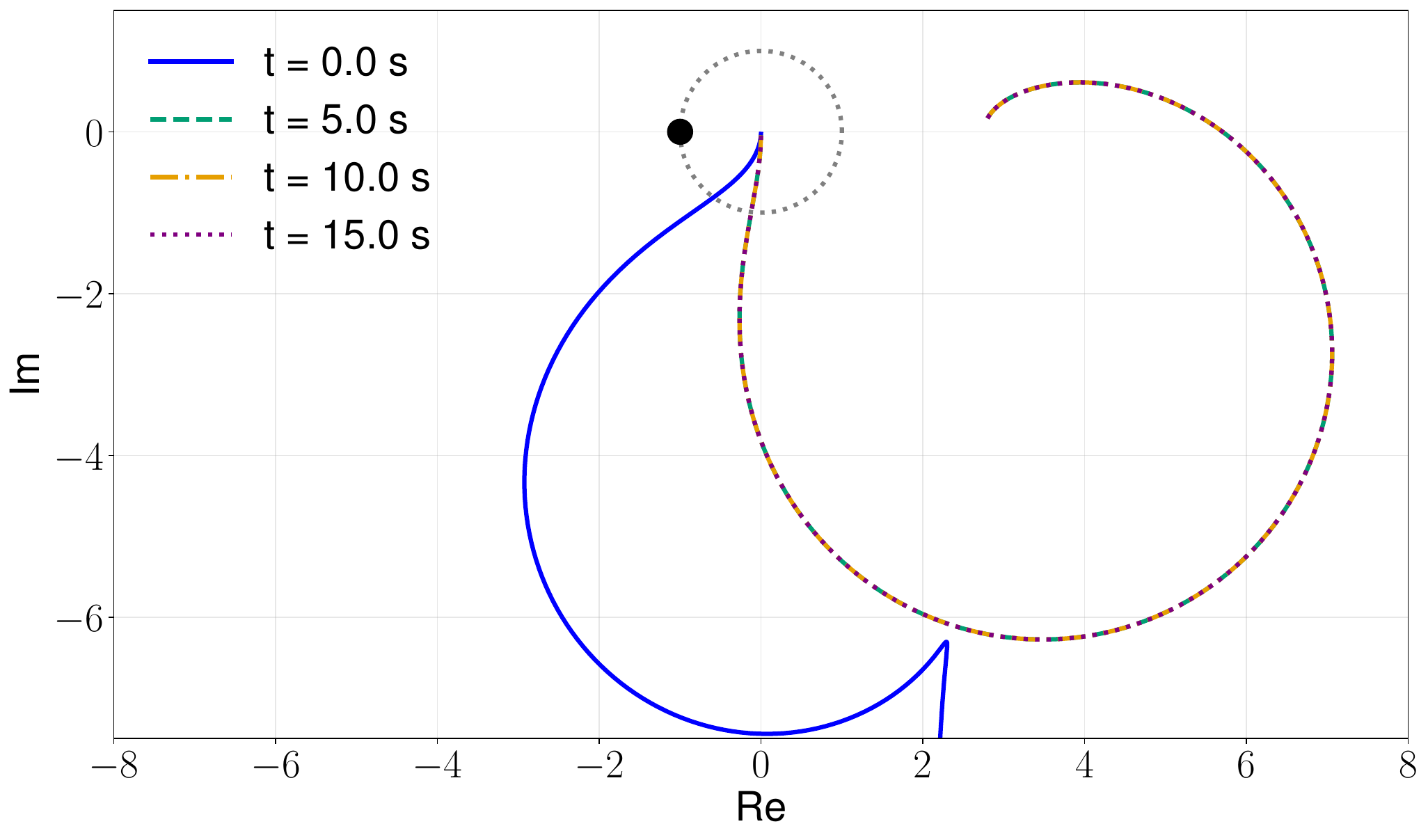}
    \subcaption{Nyquist plot for $[0, 20]$\,s.}
    \label{fig:stable_nyquist_0_20}
\end{minipage}
\hspace{0.01\linewidth}
\begin{minipage}[t]{0.48\linewidth}
\centering
    \includegraphics[clip,bb= 6 12 977 573,width=\columnwidth]{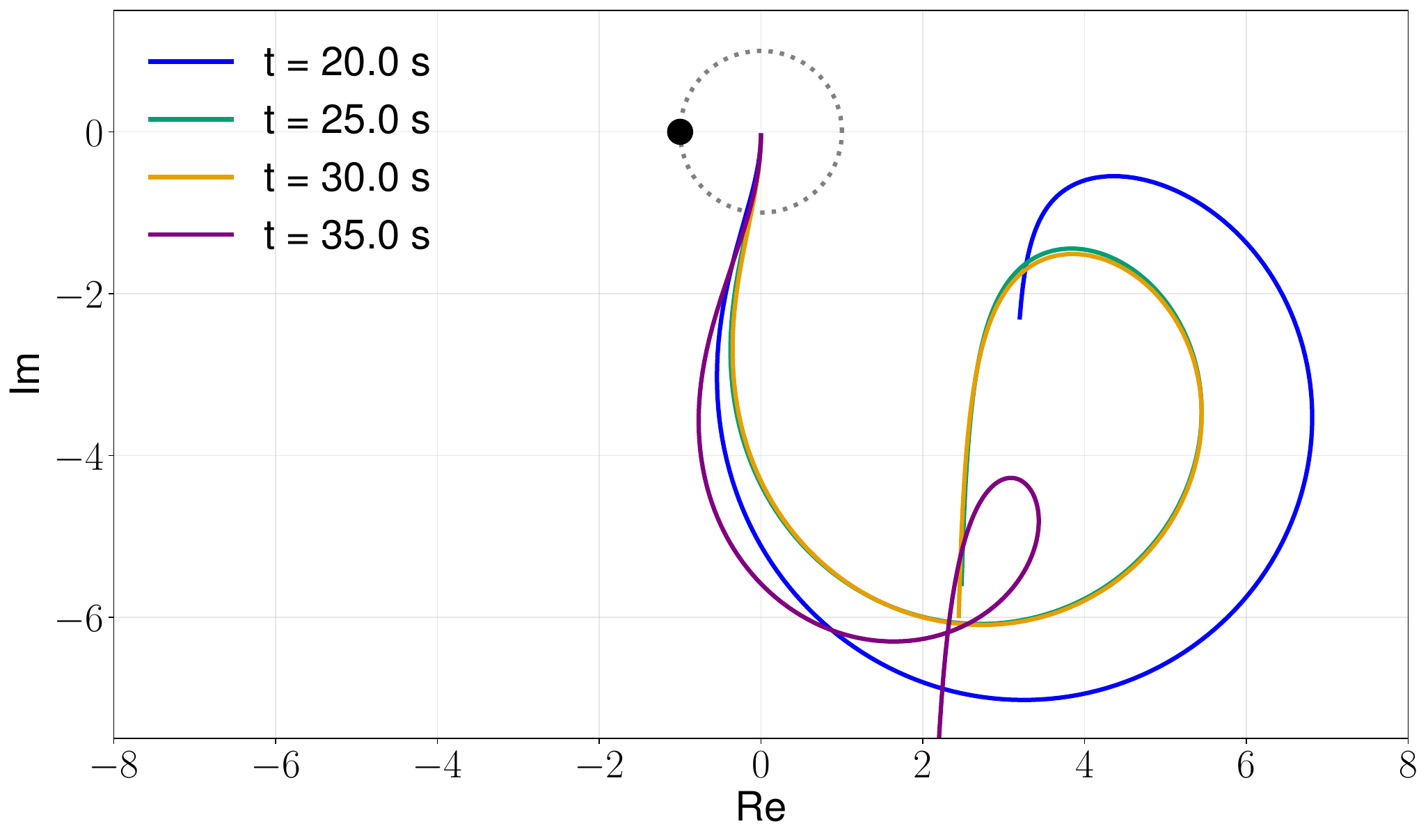}
    \subcaption{Nyquist plot for $[20, 40]$\,s.}
    \label{fig:stable_nyquist_20_40}
\end{minipage}
\\ \smallskip
\begin{minipage}[t]{0.48\linewidth}
\centering
    \includegraphics[clip,bb=6 12 977 573,width=\columnwidth]{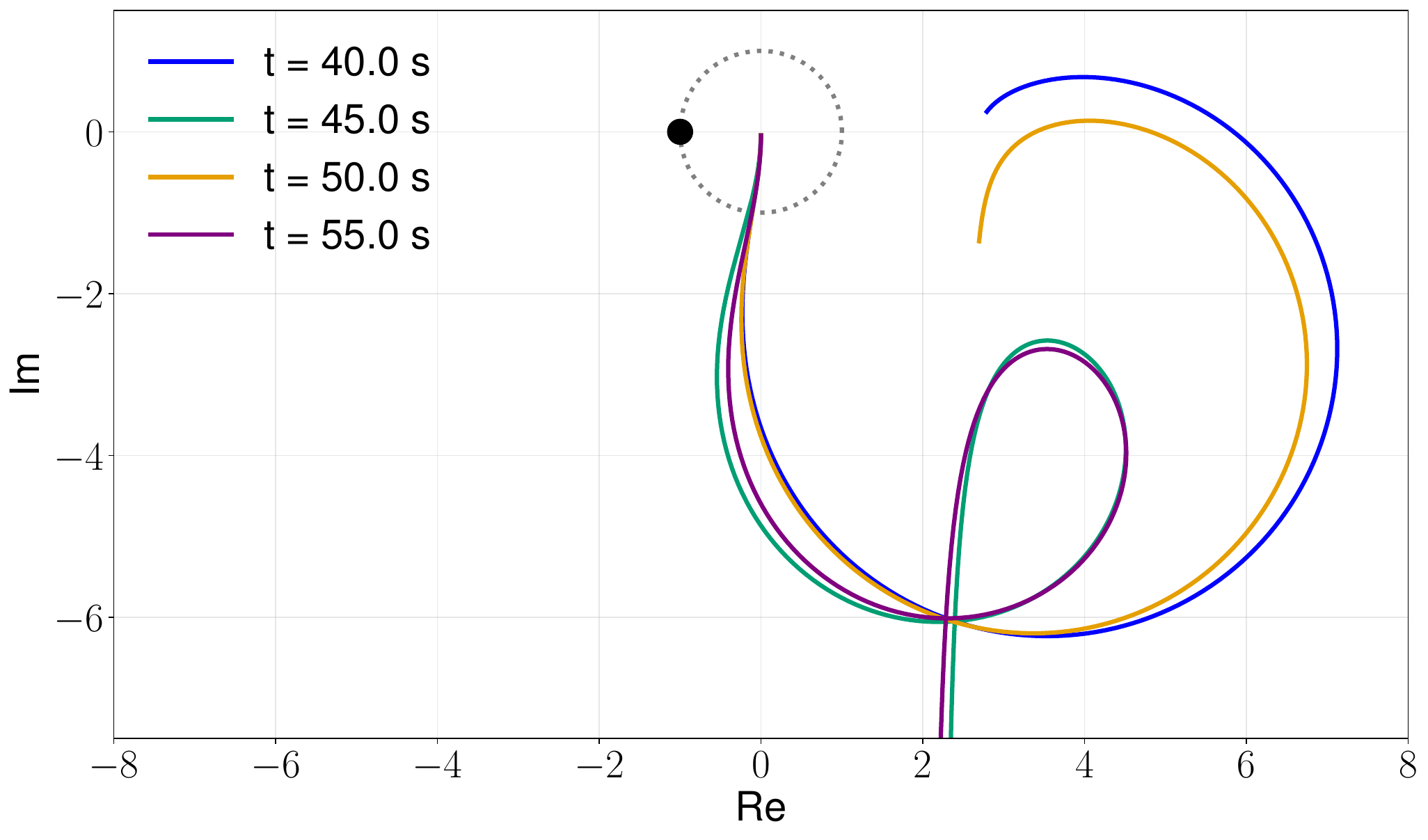}
    \subcaption{Nyquist plot for $[40, 60]$\,s.}
    \label{fig:stable_nyquist_40_60}
\end{minipage}
\hspace{0.01\linewidth}
\begin{minipage}[t]{0.48\linewidth}
\centering
    \includegraphics[clip,bb=6 7 1017 626,width=\columnwidth]{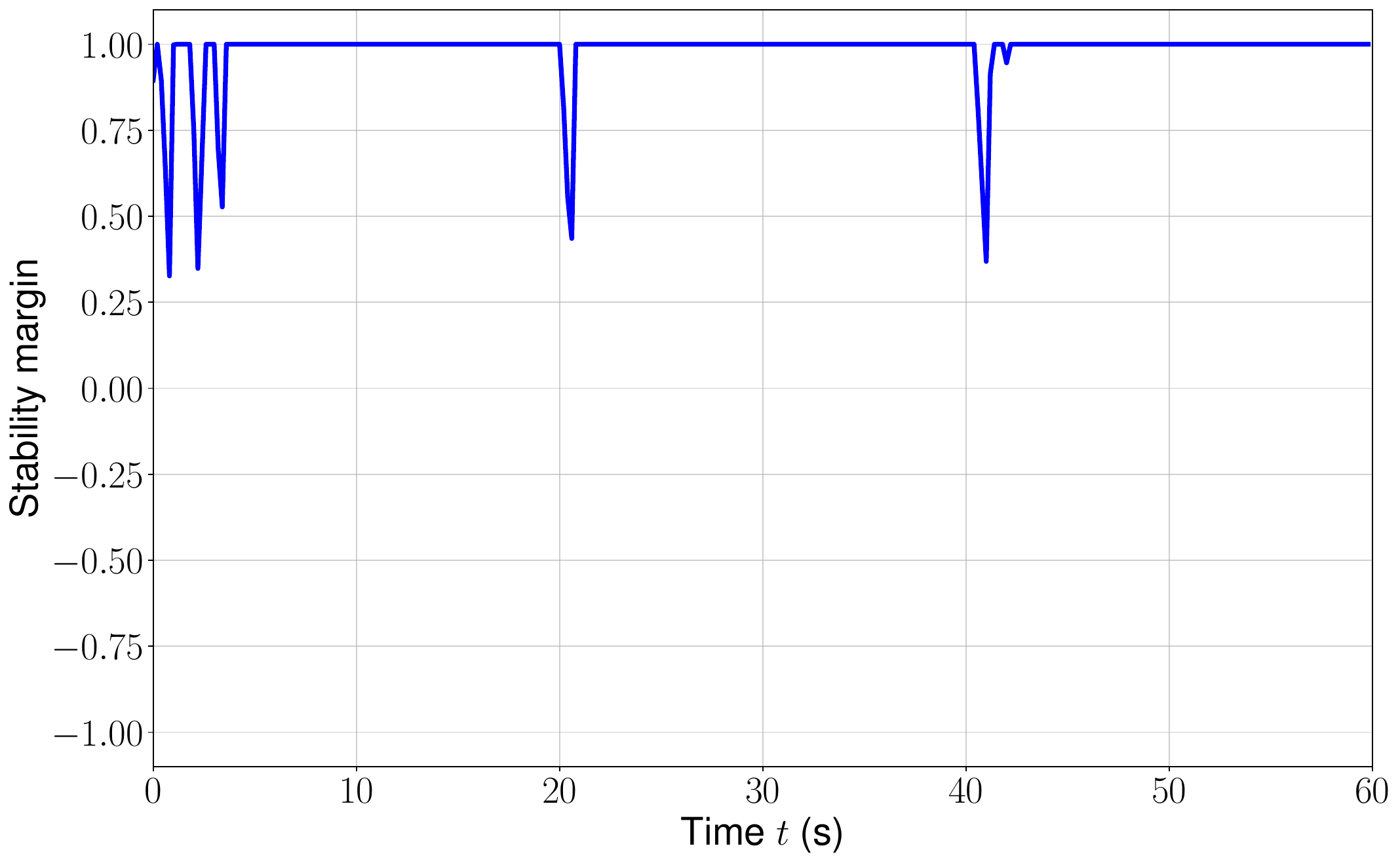}
    \subcaption{Temporal evolution of stability margins.}
    \label{fig:stable_margins}
\end{minipage}
\caption{Stability analysis of the stability-constrained adaptive PID controller: (a)--(c) Nyquist plots at different time intervals and (d) time-varying stability margins during servo control.}
\label{fig:stability_analysis_stable}
\end{figure}

\begin{figure}[!t]
\centering
\includegraphics[bb=7 7 1098 668,width=0.7\columnwidth]{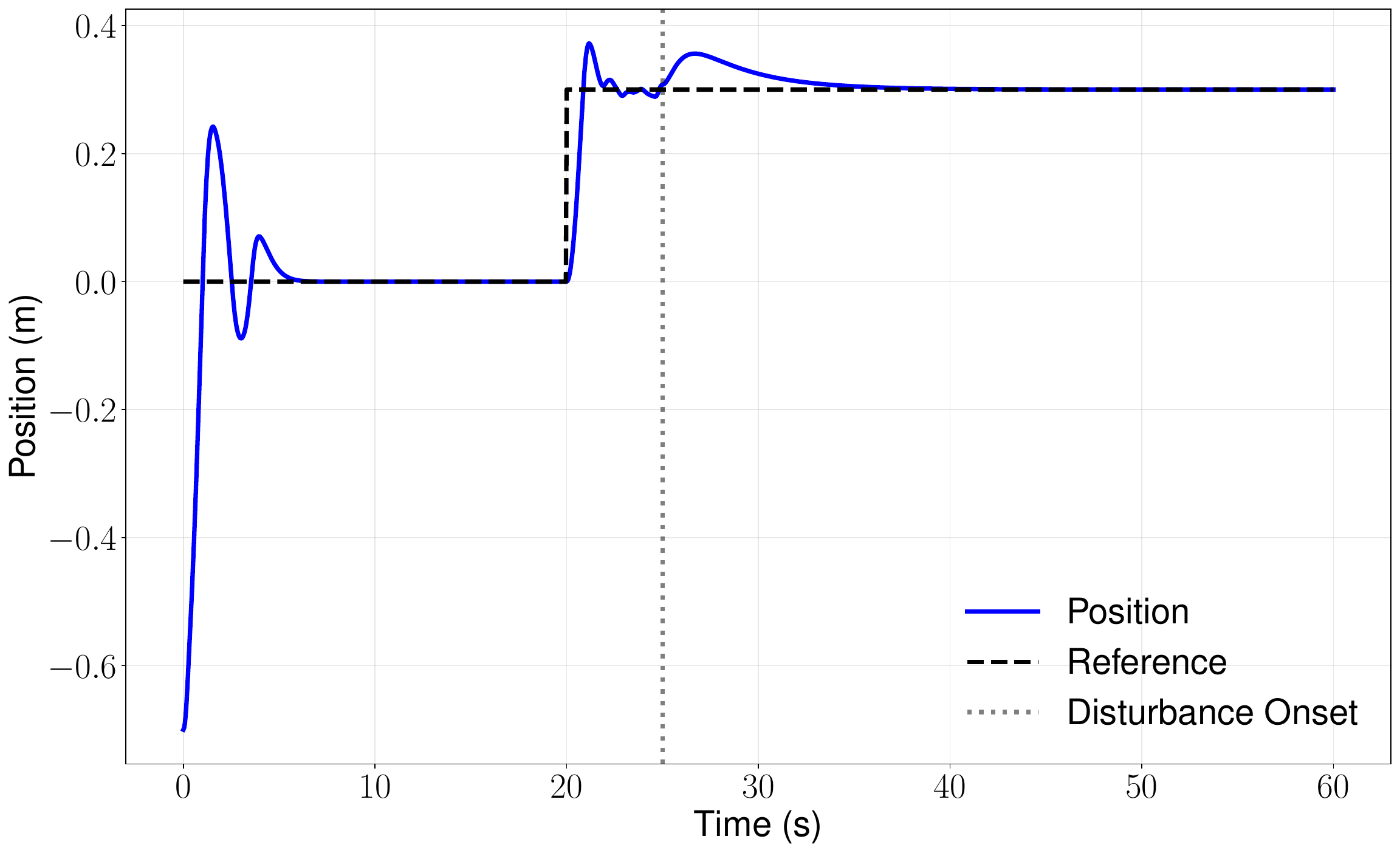}
\caption{System response with stability-constrained gains under constant disturbance. Time-varying gains ($t \leq 25$\,s) followed by frozen gains at $t=25$\,s with disturbance $d(t)=0.3$ ($t > 25$\,s).}
\label{fig:stable_disturbance_response}
\end{figure}

Under these circumstances, the resulting control performance, led by the reformulated PID gain optimization problem (\ref{prob:proposed_PID}) with the modified term (\ref{eq:modify_cost_theta}), is shown in Figs.~\ref{fig:stable_servo_results} and \ref{fig:stability_analysis_stable}. 
The time evolution of the stability margin in Fig.~\ref{fig:stable_margins} reveals that the obtained controller satisfies the stability-guaranteed constraint (\ref{eq:stability_condition-1}).
As shown in Fig.~\ref{fig:stable_servo_results}, the stability-constrained adaptive PID controller demonstrates superior tracking performance compared to the fixed-gain PID controller, consistent with the results in Section~\ref{sec:MSD_servo}. 
Moreover, when compared with the unconstrained adaptive case shown in Fig.~\ref{fig:result_MSD_servo}, the stability-guaranteed formulation achieves better tracking accuracy while maintaining guaranteed closed-loop stability.
From Fig.~\ref{fig:stable_gains}, the proportional and derivative gains maintain larger values throughout the entire time domain compared to the unconstrained case. 
The stability-guaranteed gains increase proportional and derivative action, resulting in an average gain crossover frequency of $3.27$\,rad/s, approximately $2.8$ times higher than in the unconstrained case.
The higher gain crossover frequency indicates increased closed-loop bandwidth, while the larger derivative gain introduces phase lead compensation, both contributing to the improved tracking performance observed in Fig.~\ref{fig:stable_tracking_error}.
For non-origin reference positions, the integral gain increases to eliminate steady-state errors, similar to the unconstrained approach, but within stability bounds.
Fig.~\ref{fig:stability_analysis_stable} presents the stability analysis for the stability-constrained case.
The Nyquist plots in Figs.~\ref{fig:stable_nyquist_0_20}--\ref{fig:stable_nyquist_40_60} demonstrate that the open-loop frequency response maintains a safe distance from the critical point $-1$ throughout the operation, in contrast to the unconstrained case shown in Fig.~\ref{fig:stability_analysis}.
Fig.~\ref{fig:stable_margins} confirms that the stability margin remains close to 1.0 for almost the entire time domain, confirming that the optimization successfully maintains sufficient stability margins while achieving the control objectives.
To verify the robustness of the stability-constrained controller, we conducted the same disturbance rejection test as in Fig.~\ref{fig:disturbance_response}. 
As shown in Fig.~\ref{fig:stable_disturbance_response}, the system maintains stability even under the constant disturbance, in stark contrast to the divergent behavior observed in the unconstrained case.

These results suggest that properly constraining the PID gain search space to ensure stability can significantly enhance the performance of optimization-based gain design.
While the Routh-Hurwitz criterion provides analytical stability-guaranteed constraints for linear systems, such as the MSD model, deriving these constraints for general nonlinear systems remains challenging, as demonstrated by the manipulator dynamics in Section~\ref{sec:05}, where analytical stability conditions are typically intractable.
This limitation presents a significant challenge that must be addressed for the broader application of the proposed method to complex nonlinear systems, particularly when explicitly considering stability.

\section{Conclusion}
\label{sec:07}

This paper has proposed a novel model-based adaptive PID control using PINNs, which is a promising approach to data-driven system modeling. 
The proposed control enables a gradient-based online PID gain optimization that explicitly accounts for the nonlinearities of the system dynamics by utilizing automatic differentiation of PINNs.
Therefore, once a high-accuracy PINNs model of the system is obtained, the proposed control method can easily design a PID controller widely used in industry, while considering the structure of standard or conventional control design approaches.
The effectiveness of stability, convergence, and stability margin of the proposed control method is illustrated through several simulations. 
In particular, the proposed method was confirmed to have advantages in control performance for equilibrium points that require inputs to overcome the nonlinearity of the system.

One possible direction for future work is to extend the handleable classes of the proposed control method. 
From the system structure viewpoint, the proposed control method is extended to observable and controllable systems with limited sensors, i.e., the output feedback systems. 
Another direction is to theoretically guarantee stability and robustness for the closed-loop systems using PINNs.

\begin{IEEEbiography}[{\includegraphics[width=1in,height=1.25in,clip,keepaspectratio,bb=0 0 576 769]{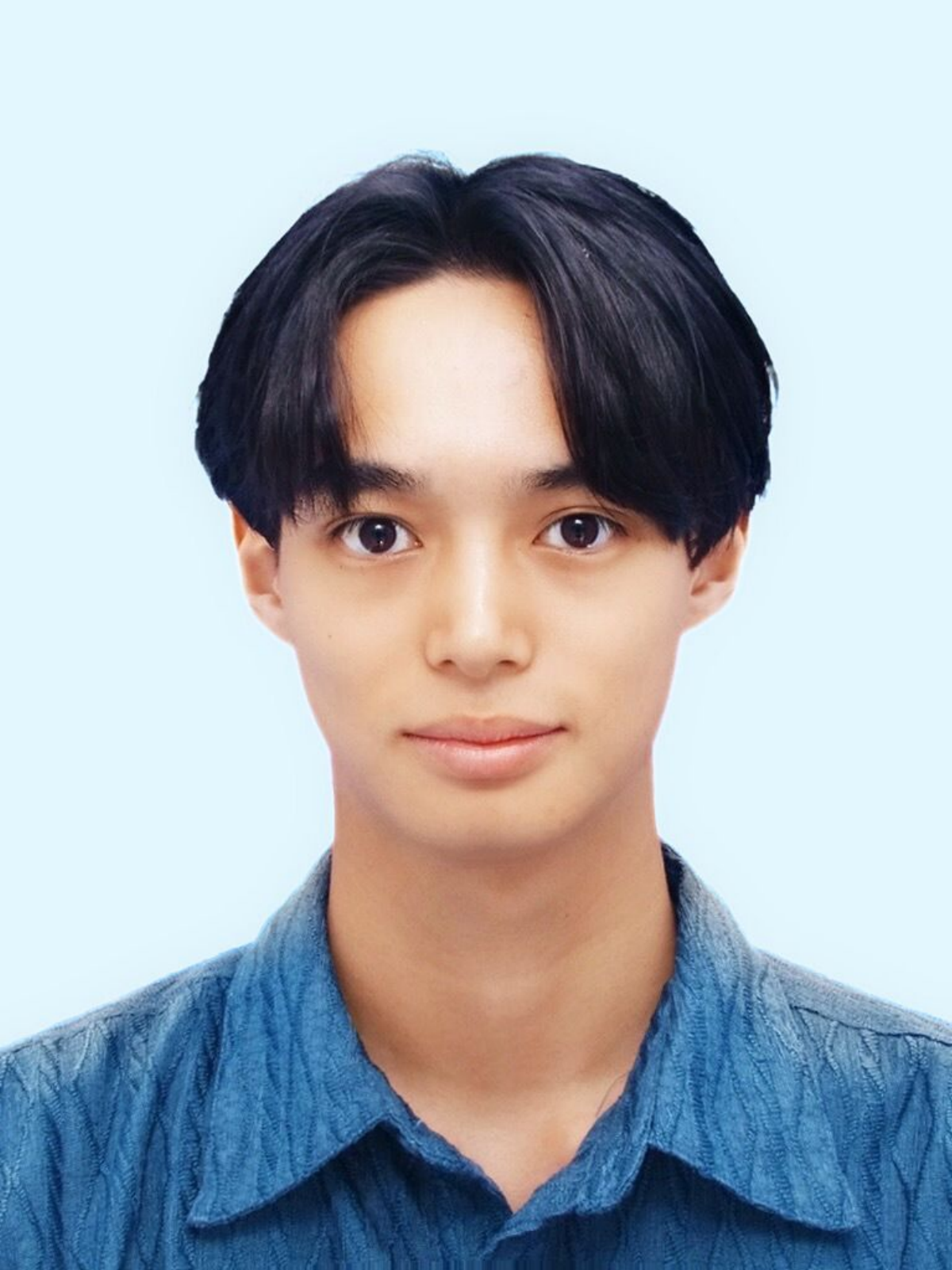}}]{Junsei Ito} received the B.Eng. degree in electrical engineering and bioscience from Waseda University, Tokyo, Japan in 2025. 
He is currently a master's student with the Department of Electrical Engineering and Bioscience, Waseda University. 
He is also a Japan Science and Technology Agency ACT-X researcher.
His research interests include machine learning based control theory and its applications. 
\end{IEEEbiography}

\begin{IEEEbiography}[{\includegraphics[width=1in,height=1.25in,clip,keepaspectratio,bb=0 0 198 247]{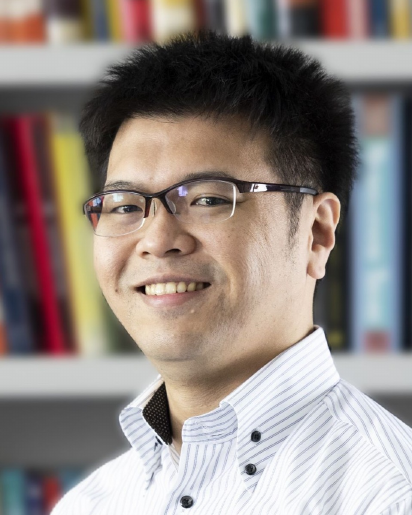}}]{Yasuaki Wasa} (Member, IEEE) received the B.Eng. degree in control and systems engineering and the M.Eng. and Ph.D.~degrees in mechanical and control engineering from Tokyo Institute of Technology, Tokyo, Japan, in 2011, 2013, and 2016, respectively. He has been an Associate Professor in the Department of Electrical Engineering and Bioscience at Waseda University, Tokyo, since April 2024. Prior to this, he was a Research Fellow with the Japan Society for the Promotion of Science, where he served as a Junior Researcher and Assistant Professor in 2016. 
He is the co-editor of Economically-Enabled Energy Management (Springer Nature, 2020).
His research interests include dynamic market mechanisms, distributed learning in smart grids, and cyber-physical human systems. He was a recipient of the Outstanding Paper Award (2015, 2023, 2025), the Control Division Pioneer Award (2025), and the Young Author Award (2018), all from the Society of Instrumental and Control Engineers (SICE). 
He was awarded the Best Student Paper Award Finalist in the 2014 IEEE Multi-Conference on Systems and Control and the Asian Control Conference Best Paper Award Finalist (2019, 2022). 
He is serving as an AE for the Asian Journal of Control and the SICE Journal of Control, Measurement, and System Integration.
\end{IEEEbiography}

\end{document}